\documentclass[12pt, a4paper]{article}
\usepackage{amsmath,amssymb}
\usepackage{graphicx}
\usepackage{bm}
\usepackage{cite}
\usepackage{color}
\usepackage{url}
\usepackage{float} 
\usepackage{booktabs}
\usepackage[normalem]{ulem}
\definecolor{blue}{rgb}{0.27, 0.42, 0.81}
\usepackage{longtable}
\usepackage{amsmath}
\usepackage{slashed}
\usepackage{ulem}
\usepackage{comment}
\usepackage{appendix}
\usepackage{ulem}
\usepackage[
colorlinks=true,
urlcolor=blue,
citecolor=blue,
linkcolor=red,
linktocpage=true,
]{hyperref}
\definecolor{mygray}{gray}{0.8}

\renewcommand{\thefootnote}{\fnsymbol{footnote}}
\usepackage[a4paper, top=1.2in, bottom=1.2in, left=0.9in, right=0.7in]{geometry}
\newcommand{\nn}{\nonumber}

\newcommand{\be}{\begin{equation}}
\newcommand{\ee}{\end{equation}}
\newcommand{\bea}{\begin{eqnarray}}
\newcommand{\eea}{\end{eqnarray}}

\def\mLb{{m_{\Lambda_b}}}
\def\mmLb{{m^2_{\Lambda_b}}}
\def\mL{{m_{\Lambda}}}
\def\mmL{{m^2_{\Lambda}}}

\def\mC{{\mathcal{C}}}

\def\Lb{{\Lambda_b}}
\def\Lst{{\Lambda^\ast}}

\allowdisplaybreaks
\renewcommand{\thefootnote}{\#\arabic{footnote}}
\begin{document}
\sloppy
\begin{center}
\vspace*{5cm}
{\LARGE\bf
\boldmath
$\Lambda_b \to \Lambda \ell^+\ell^-$ and $\Lambda_b \to \Lambda^{\ast} \ell^+\ell^-$ decays with polarized leptons
}

\vspace*{1.2cm}
{\bf Diganta Das $^{a}$, Vedant Parekh $^{a}$, Ria Sain $^{b,*}$, Dargi Shameer $^{a,*}$}
\\\vspace*{.5cm}
$^{a}$ Center for Computational Natural Sciences and Bioinformatics, International Institute of Information
Technology, Hyderabad 500 032, India\\\vspace*{.2cm}
$^{b}$Institute of Particle Physics and Key Laboratory of Quark and Lepton Physics (MOE), Central China Normal University, Wuhan, Hubei 430079, China\\\vspace*{.2cm}
\end{center}
{\let\thefootnote\relax
 \footnotetext{$^{*}$Corresponding authors: \texttt{ria.sain.2013@gmail.com} (Ria~Sain), \texttt{sha13meer@gmail.com} (D.~Shameer)}}

\vspace*{5mm}
\begin{abstract}
We present an analysis of lepton polarization observables in the rare flavor-changing neutral current decays $\Lambda_b \to \Lambda\ell^+\ell^-$ and $\Lambda_b \to \Lambda^*\ell^+\ell^-$. These decays are driven by the quark-level transition $b \to s\ell^+\ell^-$ and are sensitive to short-distance physics within and beyond the Standard Model. We work within an effective Hamiltonian framework and derive single lepton polarization asymmetries, double lepton polarization asymmetries, and polarized forward-backward asymmetries using the helicity method. The observables are expressed in terms of transversity amplitudes and evaluated using state-of-the-art lattice QCD form factors. We analyze the $q^2$ dependence of several observables for both the ground-state $\Lambda$ and the excited $\Lambda^*(1520)$ final states and discuss sensitivities of the observables to  chiral structure of the effective operators and complex Wilson coefficients.

\end{abstract}
\newpage

\section{Introduction}
Rare flavor--changing neutral current (FCNC) decays of heavy hadrons provide some of the most sensitive probes of short--distance physics and possible effects beyond the Standard Model (SM). Such transitions are forbidden at tree level in the SM and arise only through loop--induced processes, making them highly suppressed and sensitive to physics beyond the Standard Model. Among these, the quark--level transition $b \to s \ell^+ \ell^-$ has been extensively studied in both inclusive and exclusive channels, owing to its strong sensitivity to new physics (NP) encoded in the Wilson coefficients of the effective weak Hamiltonian~\cite{Buchalla:1995vs,Ali:1999mm}.

In recent years, rare semileptonic $B$--meson decays such as $B \to K^{(*)}\ell^+\ell^-$ have attracted considerable attention, motivated both by precise experimental measurements and by observed tensions with SM expectations in branching ratios and angular observables. Several observables in these FCNC transitions deviate from the Standard Model predictions. For instance, the branching ratio for $B_s \to \phi\mu^+\mu^-$ shows a $3.6\sigma$ deviation \cite{LHCb:2015wdu,LHCb:2021zwz}, while the angular observable $P_5'$ in $B \to K^*\mu^+\mu^-$ shows local deviations of about $3\sigma$ in the intermediate $q^2$ region \cite{LHCb:2015svh,LHCb:2020lmf,LHCb:2013ghj}. The deviations have been interpreted as potential Beyond the Standard Model (BSM) physics in $b \to s\mu^+\mu^-$ processes. Unlike the older measurements of the Lepton-Flavor-Universality (LFU) ratios $R_K$ and $R_{K^*}$, LHCb's updated results \cite{LHCb:2022vje, LHCb:2022qnv} are consistent with Standard Model predictions, providing no evidence of LFU violation. To explain the anomalies, various theoretical extensions beyond the SM (BSM) have been proposed. Some of the well-explored BSM theories include supersymmetry \cite{Haber:2017aci,Kazakov:2000ra,Ali:1999mm}, $Z'$ models \cite{Nasrullah:2018puc}, two-Higgs-doublet models \cite{Hu:2017qxj}, and leptoquarks \cite{Das:2016vkr}. Model-independent interpretations based on effective field theory have also been proposed \cite{Khan:2025rgc,Das:2020cpv,Das:2018sms,Alok:2010zd,Alok:2011gv,Descotes-Genon:2013wba,Hurth:2013ssa,Datta:2019zca,Kumar:2019qbv,Alok:2019ufo,Carvunis:2021jga,Alguero:2021anc,Alguero:2023jeh,Geng:2021nhg,Hurth:2021nsi,Ishaq:2013toa,Bhutta:2024zwj,Salam:2024nfv,Farooq:2024owx,Aarfi:2025qcp,Angelescu:2021lln,Alok:2022pjb,Ciuchini:2022wbq,SinghChundawat:2022ldm,Munir:2015gsp,Alok:2023yzg,Alok:2024cyq,MunirBhutta:2020ber,Altmannshofer:2021qrr}.

If there is any New Physics in $b \to s \ell^+ \ell^-$ transitions, it should appear in other decay modes driven by the exact same quark-level process. The baryonic decays $\Lambda_b \to \Lambda(1116) \ell^+ \ell^-$ and $\Lambda_b \to \Lambda^\ast(1520) \ell^+ \ell^-$ are prime examples. The $\Lambda_b$ baryon decays have been widely studied in the literature \cite{CDF:2011buy,Detmold:2012vy,Das:2018iap,Wang:2008ni,Beneke:2004dp,Boer:2014kda,Feldmann:2011xf,Yan:2019tgn,Roy:2017dum,Das:2023kch,Mahmoudi:2026aul,Khan:2025rgc}. The LHCb Collaboration reported the first observation of the $\Lambda_b^0 \to p K^- \mu^+ \mu^-$ decay, together with a search for CP violation in this mode\cite{LHCb:2017slr}. The LHCb also tested Lepton Flavor Universality (LFU) in $\Lambda_b^0 \to p K^- \ell^+ \ell^-$ decay, and found that the results agree with Standard Model predictions \cite{LHCb:2019efc}. One advantage of the $\Lambda_b \to \Lambda (\to N\pi) \ell^+ \ell^-$ cascade decay is that even if the initial $\Lambda_b$ state isn't polarized, the spin of the daughter $\Lambda$ baryon helps us probe the helicity structure of the weak effective Hamiltonian \cite{Mannel:1997xy,Hiller:2007ur,Wang:2008sm,Chen:2001ki,Gutsche:2013pp,Gutsche:2013oea,Nasrullah:2018vky,Nasrullah:2020glp,Detmold:2016pkz}. 

These decays also provide new angular coefficients in the full decay chain, offering deeper insight into the underlying kinematics. High-precision lattice QCD calculations for the $\Lambda_b \to \Lambda$ form factors were carried out in \cite{Detmold:2016pkz}, which also provided Standard Model predictions for the differential branching fraction and various angular observables in the four-fold $\Lambda_b \to \Lambda(\to p^+ \pi^-) \mu^+ \mu^-$ process (assuming unpolarized initial $\Lambda_b$ and unpolarized final muons). The full angular distribution for cases where the initial $\Lambda_b$ itself is polarized has also been thoroughly explored \cite{Blake:2017une,Das:2020qws}. Beyond the ground--state $\Lambda$ baryon, decays of the $\Lambda_b$ into excited strange baryon resonances, $\Lambda^*$, have also been investigated \cite{Das:2020cpv,Descotes-Genon:2019dbw}. These modes are particularly interesting as they probe different hadronic structures and form factor dynamics, and allow for a richer set of angular and polarization observables. Theoretical studies have examined the decay $\Lambda_b \to \Lambda^*(J^P)\ell^+\ell^-$ for various spin--parity assignments, highlighting the role of heavy quark symmetry and form factor relations in controlling theoretical uncertainties\cite{Das:2020cpv,Descotes-Genon:2019dbw,Mott:2015zma}.

While decay rates and angular distributions provide important constraints, polarization observables offer complementary and sometimes cleaner probes of the underlying dynamics. In particular, the polarization of the final--state leptons is directly sensitive to the chiral structure of the effective operators and can help disentangle different NP scenarios\cite{Aliev:2002hj}. Lepton polarization asymmetries in $b \to s \ell^+ \ell^-$ transitions have been investigated in inclusive and exclusive decays as theoretically clean observables with enhanced sensitivity to scalar, pseudoscalar, and tensor interactions\cite{Aliev:2004yf,Aliev:2004af,Aliev:2012ac,Bensalem:2002ni,Zolfagharpour:2007eh,Kamenik:2017ghi}. Subsequent studies demonstrated that single-lepton polarization asymmetries, double--lepton polarization asymmetries, and polarized forward--backward asymmetries can provide independent constraints on the Wilson coefficients, often with reduced dependence on hadronic form factors\cite{Aliev:2002hj,Aliev:2004af}. These observables have been explored in detail for mesonic decays, including $B \to K^{(*)}\ell^+\ell^-$, establishing their importance as precision tools in flavour physics\cite{Cornell:2004cp}. The baryonic sector offers additional opportunities to study polarization observables due to the intrinsic spin of the initial and final baryons. In particular, the decay $\Lambda_b \to \Lambda \ell^+ \ell^-$ has been widely studied as a promising channel to access both lepton and hadron polarization effects \cite{Aslam:2008hp,Aliev:2002hj,Aliev:2002tr,Aliev:2006xd,Aliev:2005yz,Bashiry:2007pd,Aliev:2004yf,Aliev:2012ac}. Several works have analyzed differential decay rates, forward--backward asymmetries, and single and double lepton polarization asymmetries in this mode within the SM and its extensions, showing that baryonic decays can provide complementary information to mesonic channels and are sensitive to a broad class of NP interactions\cite{Aliev:2002hj}; lepton polarization asymmetries have also been computed for $\Lb$ decays into several excited $\Lst(J^P)$ states \cite{Mott:2015zma}. 

In this work, we present an analysis of lepton polarization observables in the rare decays $\Lambda_b \to \Lambda \ell^+\ell^-$ and $\Lambda_b \to \Lambda^*(1520) \ell^+\ell^-$ and perform a numerical analysis for muons. To the best of our knowledge, the lepton polarization observables for $\Lambda_b \to \Lambda^* \ell^+\ell^-$ are presented here for the first time. The $\Lambda^*$ channel differs from the $\Lambda$ channel in some structural aspects. The $\Lambda$ has negative parity, and the spin-3/2 final state introduces an additional set of amplitudes and from the $f_g$ form factors. The richer helicity structure of the spin-$3/2$ final state leads to different interference patterns among the transversity amplitudes, allowing the observables to probe combinations of Wilson coefficients that are complementary to those accessible in the $\Lambda$ mode.  An important point to note is that the charged-lepton polarization is reconstructed from the angular distribution of the lepton's decay products. This makes it difficult to measure muon polarization at the LHCb. We therefore present muon results as a theoretical benchmark. We compute single and double lepton polarization asymmetries, as well as polarized forward--backward asymmetries, within an effective Hamiltonian framework describing the $b \to s \ell^+ \ell^-$ transition. The paper is organized as follows: In Sec.~\ref{subsec:effHam} we give the effective Hamiltonian for $b\to s\ell^+\ell^-$ transition. In Sec.~\ref{sec:kin} we describe the decay kinematics and give the definitions of the leptonic amplitudes. In Sec.~\ref{sec: intro} the definitions of observables are introduced. We perform the numerical analysis of  $\Lambda_b \to \Lambda \ell^+\ell^-$ in Sec.~\ref{subsec:Lb2L}, $\Lambda_b \to \Lambda^* \ell^+\ell^-$ in Sec.~\ref{subsec:Lb2Lst}, and the $q^2$-binned predictions are collected in Tables \ref{tab:LbToL-bin-avg} and \ref{tab:LbToLSt-bin-avg}. The hadronic matrix elements, form factors and the observables are defined in the appendices.

\section{Effective Hamiltonian \label{subsec:effHam}}
We work in an effective field theory framework where the Hamiltonian for the $b\to s\ell^+\ell^-$ transition is
\be\label{eq:hamiltonian}
H_{\rm eff} = -\frac{4G_F}{\sqrt{2}} V_{tb} V_{ts}^{\ast}\frac{\alpha_e}{4\pi} 
\left( 
  \sum_{i}  C_i O_i + \sum_{j} C_j^{\prime} O_j^{\prime}
\right)\, .
\ee
Here $G_F$ is the Fermi constant, $\alpha_e=e^2/4\pi$ is the fine structure constant, and $V_{tb}, V_{ts}^\ast$ are the Cabibbo-Kobayashi-Maskawa matrix elements, and $P_{L(R)} = (1\mp \gamma_5)/2$ are the chirality projectors. In the Standard Model the dominant operators contributing are $\mathcal{O}_{7,9,10}$, whereas $\mathcal{O}_{7,9,10}^\prime$ are the chirality-flipped counterparts that may be present in beyond the SM scenarios. The Dirac structures of the operators are
\be
\mathcal{O}_{7}^{(\prime)} = \frac{m_b}{e} (\bar{s} \sigma_{\mu\nu} P_{R(L)} b) F^{\mu\nu},\, \quad \mathcal{O}_{9}^{(\prime)}=(\bar{s}\gamma^{\mu}P_{L(R)}b)(\bar{\ell} \gamma_{\mu}\ell),\, \quad
\mathcal{O}_{10}^{(\prime)}=(\bar{s}\gamma^{\mu}P_{L(R)}b)(\bar{\ell} \gamma_{\mu}\gamma_5\ell)
\ee
The short-distance physics associated with the operators are denoted by $C_i^{(\prime)}$ where the primed coefficients vanish in the absence of NP. NP with same Dirac structures as $\mathcal{O}_{9,10}$ may also appear, but can be trivially included in our analysis by ${C}_{9,10}\to {C}_{9,10} + {C}_{9,10}^{\rm NP}$. In the SM $C_7(m_b) \simeq -0.3$, and $C_9(m_b)\approx - C_{10}(m_b)\simeq  4.2$. Note that in the SM four-quark operators related to $b\to sq\bar{q}$ as well as the $b\to sg$ chromomagnetic operator $\mathcal{O}_8$ are also present \cite{Buchalla:1995vs,Grinstein:1988me,Chetyrkin:1996vx}. Their contributions are usually absorbed in the definition of the $C_{7,9}$ itself by redefining $C_{7, 9} \to C_{7, 9}^{\rm eff}$ \cite{Grinstein:1988me, Buchalla:1995vs, Chetyrkin:1996vx}. For the expressions of $C_{7,9}^{\rm eff}$ and the values of other SM Wilson coefficients used in our numerical analysis, we refer to \cite{Das:2022xjg}. For notational simplicity we denote the Wilson coefficients by $C_{7, 9}$. The hadronic matrix elements of the operators between the initial and final state hadron are expressed in terms of form factors detailed in Appendix \ref{eq:tas}.

\section{Theoretical Framework \label{sec:kin}}
\subsection{Kinematics and spin projection}
In this section we describe the theoretical framework of calculating observables characterizing polarized on-shell massive leptons. We work in a frame where the parent $\Lambda_b$ is at rest, the $\ell^+\ell^-$ pair moves in the $+z$ direction, and the $\Lambda^{(\ast)}$ moves in the $-z$ direction. Note that from this point onward, unless otherwise mentioned, $\Lambda$ stands for both types of daughter hadron, $\Lambda,\Lambda^\ast$. In this frame, the four-momentum of parent $\Lambda_b$, daughter $\Lambda$ and the $\ell^+\ell^-$ pair respectively are
\be
p^\mu|_{\Lambda_b} = (\mLb,0,0,0),\quad k^\mu |_{\Lambda_b} = ((\mLb-q^0),0,0,-|q|)\, ,\quad q^\mu |_{\Lambda_b} = (q^0,0,0,|q|)
\ee
where 
\be
q^0 \big|_{\Lambda_b} = \frac{\mLb^2 - \mL^2 + q^2}{2\mLb}\, ,\quad\quad |q| \big|_{\Lambda_b} = \frac{\sqrt{\lambda(\mLb^2, \mL^2, q^2)}}{2\mLb}\, .
\ee
Here $\lambda(a,b,c) = a^2+b^2+c^2 - 2(ab + bc + ca)$. In the $\ell^+\ell^-$ rest frame the four-momentum of the $\ell^-$, $\ell^+$, and $\Lambda$ are 
\bea
p^\mu_-\big|_{\ell^+\ell^-} &=& \bigg(E_\ell,|q_{2\ell}|\sin{\theta_\ell},0,|q_{2\ell}|\cos{\theta_\ell}\bigg)\ \\
p^\mu_+\big|_{\ell^+\ell^-} &=& \bigg(E_\ell,-|q_{2\ell}|\sin{\theta_\ell},0,-|q_{2\ell}|\cos{\theta_\ell}\bigg)\  \\
k^\mu \big|_{\ell^+\ell^-} &=& \bigl(E_{\Lambda},0,0,-|\vec{p}| \bigr) 
\eea
where $q_{2\ell}$ is given as
\be
q_{2\ell} = \beta_{\ell}\frac{\sqrt{q^2}}{2} ,\quad \beta_{\ell} =\sqrt{1-\frac{4m^2_\ell}{q^2}} , \quad E_\ell = \frac{\sqrt{q^2}}{2}
\ee
and $q^\mu = p_-^\mu + p_+^\mu$.

Each of the final state leptons can be either in longitudinal ($L$), transverse ($T$), or normal ($N$) polarization state. One can therefore construct observables with the final state leptons in either of these states. The polarizations are defined with respect to a spacelike orthonormal direction attached to the lepton. The directions are defined in terms of the momentum of daughter particles defined on the $\ell^+\ell^-$ rest frame. The lepton spinors are projected onto three spacelike directions
\bea\label{eq:sLNT}
s^{\mp \mu}_L = \bigg(0, \frac{\vec{p}_\mp }{|\vec{p}_\mp |}  \bigg)\, ,\quad s^{\mp \mu}_N = \bigg( 0, \frac{\vec{p} \times \vec{p}_\mp }{|\vec{p}\times \vec{p}_\mp |}   \bigg)\, ,\quad s^{\mp \mu}_T = \bigg( 0,  \frac{\vec{p} \times \vec{p}_\mp }{|\vec{p}\times \vec{p}_\mp|} \times  \frac{\vec{p}_\mp }{|\vec{p}_\mp |}    \bigg)
\eea
where $\vec{p}_-, \vec{p}_+$ and $\vec{p}$ are the 3-momentum of $\ell^-, \ell^+$ and $\Lambda$ defined in the $\ell^-\ell^+$ rest frame above. As the daughter hadron momentum $\vec{p}$ has only the $z$ component, the three vectors in the above definitions can be replaced by their corresponding unit vectors
\be\label{eq:unit-vec-10}
\hat{p}_- = \bigl(\sin\theta_{\ell},0,\cos\theta_{\ell} \bigr) \, ,\qquad
\hat{p}_+ = \bigl(-\sin\theta_{\ell},0,-\cos\theta_{\ell} \bigr) \, ,\qquad
\hat{p} = (0,0,-1)
\ee
The unit vectors \eqref{eq:unit-vec-10} are constructed from momenta in the $\ell^+\ell^-$ rest frame and define the spin quantization axes in the rest frame of each lepton, 
\be
s_{L}^{-\mu} = [0,\sin\theta _{\ell},0,\cos \theta_{\ell}] \, ,\qquad
s_{N}^{-\mu} = [0,0,-1,0] \, ,\qquad
s_{T}^{-\mu} = [0,-\cos\theta_{\ell},0,\sin\theta_{\ell}]  
\ee
Boosting these vectors to the  $\ell^+\ell^-$ rest frame, we obtain the $s^{-\mu}_{L,N,T}$ for the negatively charged lepton in the  $\ell^+\ell^-$ rest frame as
\be\label{eq:smi-mu}
 s_{L}^{-\mu} = \biggl[\beta_\ell\frac{\sqrt{q^2}}{2m_{\ell}} ,\frac{\sqrt{q^{2}}}{2m_{\ell}}\sin\theta_{\ell} ,0,\frac{\sqrt{q^{2}}}{2m_{\ell}}\cos\theta_{\ell}\biggr]  \, \qquad
 s_{N}^{-\mu} = [0,0,-1,0] \, ,\qquad 
 s_{T}^{-\mu} = [0,-\cos\theta_{\ell},0,\sin\theta_{\ell}]  \,.
\ee
As expected, only the $s^{-\mu}_L$ changes and the other two remains same. A similar calculation yields the following vectors for the positively charged lepton in the  $\ell^+\ell^-$ rest frame as
\be\label{eq:spl-mu}
s_{L}^{+\mu} = \biggl[\beta_\ell\frac{\sqrt{q^2}}{2m_{\ell}} ,\frac{-\sqrt{q^{2}}}{2m_{\ell}}\sin\theta_{\ell} ,0,\frac{-\sqrt{q^{2}}}{2m_{\ell}}\cos\theta_{\ell}\biggr]\, ,\quad
s_{N}^{+\mu} = [0,0,1,0] \,,\quad
s_{T}^{+\mu} = [0,-\cos\theta_{\ell},0,\sin\theta_{\ell}] \, .
\ee
For completeness, the Lorentz boost matrix is given in Appendix \ref{sec:LB}.

The four-vectors $s_{L,N,T}^{\pm\mu}$ in Eqs. \eqref{eq:smi-mu} and \eqref{eq:spl-mu} can now be used to construct spin projection operators
\be
P_\pm({s}^{\pm}_A) = \frac{1}{2}(1 \pm  \gamma_5 \slashed{s}^{\pm}_A)\, ,
\ee
where $A=L,N,T$. Since $\slashed{s}^\pm \slashed{s}^\pm = -1$, the operators satisfy the following properties: $P_+(s_A^\pm)+P_-(s_A^\pm)=1$, $P_+(s_A^\pm)P_-(s_A^\pm)=0$ and $P^2=P$. The lepton spinors projected onto the spacelike directions are defined as follows
\bea
\bar{u}_{\ell^-} (\lambda^-, s_A^-) &=& \bar{u}_{\ell^-}(\lambda^-) P_+(s_A^-) \, ,\qquad
\bar{u}_{\ell^-} (\lambda^-, -s_A^-) = \bar{u}_{\ell^-}(\lambda^-) P_-(s_A^-)\, ,\nn\\
v_{\ell^+}(\lambda^+,s_A^+ ) & = & P_+(s_A^+) v_{\ell^+}(\lambda^+ ) \, ,\qquad
v_{\ell^+}(\lambda^+,-s_A^+ ) =  P_-(s_A^+) v_{\ell^+}(\lambda^+ ) \, ,
\eea
where $\lambda^\pm$ are spin labels used for the positively and negatively charged leptons. 

Using the spinors projected onto the three spacelike directions $L,N,T$ we calculate the decay amplitude. Assuming the lepton and hadronic parts can be factorized, the decay amplitude $\Lambda_b \to \Lambda^{(\ast)}\ell^+\ell^-$ 
%
can be split into leptonic and hadronic parts as 
\begin{eqnarray}
\mathcal{M}_{\Lambda_b}^{\lambda^-,\lambda^+}(s_{\Lambda_b},s_{\Lambda^{(*)}},s_A^-,s_B^+) &=& -\frac{4G_F}{\sqrt2}V_{tb}V^*_{ts}\frac{\alpha_e}{4\pi} \sum_{L(R)} \frac{1}{4}
\bigg( \sum_{\lambda}\eta_{\lambda} H_{VA,\lambda}^{L,(R)}(s_{\Lambda_b},s_{\Lambda^*}) L^{\lambda^-,\lambda^+}_{L(R),\lambda}(s^-_A,s^+_B) \bigg)\, .\nonumber
\end{eqnarray}
Here $\eta_t=1$ and $\eta_{\pm 1,0}=-1$. The hadronic helicity amplitude $H_{VA,\lambda}^{L,(R)}(s_{\Lambda_b},s_{\Lambda^*})$ and the leptonic helicity amplitude $L^{\lambda^-,\lambda^+}_{L(R),\lambda}(s^-_A,s^+_B)$ are calculated by taking projections of the full amplitudes onto the polarization direction of virtual gauge bosons decaying to the lepton pair. Denoting the polarization vector of the virtual gauge boson by $\bar{\epsilon}(a)$, where $a=0, \pm 1, t$ we write the lepton helicity amplitudes as
\bea
L_{L(R),\lambda}^{\lambda^-,\lambda^+}(s^-_A,s^+_B) &=& \bar{\epsilon}^\mu(\lambda) \langle \bar{\ell}(\lambda^-, s^-_A) \ell(\lambda^+, s^+_B) |\bar{\ell} \gamma_\mu (1\mp\gamma_5) \ell |0\rangle\, \nn \\& =& \bar{\epsilon}^\mu(\lambda) \bar{u} (\lambda^-, s^-_A) \gamma_\mu (1\mp\gamma_5)v(\lambda^+,s^+_B)\, .
\eea 
The details of hadronic helicity amplitudes and their calculations can be found in \cite{Das:2018sms} \cite{Das:2018iap} \cite{Das:2020cpv}. However, it is customary to express the modulus squared of the amplitudes in terms of transversity amplitudes which can be found in \cite{Das:2018sms,Das:2018iap,Das:2020cpv}. For $\Lambda_b\to\Lambda\ell^+\ell^-$ decay there are 12 transversity amplitudes $A^{L(R)}_{\perp 1,\parallel 1}$,$A^{L(R)}_{\perp 0,\parallel 0}$,$A^{L(R)}_{\perp t,\parallel t}$ which can be written in terms of 10 form factors $f^{V(A)}_{\perp, 0, t},f^{T(T5)}_{\perp, 0}$. For $\Lambda_b\to\Lambda^\ast\ell^+\ell^-$ decay there are 16 transversity amplitudes $B^{L(R)}_{\perp 1,\parallel 1}$,$A^{L(R)}_{\perp 1,\parallel 1}$,$A^{L(R)}_{\perp 0,\parallel 0}$,$A^{L(R)}_{\perp t,\parallel t}$ which can be written in terms of 14 form factors $f^{V(A)}_{\perp,0,t,g},f^{T(T5)}_{\perp,0,g}$. For numerical analysis the form factors have been taken from Lattice QCD calculation \cite{Meinel:2021mdj,Detmold:2016pkz}.



\section{Observables for $\Lambda_b\to\Lambda^{(\ast)}\ell^+\ell^-$} \label{sec: intro}
Based on the formalism we first calculate the differential decay rates $d\Gamma (s_A^-,s_B^+)/dq^2$ where the $\ell^-$ spin is projected onto the direction of $A$ and the $\ell^+$ spin is projected onto the direction of $B$. From the differential rate with both lepton's spin projected onto $A,B\in\{L,N, T\}$ direction, we define single lepton polarization asymmetry  
\bea\label{eq:def-single-lep-pol-asym}
\mathcal{P}^{-}_A &=& \frac{\biggr[\frac{d\Gamma}{dq^2} (s^{-}_A,s^{+}_A)+\frac{d\Gamma}{dq^2}(s^{-}_A,-s^{+}_A) \biggl]-\biggr[\frac{d\Gamma}{dq^2}(-s^{-}_A,s^{+}_A)+\frac{d\Gamma}{dq^2}(-s^{-}_A,-s^{+}_A)\biggl]}{\biggr[\frac{d\Gamma}{dq^2} (s^{-}_A,s^{+}_A)+\frac{d\Gamma}{dq^2}(s^{-}_A,-s^{+}_A) \biggl]+\biggr[\frac{d\Gamma}{dq^2}(-s^{-}_A,s^{+}_A)+\frac{d\Gamma}{dq^2}(-s^{-}_A,-s^{+}_A)\biggl]}\, ,\\
\mathcal{P}^{+}_A &= &\frac{\biggr[\frac{d\Gamma}{dq^2} (s^{-}_A,s^{+}_A)+\frac{d\Gamma}{dq^2}(-s^{-}_A,s^{+}_A) \biggl]-\biggr[\frac{d\Gamma}{dq^2}(s^{-}_A,-s^{+}_A)+\frac{d\Gamma}{dq^2}(-s^{-}_A,-s^{+}_A)\biggl]}{\biggr[\frac{d\Gamma}{dq^2} (s^{-}_A,s^{+}_A)+\frac{d\Gamma}{dq^2}(-s^{-}_A,s^{+}_A) \biggl]+\biggr[\frac{d\Gamma}{dq^2}(s^{-}_A,-s^{+}_A)+\frac{d\Gamma}{dq^2}(-s^{-}_A,-s^{+}_A)\biggl]}
\eea
To study correlation between the spin of the two leptons, a double lepton polarization asymmetry can be defined as:
\bea
\mathcal{P}_{AB}=\frac{\biggr[\frac{d\Gamma}{dq^2}(s_A^-,s_B^+)-\frac{d\Gamma}{dq^2}(-s_A^-,s_B^+)\biggl]-\biggr[\frac{d\Gamma}{dq^2}(s_A^-,-s_B^+)-\frac{d\Gamma}{dq^2}(-s_A^-,-s_B^+)\biggl]}{\biggr[\frac{d\Gamma}{dq^2}(s_A^-,s_B^+)+\frac{d\Gamma}{dq^2}(-s_A^-,s_B^+)\biggl]+\biggr[\frac{d\Gamma}{dq^2}(s_A^-,-s_B^+)+\frac{d\Gamma}{dq^2}(-s_A^-,-s_B^+)\biggl]}
\eea
We also study the polarized forward-backward asymmetry of the lepton, which is defined as
\bea
\mathcal{A}_{\rm FB}^{AB} &= \left( \frac{d \Gamma}{dq^2} \right)^{-1} \left \{ \int_0^1 d \cos\theta_\ell - \int_{-1}^0 d \cos\theta_\ell \right \} \left\{ \biggl[ \frac{d^2 \Gamma}{d q^2 \, d \cos\theta_\ell}(s_A^-,s_B^+) - \frac{d^2 \Gamma}{d q^2 \, d \cos\theta_\ell}(s_A^-,-s_B^+) \biggr] \right. \notag \\
&\quad \left. - \biggl[ \frac{d^2 \Gamma}{dq^2 \, d\cos\theta_\ell}(-s_A^-,s_B^+) - \frac{d^2 \Gamma}{dq^2 \, d \cos\theta_\ell}(-s_A^-,-s_B^+) \biggr] \right\}, \notag \\
&= \mathcal{A}_{FB}(s_A^-,s_B^+) - \mathcal{A}_{FB}(-s_A^-,s_B^+) - \mathcal{A}_{FB}(s_A^-,-s_B^+) + \mathcal{A}_{FB}(-s_A^-,-s_B^+).
\eea

We have expressed the observables in terms of transversity amplitudes which are given in the Appendix \ref{eq:tas}. 
%
%

\section{Observables for $\Lambda_b \to \Lambda \mu^+\mu^-$ decay \label{subsec:Lb2L}}
In this section, we present the Standard Model estimate and new-physics sensitivities for the $\Lambda_b \to \Lambda \mu^+\mu^-$ observables. The $\Lambda_b \to \Lambda^\ast \mu^+\mu^-$ observables are discussed in the next section. Before we proceed, we clarify the inputs of our analysis. The decay rates are expressed in terms of transversity amplitudes. For both $\Lambda_b\to \Lambda\ell^+\ell^-$ and $\Lambda_b \to \Lambda^\ast\ell^+\ell^-$ decay the form factors are taken from calculations in lattice QCD \cite{Detmold:2016pkz} \cite{Meinel:2021mdj}. The new physics benchmark points have been presented in table \ref{tab:NP_benchmarks}.  We have performed our numerical analysis in Flavio \cite{Straub:2018kue}. In our analysis we have only included the form factor uncertainties.
\begin{table}[t]
\centering
\begin{tabular}{lccc}
\hline\hline
Benchmark & $ C^{\rm NP}_9$ & $C_9'$ & $ C^{\rm NP}_{10}$ \\
\hline
Benchmark I  \cite{LHCb:2024onj}
& $-0.71$ 
& $0$ 
& $0$ \\

Benchmark II  \cite{LHCb:2024onj}
& $-0.67$ 
& $0.9$ 
& $0$ \\

Benchmark III  \cite{Fleischer:2022klb,Fleischer:2025ucq}
& $0.30\,|C_9^{\mathrm{SM}}|\,e^{i\,220^\circ}$ 
& $0$ 
& $-0.30\,|C_9^{\mathrm{SM}}|\,e^{i\,220^\circ}$ \\

\hline\hline
\end{tabular}
\caption{New Physics benchmark scenarios considered in this work. 
For Benchmark II, the parameters correspond to the upper edge of the $1\sigma$ region allowed by data. }
\label{tab:NP_benchmarks}
\end{table}

\subsection{Single lepton polarization asymmetries}
\begin{figure}[H]
    \centering
    \includegraphics[width=0.4\textwidth]{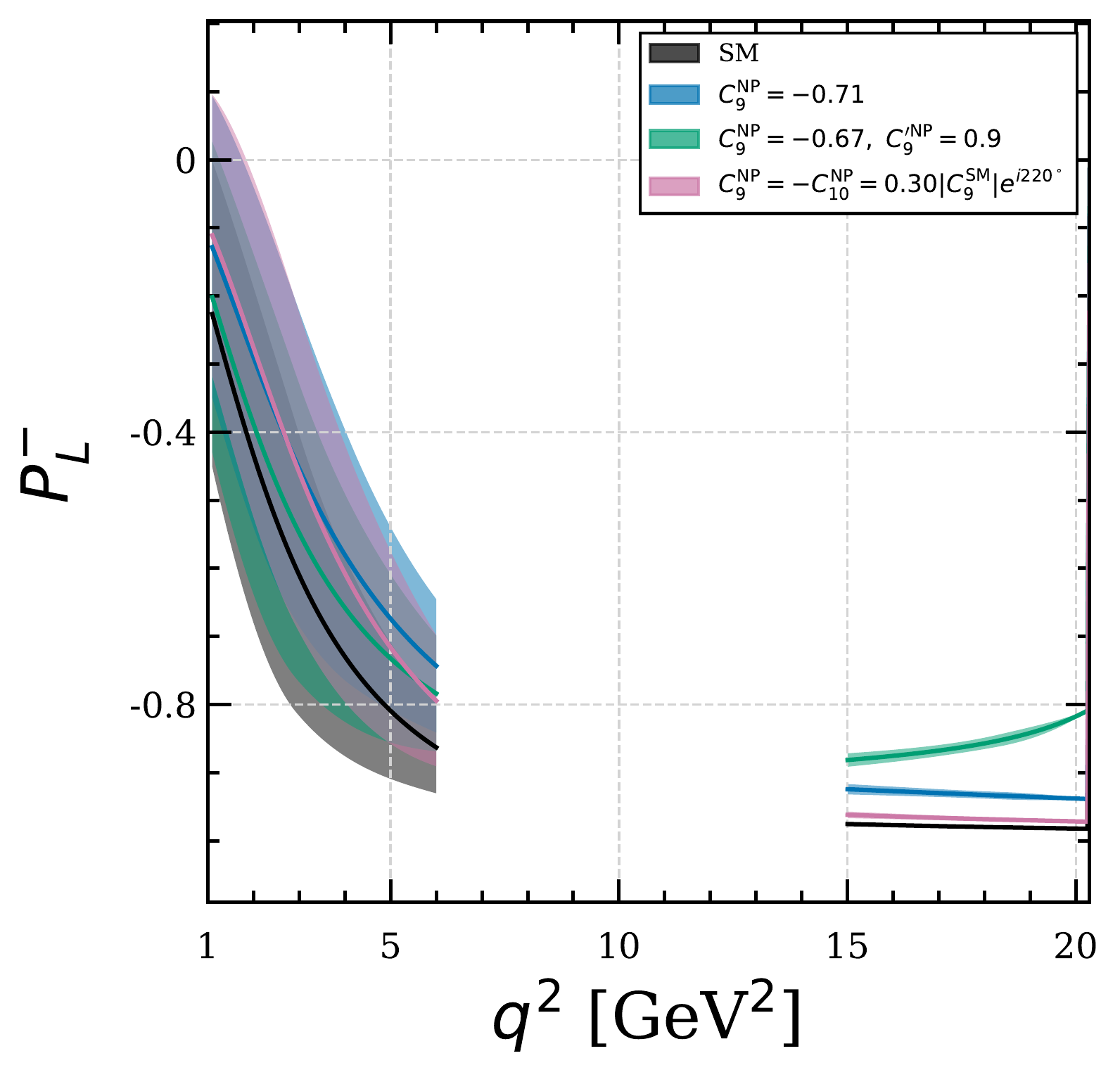}
    \includegraphics[width=0.4\textwidth]{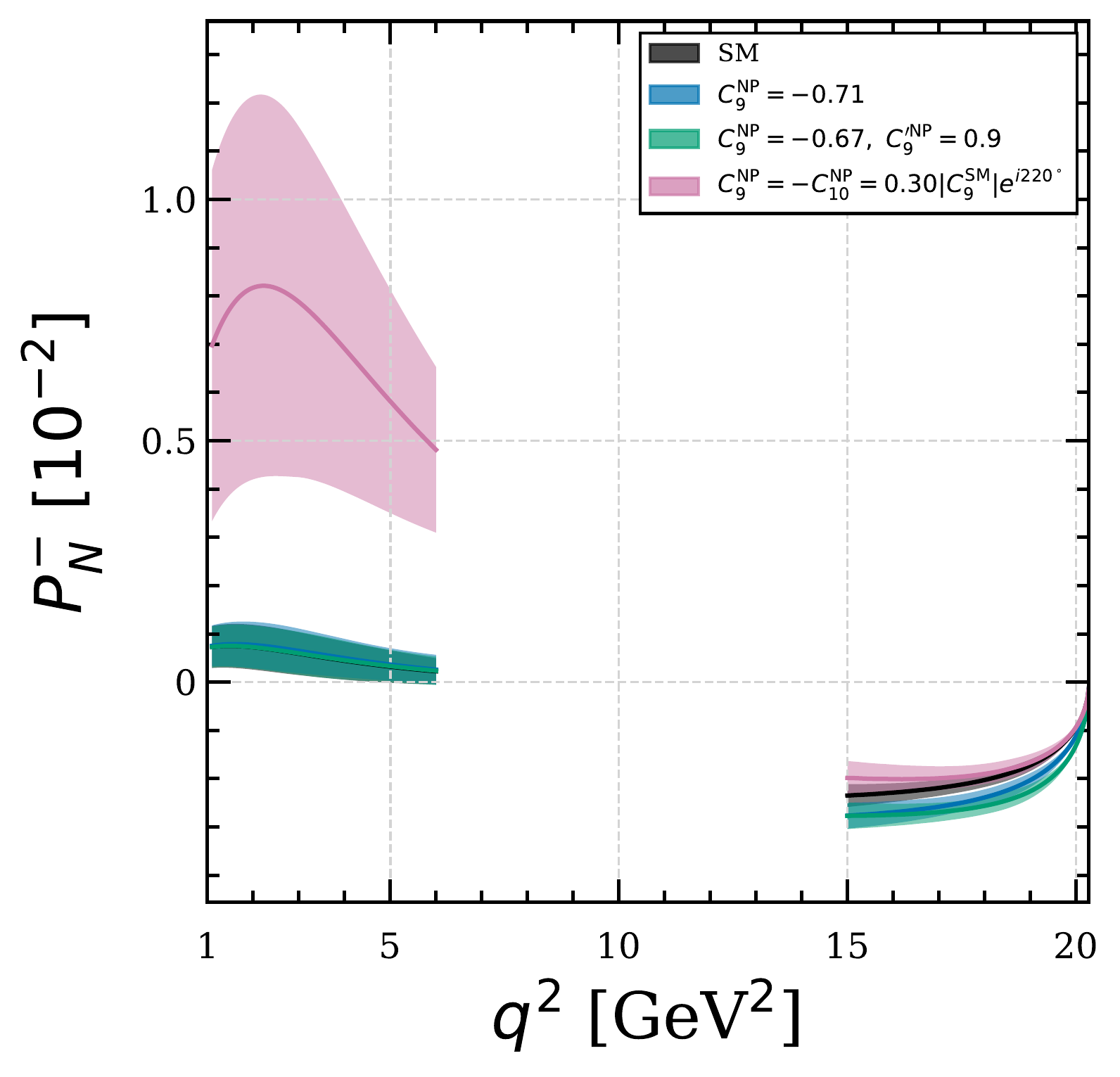}
    \caption{Single lepton polarization asymmetries
    $P_L^{-}$ and $P_N^{-}$
    as a function of $q^2$ for $\Lambda_b\to\Lambda\mu^+\mu^-$
    decay in the SM and beyond.}   
    \label{fig:Lb2Lmumu_PLmiPNmi}
\end{figure}
The longitudinal polarization asymmetries $P_L^\pm$ satisfy the exact relation $P_L^- = -P_L^+$. The numerator is proportional to $\sqrt{q^2}\sqrt{q^2-4m_\ell^2}\,\sum_\lambda(|A^R_\lambda|^2 - |A^L_\lambda|^2)$, (see Appendix~\ref{sec:pol-assym-obs-L} for analytical expressions) making $P_L^\pm$ a direct measure of the chiral asymmetry between left and right handed transversity amplitudes. In the massless limit, $m_\ell \to 0$, the longitudinal asymmetry reduces to the left-right helicity asymmetry and is therefore $\mathcal{O}(1)$ for muons. In the $q^2\to 4m_\ell^2$ limit, the asymmetries vanish through the phase-space factor. It is further reduced in magnitude at low $q^2$, where the photon pole contribution, which enters the left and right-handed amplitudes equally, dilutes the chiral asymmetry. Compared to the low-$q^2$, at high-$q^2$ hadronic uncertainties are vastly reduced since the dipole term in the amplitude is suppressed and each transversity amplitude is dominated by one form factor, leading to better cancellations. At high $q^2$, the SM and three benchmark scenarios are distinguishable from each other as shown in figure Fig.~\ref{fig:Lb2Lmumu_PLmiPNmi}
\begin{figure}[H]
    \centering
    \includegraphics[width=0.4\textwidth]{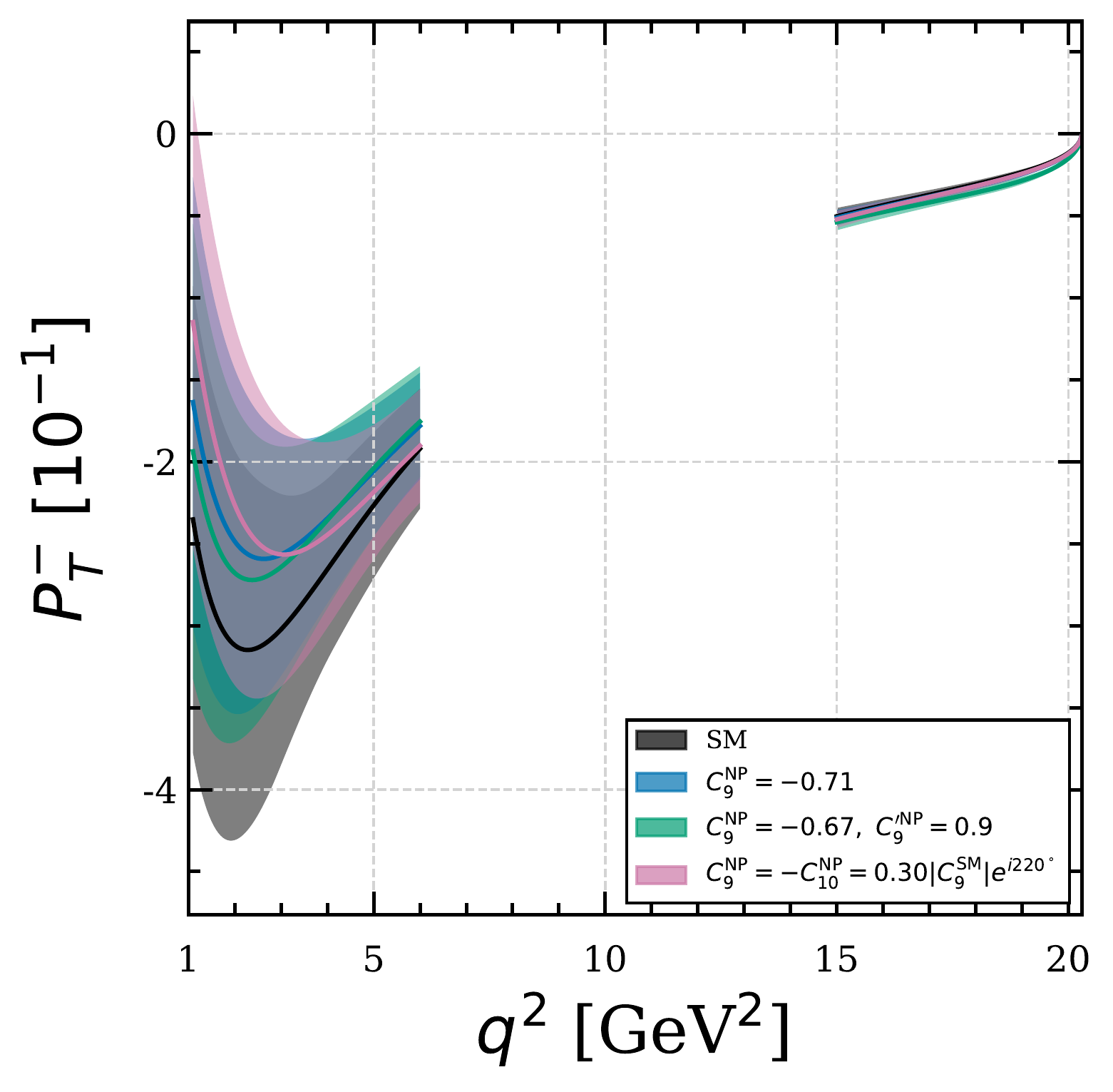}
    \includegraphics[width=0.4\textwidth]{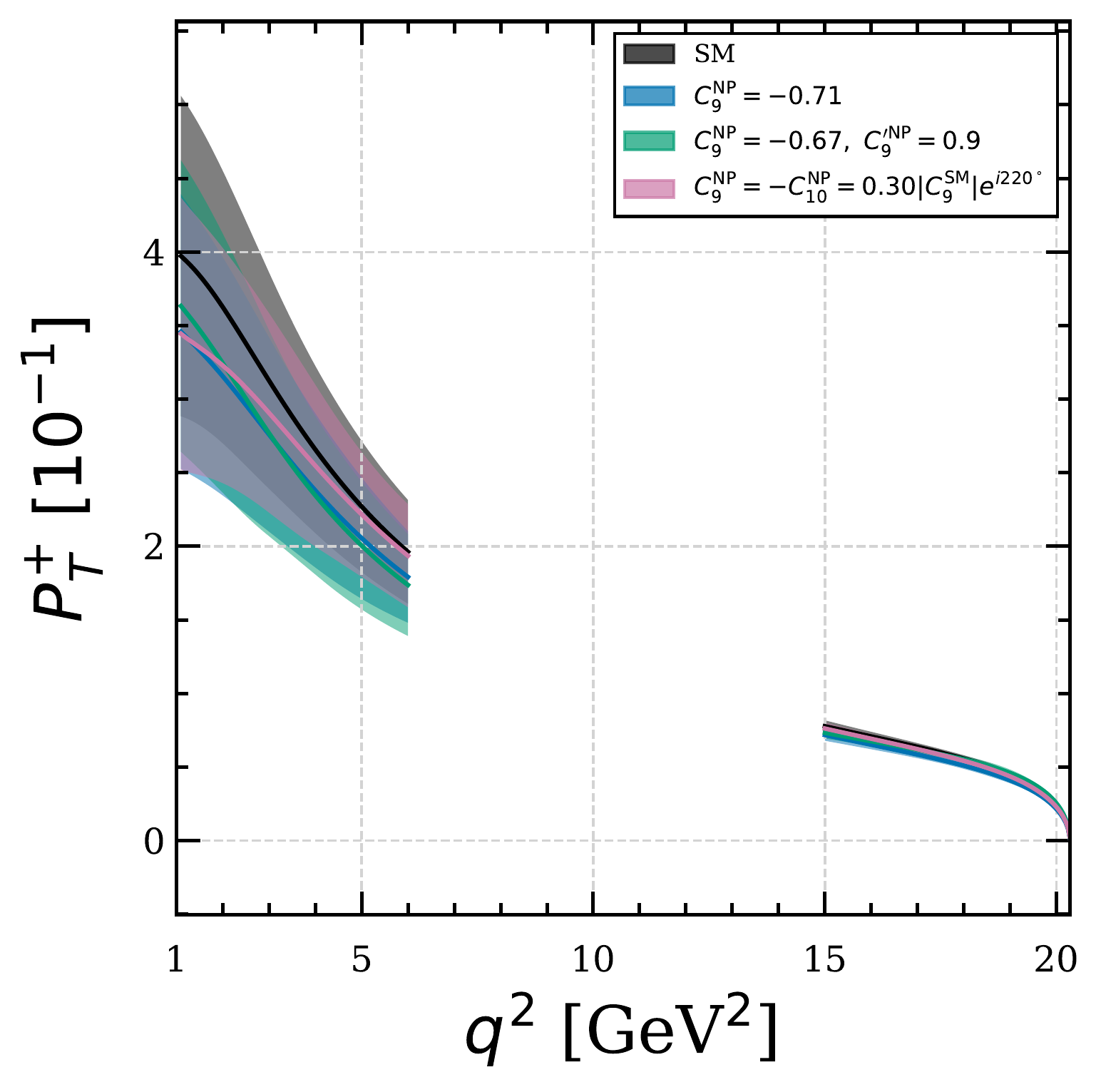}
    \caption{Single lepton polarization asymmetries
    $P_T^{-}$ and $P_T^{+}$
    as a function of $q^2$ for $\Lambda_b\to\Lambda\mu^+\mu^-$
    decay in the SM and beyond.}
    
    \label{fig:Lb2Lmumu_PTmiPTpl}
\end{figure}

The normal polarization asymmetries $P_N^\pm$ are proportional to
imaginary parts of interference terms between transversity amplitudes,
\be
P_N^{\mp} \propto m_\ell\sqrt{q^2 - 4m_{\ell}^2} \, \text{Im} \left[ \pm A_{\parallel 1}^R (A_{\perp 1}^L)^* + (A_{\parallel t}^L - A_{\parallel t}^R)((A_{\parallel 0}^L)^* - (A_{\parallel 0}^R)^*) + \{ \parallel \leftrightarrow \perp \} \right]\, .
\ee
In the second term $(A_{\parallel t}^L - A_{\parallel t}^R)((A_{\parallel 0}^L)^* - (A_{\parallel 0}^R)^*)$, the imaginary part cancels and 
\begin{equation}
P_N^- = -P_N^+.
\end{equation}
Therefore, in Fig.~\ref{fig:Lb2Lmumu_PLmiPNmi} we show only $P_N^-$. As the imaginary part of the SM Wilson coefficients are small, and there is an additional $m_\ell$ factor, the observable is of the order $\mathcal{O}(10^{-4}-10^{-3})$. The new-physics benchmarks I and II, which contain no complex phase, are numerically indistinguishable from the SM within the hadronic uncertainties. In contrast, benchmark III, which has a complex phase in $C_9^{\mathrm{NP}}$ and $C_{10}^{\mathrm{NP}}$, produces a pronounced enhancement of $P_N^\pm$ in the low-$q^2$ region reaching about $8\times 10^{-3}$, more than an order of magnitude above the SM. However, at high $q^2$ it overlaps with the SM. Another feature of $P_N^{\pm}$, visible in Fig.~\ref{fig:Lb2Lmumu_PLmiPNmi} and reflected in Table~\ref{tab:LbToL-bin-avg}, is a change of sign between the two $q^2$ bins: $P_N^{-} = (4.67 \pm 3.44)\times 10^{-4}$ in the low-$q^2$ bin and $(-1.81 \pm 0.14)\times 10^{-3}$ at high $q^2$. Note that $P_N^\pm$ as well as observables carrying an odd number of projections along the normal direction, $P_{LN}$, $P_{NL}$, $P_{NT}$, $P_{TN}$, $A_{FB}^{LN}$ and $A_{FB}^{NT}$ - are $T$-odd, since $\vec{s}_N \propto \vec{p} \times \vec{p}_{\mp}$ contributes a triple product for each normal projection. The asymmetry $P_{NN}$, with two such projections, that we subsequently discuss, is $T$-even and correspondingly involves real parts of interference terms. A nonzero value of a $T$-odd observable does not by itself establish $CP$ violation, since absorptive strong phases coming from the strong phases in ${\rm Im}\,C_9^{\rm eff}$ and ${\rm Im}\,C_7^{\rm eff}$ generate $T$-odd effects with real Wilson coefficients. By $CPT$ and unitarity, a weak phase changes sign under $CP$ conjugation while an absorptive strong phase does not, so a genuine $CP$-violation test requires the $CP$-odd combination $
\frac{1}{2}\left[\,P_N(\Lambda_b) - \bar{P}_N(\bar{\Lambda}_b)\,\right], $ in which the strong-phase contribution cancels. The separation of the benchmark III from the SM should therefore be read as sensitivity to a weak phase in the presence of a known strong phase.

The transverse polarization asymmetries $P_T^\pm$ arise from real parts of interference terms between different transversity amplitudes. For muons, the qualitative behavior of $P_T^{-}$ and $P_T^+$ differ noticeably in both sign and magnitude at high $q^2$ (also see table \ref{tab:LbToL-bin-avg}) so we show both observables   in Fig. \ref{fig:Lb2Lmumu_PTmiPTpl}.  The origin is traced from the expressions presented in Appendix \ref{sec:pol-assym-obs-L}. Symbolically, they can be written as
\begin{equation}
P_T^{\mp}=-\frac{4\pi m_\ell\sqrt{q^2}}{\Delta}\,\mathrm{Re}\Big[\mathcal{T}\pm\mathcal{X}_t\Big],\qquad
\mathcal{X}_t\equiv\sum_{i=\parallel,\perp}\big(A^L_{it}-A^R_{it}\big)\big(A^L_{i0}+A^R_{i0}\big)^*,
\label{eq:Xt}
\end{equation}
where $\mathcal{T}$ collects the purely transverse terms common to both. The timelike--longitudinal term $\mathcal{X}_t$ enters with opposite signs. It dominates at low $q^2$, where the timelike and longitudinal amplitudes grow as $1/\sqrt{q^2}$, giving $P_T^+\simeq-P_T^-$. At high $q^2$ the common term $\mathcal{T}$ contributes significantly compared to $\mathcal{X}_t$ and splits the two magnitudes. Being suppressed by a single power of $m_\ell$, the transverse asymmetries are significantly smaller than the longitudinal ones, reaching $\mathcal{O}(10^{-1})$ at low $q^2$. The hadronic uncertainties reduce at high $q^2$ as the dipole term is suppressed leading to better form factor cancellations.  We find that in the entire $q^2$ region, the distinguishability of the three New Physics benchmark scenarios from the SM is poor, as shown in Fig.~\ref{fig:Lb2Lmumu_PTmiPTpl}.

The three observables therefore exhibit a clear hierarchy, $|P_L^{\pm}| \sim \mathcal{O}(1) \gg |P_T^{\pm}| \gg |P_N^{\pm}|$, with $|P_T^{\pm}| \sim \mathcal{O}(10^{-1})$ and $|P_N^{\pm}| \sim \mathcal{O}(10^{-4})$ at low $q^2$, falling to $\mathcal{O}(10^{-2})$ and rising to $\mathcal{O}(10^{-3})$ respectively at high $q^2$.

\subsection{Double lepton polarization asymmetries
            in $\Lambda_b \to \Lambda\mu^+\mu^-$}
\begin{figure}[H]
\centering
\includegraphics[width=0.33\textwidth]{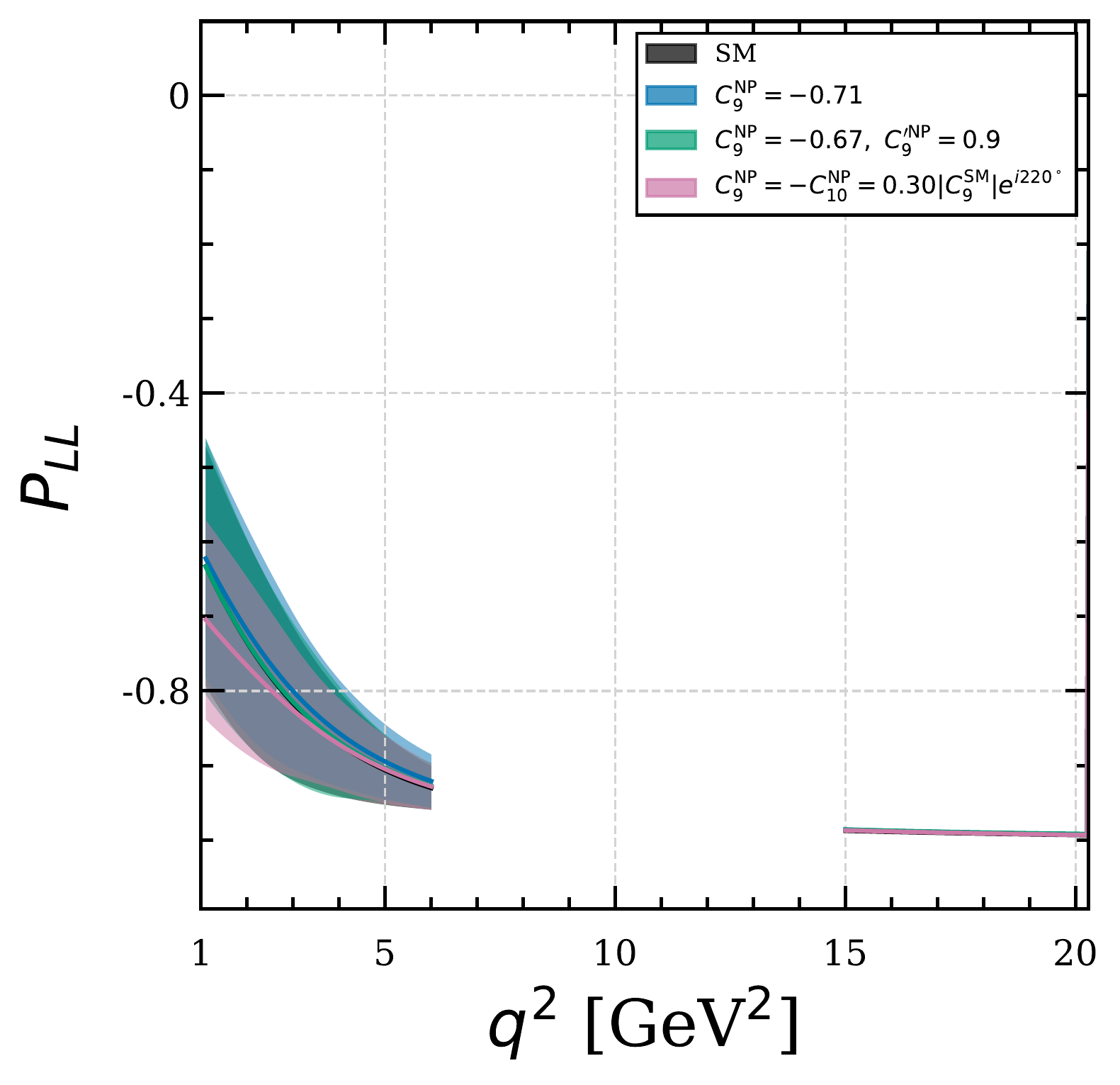}
\includegraphics[width=0.32\textwidth]{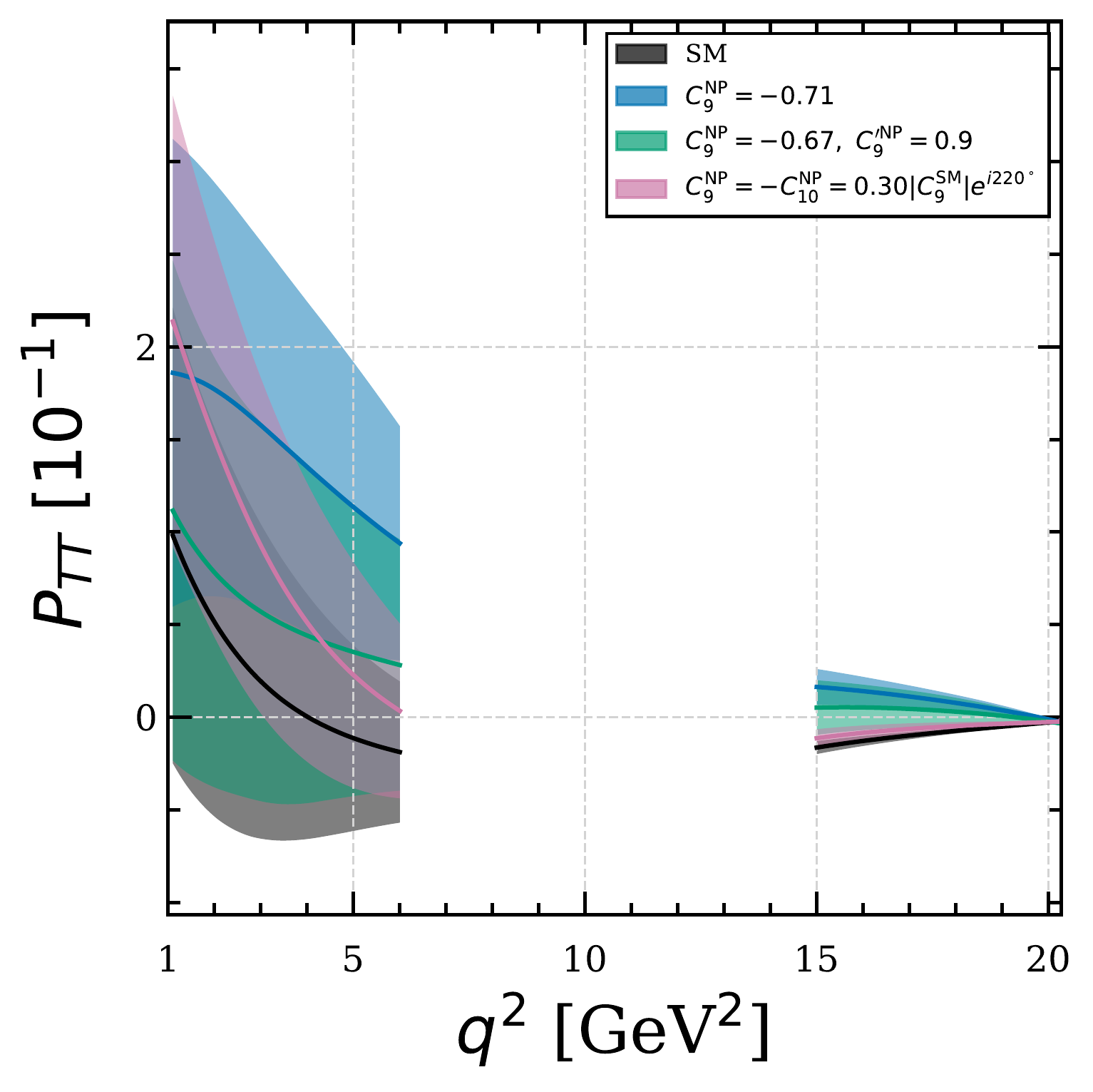}
\includegraphics[width=0.32\textwidth]{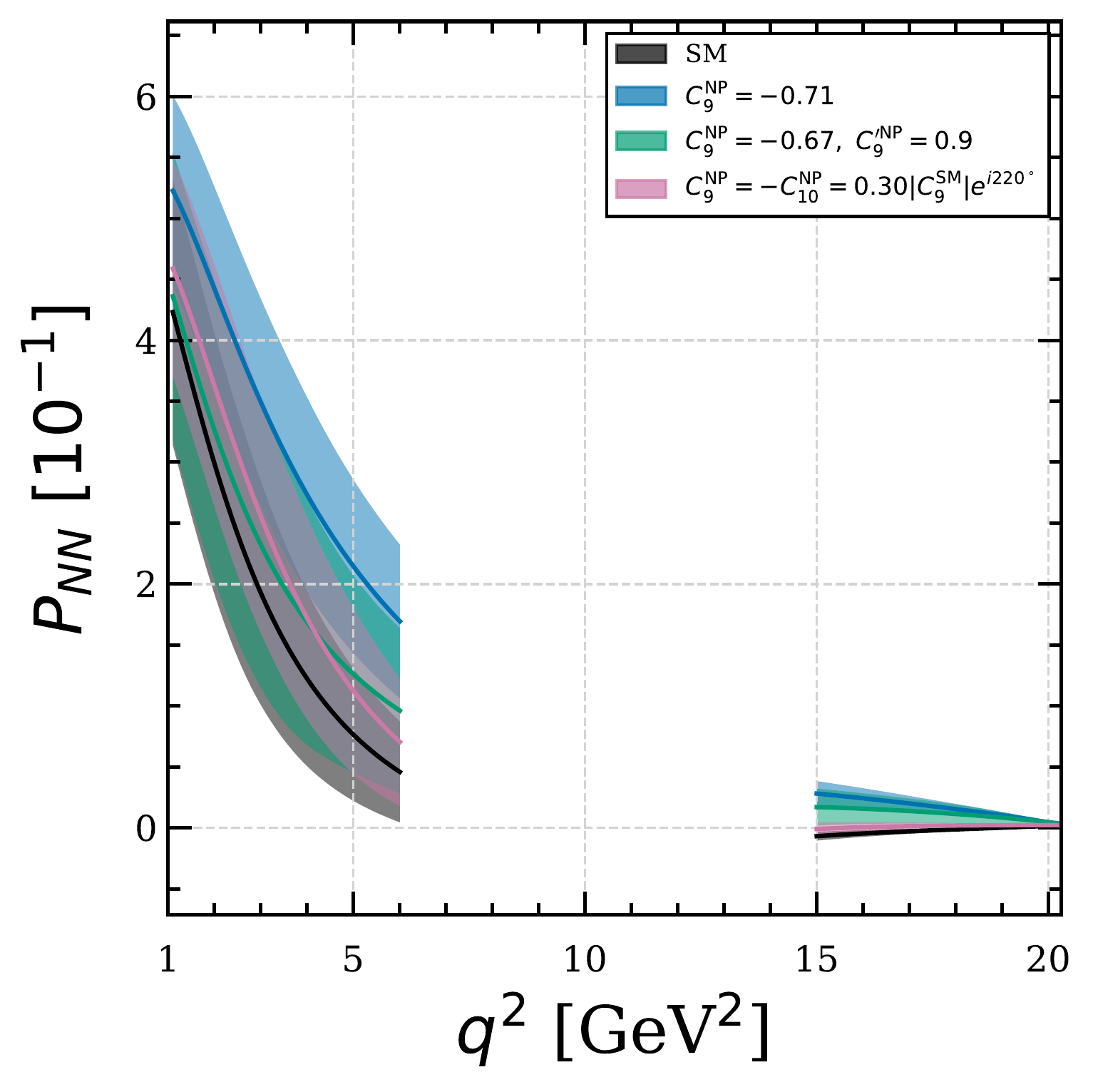}
\caption{Double lepton polarization asymmetries $P_{LL}$,$P_{TT}$ and $P_{NN}$ of $\Lambda_b \to \Lambda \mu^+\mu^-$ decay as a function of $q^2$ in the SM and new physics scenarios.}
\label{fig:Lb2Lmumu_PLLPTTPNN}
\end{figure}

\begin{figure}[H]
    \centering
    \includegraphics[width=0.4\textwidth]{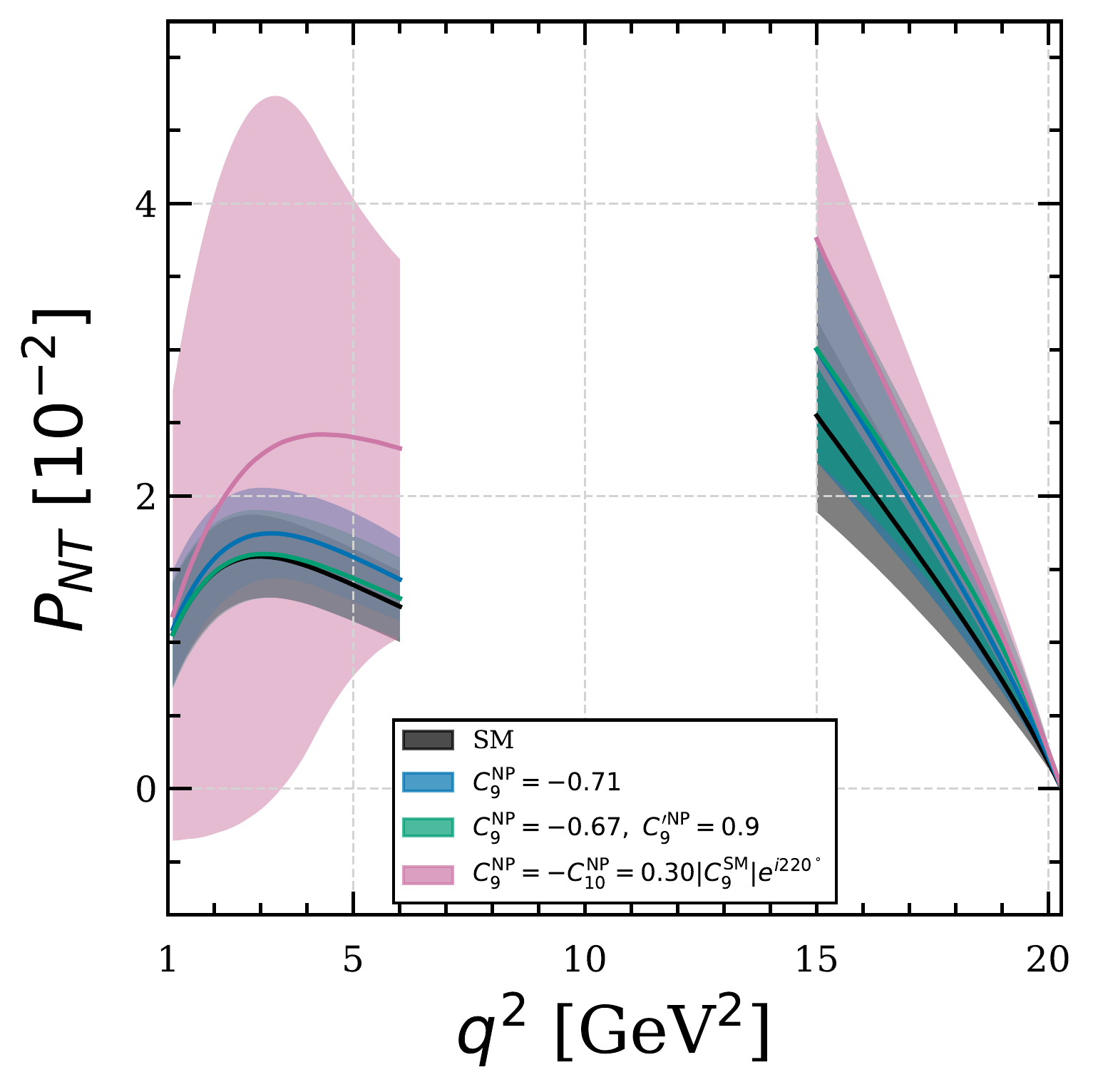}
    \includegraphics[width=0.4\textwidth]{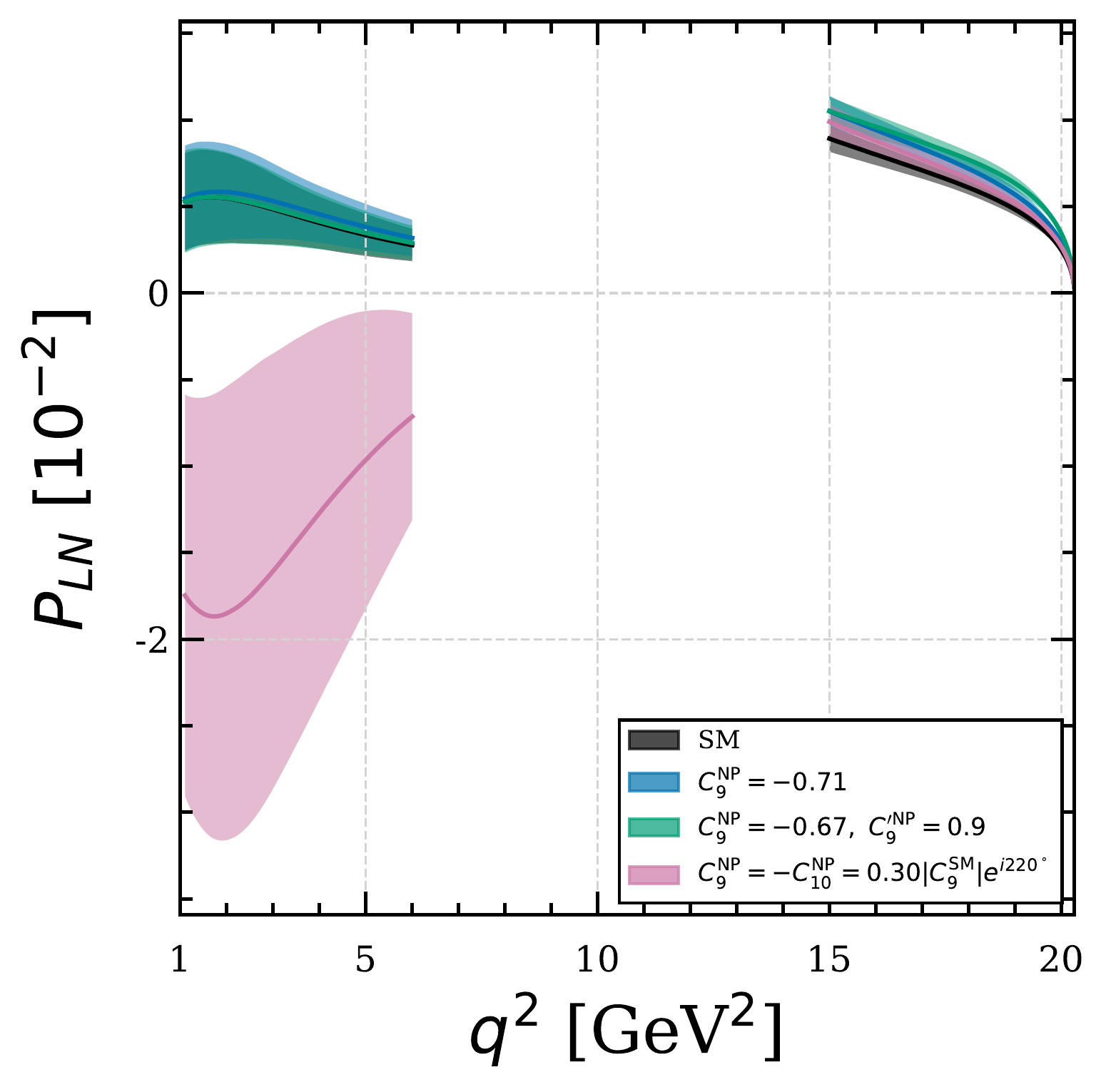}
    \caption{Double lepton polarization asymmetries $P_{NT}$ and $P_{LN}$ as a function of $q^2$ for $\Lambda_b \to \Lambda \mu^+\mu^-$ in the SM and beyond.}
    \label{fig:Lb2Lmumu_PNTPLN}
\end{figure}

\begin{figure}[H]
    \centering
    \includegraphics[width=0.4\textwidth]{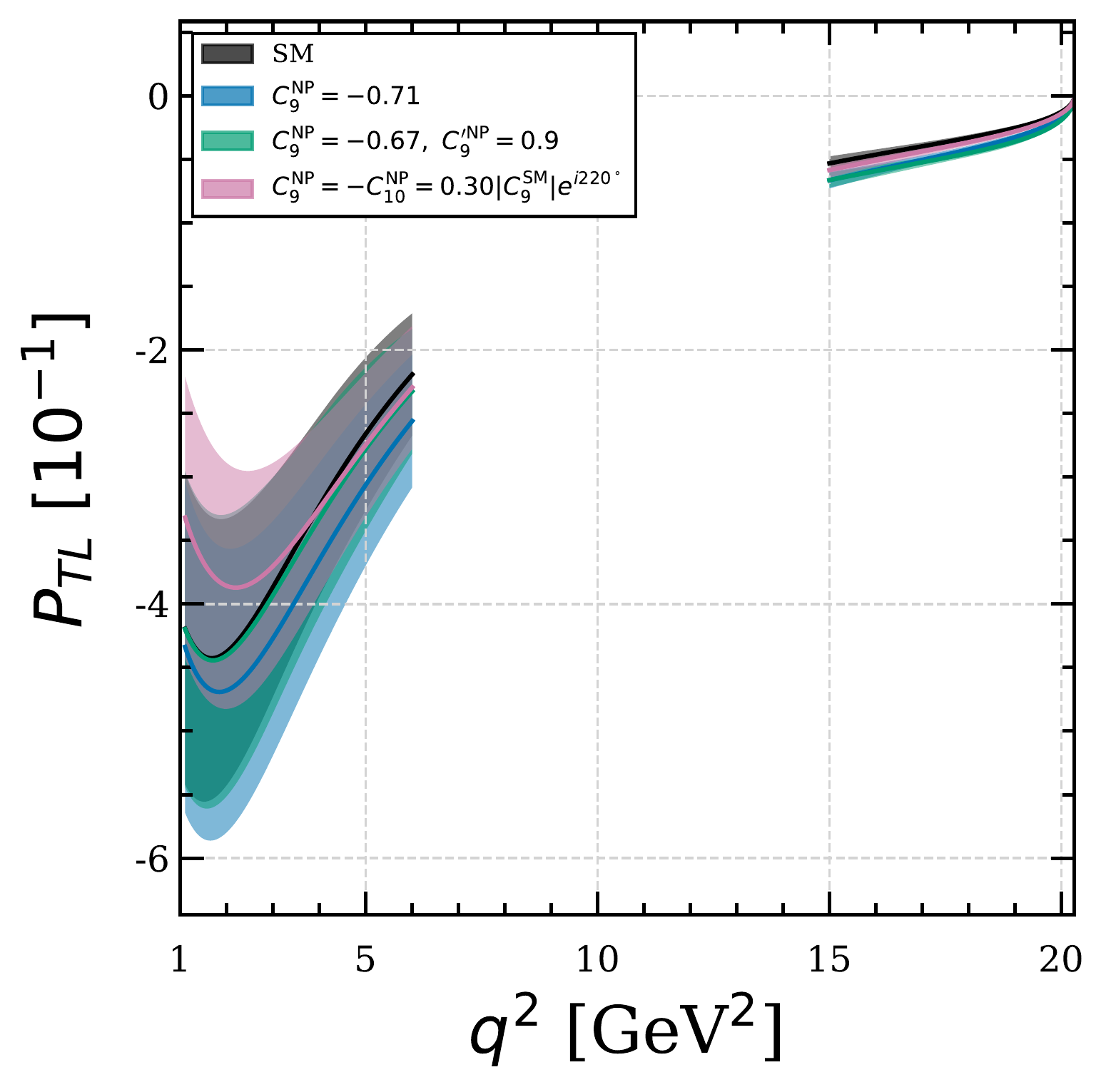}
    \includegraphics[width=0.4\textwidth]{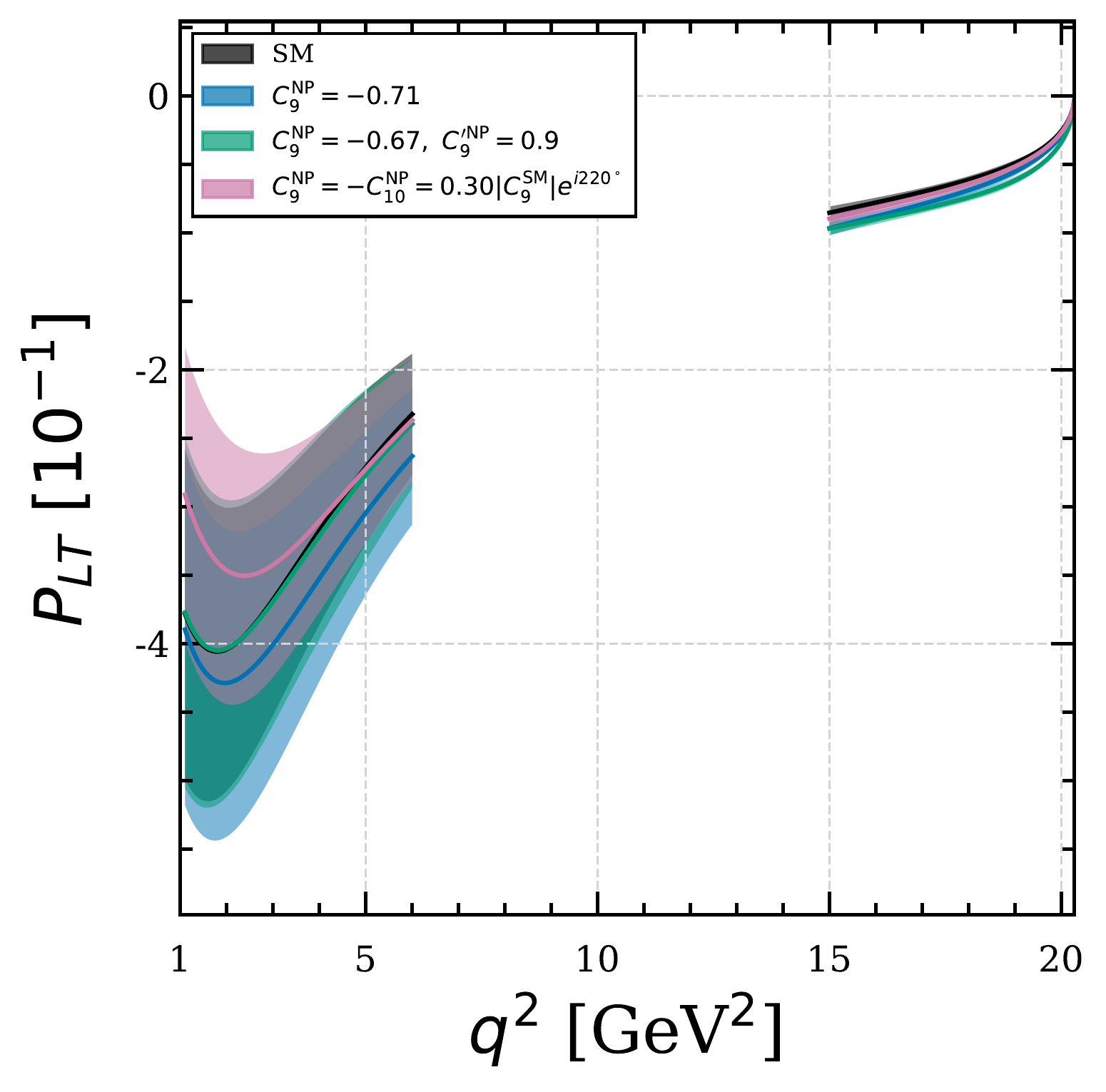}
    \caption{Double lepton polarization asymmetries $P_{TL}$ and $P_{LT}$ as a function of $q^2$ for $\Lambda_b \to \Lambda \mu^+\mu^-$ in the SM and beyond.}
    \label{fig:Lb2Lmumu_PTLPLT}
\end{figure}

The double lepton polarization asymmetries denoted as $P_{AB}$, where $A,B=L,N,T$, measure the simultaneous polarization of the $\ell^-$ along direction $A$ and the $\ell^+$ along direction $B$. They provide complementary probes of the chiral structure of the underlying short-distance dynamics beyond what is accessible from the single-lepton polarizations.  The analytical expressions are given in Appendix~\ref{sec:pol-assym-obs-L}, and the numerical results are shown in Figs.~\ref{fig:Lb2Lmumu_PLLPTTPNN}-\ref{fig:Lb2Lmumu_PTLPLT}.

In the massless limit, the longitudinal-longitudinal asymmetry $P_{LL}\to -1$ at both low and high $q^2$ independently of the Wilson coefficients. In this limit the vector and axial-vector currents produce $\ell^-$ and $\ell^+$ only with opposite helicities, so that their spin projections along $\hat{p}_-$ and $\hat{p}_+$ respectively, always have opposite signs. In the massive limit the correction is of the order of $\mathcal{O}(m_\ell^2/q^2)$ and significant only at low $q^2$. Therefore, $P_{LL}$ carries essentially no discriminating power. The bin-averaged values in the low- and high-$q^2$ regions are shown in table \ref{tab:LbToL-bin-avg}.

The numerators of the transverse-transverse and normal-normal asymmetries $P_{TT}$ and $P_{NN}$ are governed by the same left--right interference structure,
\begin{equation}
P_{TT},\,P_{NN}
\;\propto\;
\operatorname{Re}\!\left(
  q^2\bigl(
    -2A^R_{\parallel 0}A^{L\ast}_{\parallel 0}
    + A^R_{\parallel 1}A^{L\ast}_{\parallel 1}
    + \{\parallel\leftrightarrow\perp\} \bigr) + m_\ell^2 (\cdots)
\right),
\end{equation}
The coefficients of $q^2$ are common to both, and only the coefficients of $m_\ell^2$ differ. So $P_{TT} = P_{NN}$ in the massless-lepton limit. The difference between them is $\mathcal{O}(m_\ell^2/q^2)$ and is therefore visible only at low $q^2$, as can be seen in Fig.~\ref{fig:Lb2Lmumu_PLLPTTPNN}. Both $P_{TT}$ and $P_{NN}$ converge to a common value at moderate and high $q^2$. Both observables are of  $\mathcal{O}(10^{-3})$ at high $q^2$. While at the low-$q^2$ bins, hadronic uncertainties dominate, the uncertainties largely cancel at high $q^2$(Fig.~\ref{fig:Lb2Lmumu_PLLPTTPNN}). In the SM, both observables change sign between the low- and high-$q^2$ regions, i.e., the zero crossing lies in the intermediate $q^2$ region that is not shown. Since the numerators differ by $m_\ell^2$, which is small for muons, the zero crossings are expected at the same $q^2$ up to the mass-squared corrections for both $P_{TT}$ and $P_{NN}$. The SM zero-crossing in $P_{TT}$ relies on a delicate balance: since $\vert{}C_9^{\rm SM}\vert{} \approx \vert{}C_{10}^{\rm SM}\vert{}$ the leading combination $(C_9^{\rm eff}+2\kappa_iC_7^{\rm eff})^2 - \vert{}C_{10}\vert{}^2$ (with $\kappa_i$ defined below) can change sign as the dipole contribution varies with $q^2$. The NP benchmarks I and II break this near-degeneracy—via a reduced $C_9$, a large right-handed $C_9^\prime$ — and keep the central values of $P_{TT}$ and $P_{NN}$ positive over the $q^2$ ranges shown.

The normal-transverse asymmetry $P_{NT}$ is the only T-odd double asymmetry without $m_\ell$ suppression. Its numerator
\begin{equation}
P_{NT} \propto
\sqrt{q^2}\sqrt{q^2-4m_\ell^2}\;
\operatorname{Im}\!\left(
  2A^R_{\parallel 0}A^{L\ast}_{\parallel 0}
  - A^R_{\parallel 1}A^{L\ast}_{\parallel 1}
  + \{\parallel\leftrightarrow\perp\}
\right),
\end{equation}
carries no explicit $m_\ell$ prefactor beyond the phase-space factor $\sqrt{q^2-4m_\ell^2}$. In the SM it is non-zero due to the non-negligible absorptive part of $C_9^{\rm eff}$. The SM is however, indistinguishable from the benchmarks I and II. The benchmark III which has a complex phase generates a pronounced enhancement, with the central value of $P_{NT}$ reaching about $2.3\times10^{-2}$ at low $q^2$, compared with about $1.5\times 10^{-2}$ in the SM, but the large hadronic uncertainty prevents a clear separation in either bin. Since $P_{NT}= -P_{TN}$, we show only one of them in Fig.~\ref{fig:Lb2Lmumu_PNTPLN}.

The asymmetry $P_{LN}$ is, like $P_{NT}$, $T$-odd. Its numerator is the imaginary part of the same timelike--longitudinal combination whose real part splits $P_T^-$ from $P_T^+$. In the SM it is small but non-zero, $(4.14\pm1.65)\times10^{-3}$ and $(5.61\pm0.37)\times10^{-3}$ in the low- and high-$q^2$ bins. For $C'_{10}=0$, the numerator is combination of $\mathrm{Im}\big[C_{10}(C_9^{\rm eff}\pm C_9'+2\kappa_i C_7^{\rm eff})^*\big]$ where $\kappa_{\parallel}=m_bf_0^{T5}/[(\mLb-\mL)f_0^A]$ and $\kappa_{\perp}=m_bf_0^{T}/[(\mLb+\mL)f_0^V]$, with the upper(lower) sign for $i=\perp(\parallel)$. So, for real $C_{10}$ it is unchanged by a real shift of $C_9$ or a real $C_9'$. That's why the benchmarks I and II modify $P_{LN}$ only through the denominator and cannot be distinguished from the SM. In benchmark III the weak phase nearly cancels in $\mathrm{Im}(C_{10}C_9^{\rm eff*})$, since $\mathrm{Re}\,C_9^{\rm eff}\simeq -C_{10}$, but it enters directly through the dipole interference $2\kappa_i C_7^{\rm eff}\,\mathrm{Im}\,C_{10}^{\rm NP}<0$; where $\mathrm{Im}\,C_9^{\rm eff}$ is small this term dominates, so $P_{LN}$ flips sign in the low-$q^2$ bin and is enhanced by a factor of a few relative to the SM. At high $q^2$, where Im $C_9^{\rm eff}$ is larger, benchmark III is close to the SM. $P_{LN}$ is thus the most sensitive double-polarization probe of the weak phase in this benchmark (Fig.~\ref{fig:Lb2Lmumu_PNTPLN}).

The mixed asymmetries $P_{LT}$ and $P_{TL}$ involve real parts of timelike longitudinal interference terms of the form $\operatorname{Re}(A_{\parallel t}A^{\ast}_{\parallel 0})$, together with same chirality transverse terms $\operatorname{Re}(A^L_{\perp 1}A^{L\ast}_{\parallel 1}) - \operatorname{Re}(A^R_{\perp 1}A^{R\ast}_{\parallel 1})$. In the SM, $P_{TL} \approx P_{LT}$ at low $q^2$. At high $q^2$ the degeneracy is lifted as can be seen from the table \ref{tab:LbToL-bin-avg}. The sensitivity to the new-physics benchmarks is moderate, with the scenarios showing visible but overlapping separation within the hadronic uncertainties in the low $q^2$ region and the sensitivity decreases in the high $q^2$ region, as shown in Fig.~\ref{fig:Lb2Lmumu_PTLPLT}.

\subsection{$A_{FB}$ for $\Lambda_b \to \Lambda \mu^{+}\mu^{-}$}
\begin{figure}[H]
    \centering
    \includegraphics[width=0.4\textwidth]{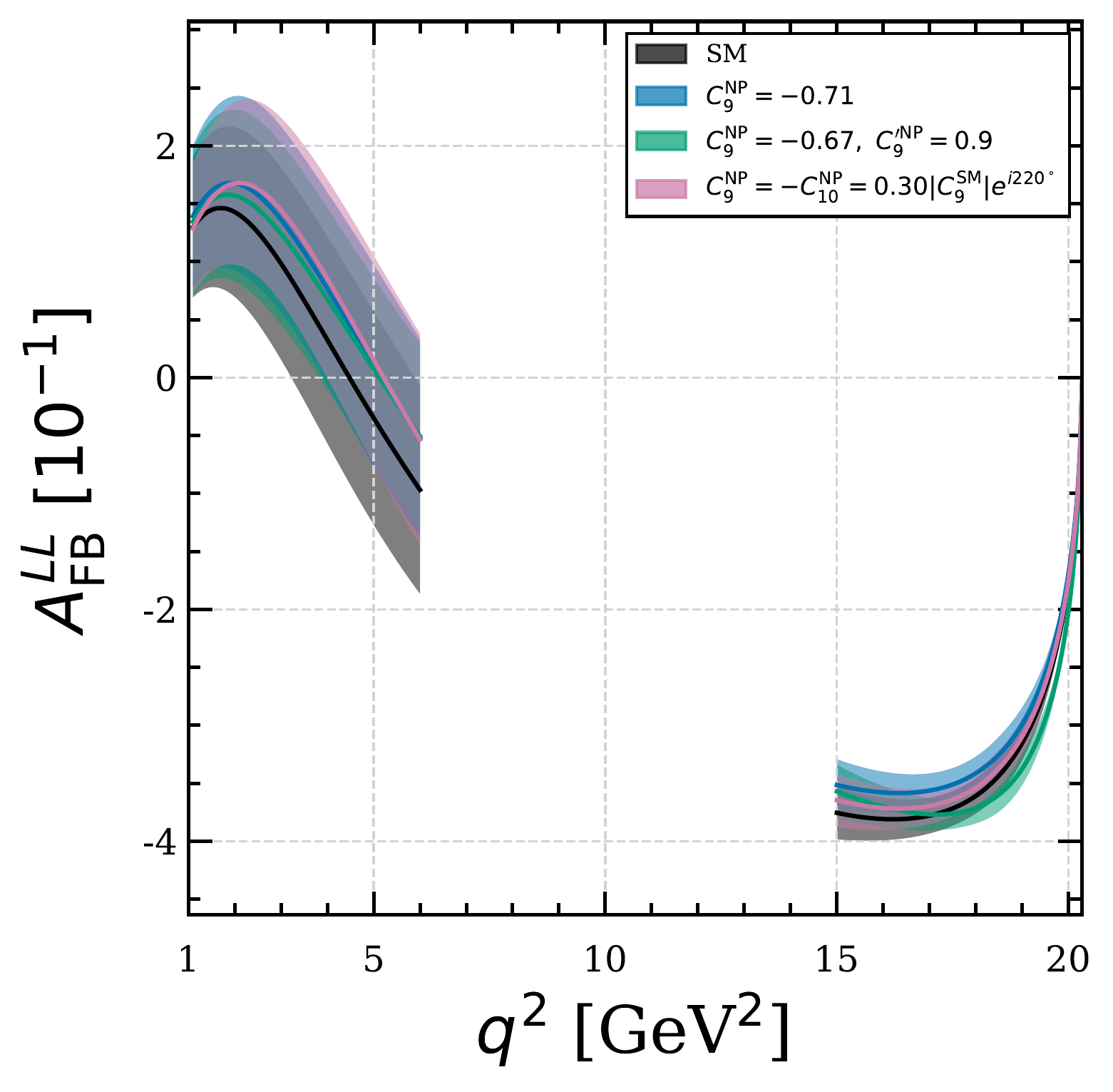}
    \includegraphics[width=0.4\textwidth]{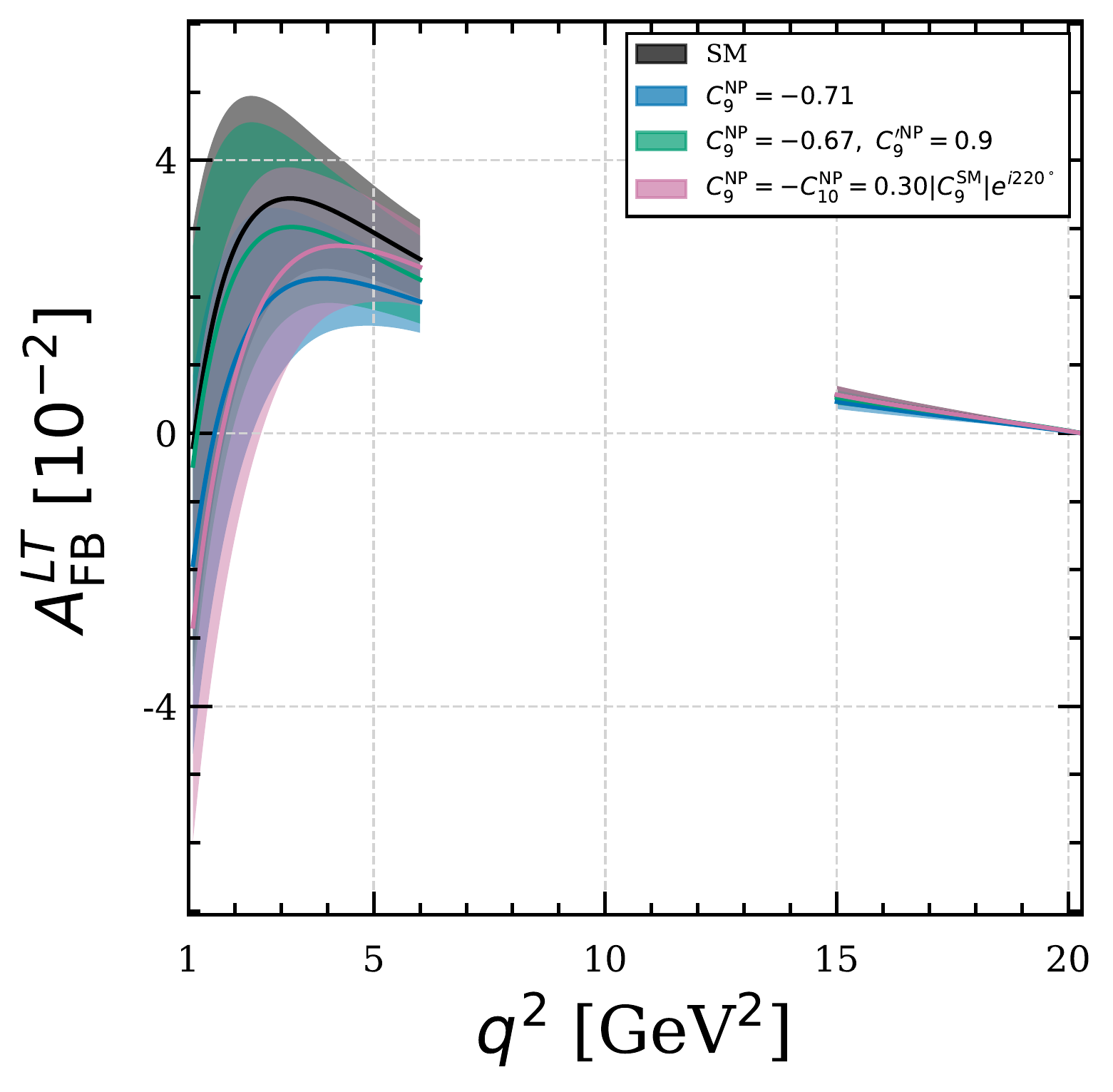}
    \caption{Doubly polarized forward-backward asymmetries  $A_{FB}^{LL}$ and $A_{FB}^{LT}$ as a function of $q^2$ for $\Lambda_b \to \Lambda \mu^+\mu^-$ in the SM and beyond.}
    \label{fig:Lb2Lmumu_AFBLLLT}
\end{figure}

The double-polarized forward–backward asymmetries $A_{\rm FB}$ depend on specific combinations of transversity amplitudes and allow one to disentangle the chiral structure of the underlying interaction. The longitudinal–longitudinal asymmetry $A_{\rm FB}^{LL}$ coincides with the standard leptonic forward–backward asymmetry, $ A_{\rm FB}^{LL} = A_{\rm FB}^\ell\,$ in the massless lepton limit and is governed by the same interference structure between transverse amplitudes, $A_{\rm FB}^{LL} \propto \mathrm{Re}\!\left(A_{\perp 1}^{L} A_{\parallel 1}^{L*}-A_{\perp 1}^{R} A_{\parallel 1}^{R*}\right)\,$ (see Appendix~\ref{sec:pol-assym-obs-L} for analytical expressions). Similar to the lepton forward-backward asymmetry $A_{\rm FB}^\ell $, the $A_{\rm FB}^{LL}$ exhibits a characteristic zero-crossing behavior. The zero-crossing position is controlled by the interference of $C_7^{\rm eff}$ and $C_9^{\rm eff}$ and is largely free of form-factor dependence at leading order in the heavy-quark limit. The NP benchmarks shift $A_{FB}^{LL}$ only within the hadronic uncertainties as shown in Fig.~\ref{fig:Lb2Lmumu_AFBLLLT}. The transverse–longitudinal asymmetry $A_{\rm FB}^{LT}$ is proportional to $m_\ell$, so it is suppressed compared to $A_{FB}^{LL}$ but has zero-crossing point at low $q^2$. The SM and NP benchmarks scenarios are indistinguishable as seen from the figure Fig.~\ref{fig:Lb2Lmumu_AFBLLLT}.

\begin{figure}[H]
    \centering
    \includegraphics[width=0.4\textwidth]{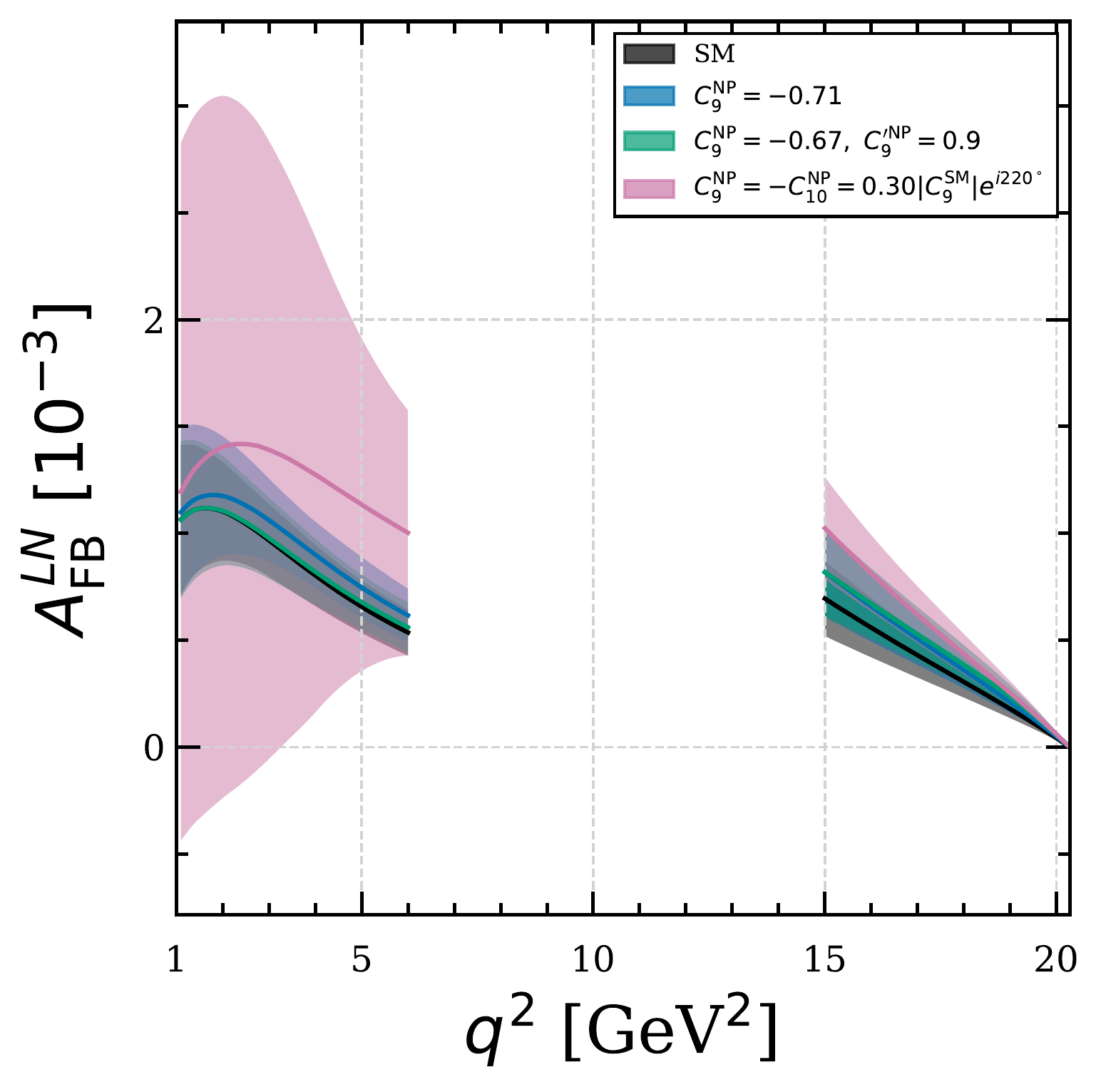}
    \includegraphics[width=0.4\textwidth]{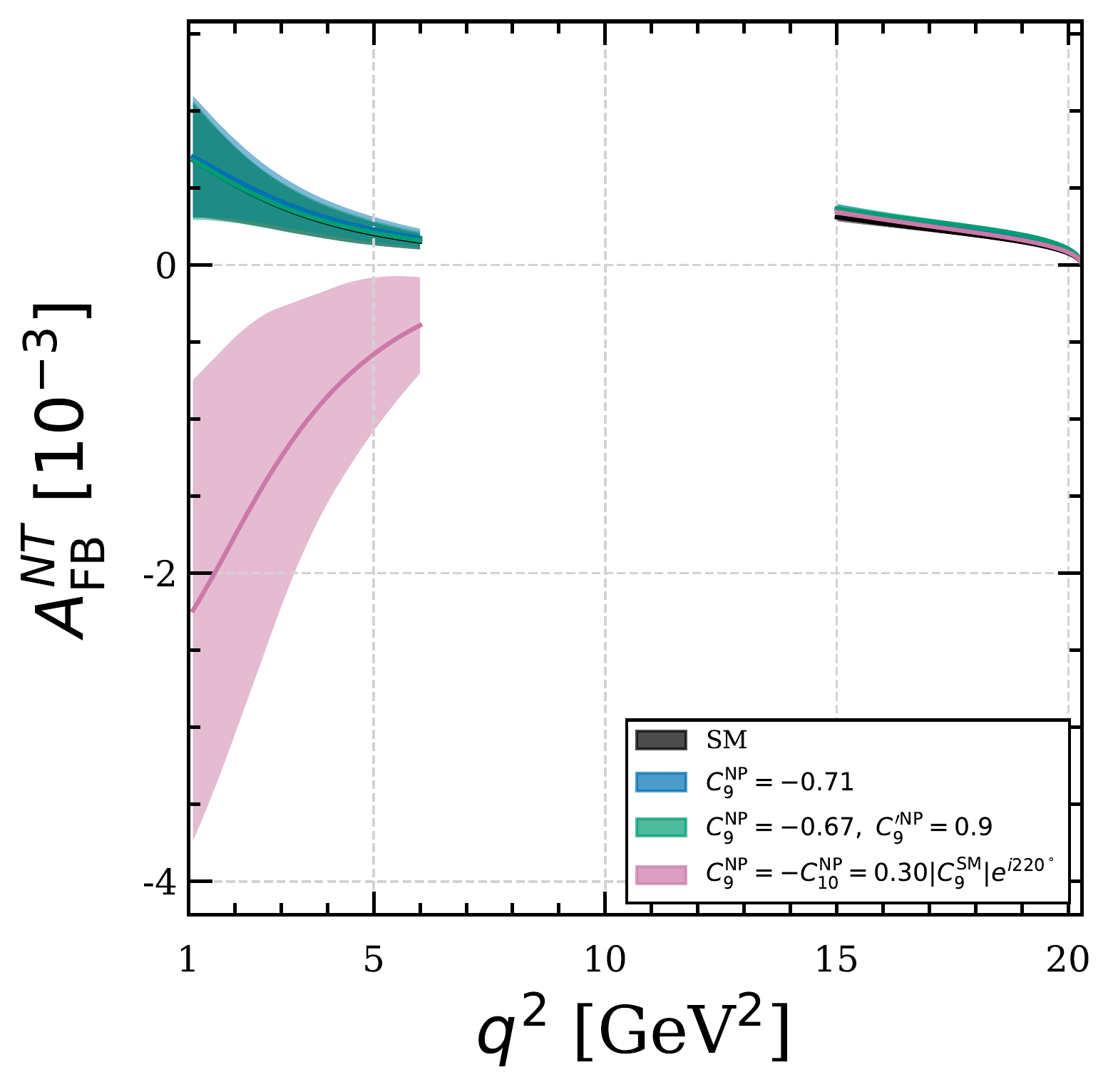}
    \caption{Doubly polarized forward-backward asymmetry   $A_{FB}^{LN}$ and $A_{FB}^{NT}$ as a function of $q^2$ for $\Lambda_b \to \Lambda \mu^+\mu^-$ in the SM and beyond.}
    \label{fig:Lb2Lmumu_AFBLNNT}
\end{figure}

The longitudinal--normal asymmetry $A_{FB}^{LN}$ is proportional to
imaginary parts of interference terms,
\begin{equation}
A_{FB}^{LN} \propto m_\ell \sqrt{q^2 - 4m_\ell^2} {\rm Im} \bigg( 2 A^R_{\|0} A^{L\ast}_{\|0} - A^R_{\|1} A^{L\ast}_{\|1} + \{ \| \leftrightarrow \perp \} \bigg)\, ,
\end{equation}
and is proportional to $P_{NT}$ therefore it shows same limited sensitivity to benchmark III as shown in Fig.~\ref{fig:Lb2Lmumu_AFBLNNT}. It receives contributions from both helicity-0 and helicity-1 amplitudes, but carries a single power of $m_\ell$ (Eq.~(\ref{eq:AFBLN_L})) which, together with the imaginary $C_9^{\rm eff}$, pushes it to the $10^{-4}$ level. Finally, the normal--transverse asymmetry $A_{FB}^{NT}$ is given by
\begin{equation}
A_{FB}^{NT} \propto m_\ell^2\, \mathrm{Im}\Big[\big(A_{\parallel t}^{L} - A_{\parallel t}^{R}\big)\big(A_{\parallel 0}^{L*} + A_{\parallel 0}^{R*}\big) + \{\parallel \leftrightarrow \perp\}\Big]\,,
\end{equation}
and is therefore suppressed for light leptons. In contrast to the SM and benchmarks I and II, the benchmark III generates a pronounced enhancement in the low $q^2$ region. At high $q^2$ the SM and the NP scenarios with real Wilson coefficients are indistinguishable as shown in Fig.~\ref{fig:Lb2Lmumu_AFBLNNT}.

\section{Numerical results for $\Lambda_b \to \Lambda^*\mu^+\mu^-$ \label{subsec:Lb2Lst}}
\label{sec:Lstar_results}
In this section we present the numerical results for the lepton
polarization observables in $\Lambda_b \to \Lambda^*\mu^+\mu^-$.
The kinematic
range of our analysis is restricted to the high-$q^2$ region
$q^2 \in [16, 16.8]\,\text{GeV}^2$, where the lattice QCD form
factor determinations are directly applicable. The three benchmark
new-physics scenarios defined in Table~\ref{tab:NP_benchmarks} are
the same as those used for the $\Lambda$ channel. The results are
displayed in Figs.~\ref{fig:Lb2LStmumu_PLmiPNmi}--\ref{fig:Lb2LStmumu_AFBLNNT}
and the $q^2$-integrated SM predictions in the kinematic bin are
collected in Table~\ref{tab:LbToLSt-bin-avg}. The observablea are denoted as $P^\ast$ and $A_{FB}^\ast$.

\subsection{Single lepton polarization asymmetries}
The longitudinal polarisation asymmetries satisfy the exact relation
\begin{equation}
P_{L}^{\ast-} = -P_{L}^{\ast+}
\end{equation}
as in the $\Lambda$ channel. This relation follows from the opposite helicity projections of the lepton and antilepton and is exact for the vector and axial-vector operators of the Hamiltonian \eqref{eq:hamiltonian}. Since scalar and pseudo-scalar operators can break this equality, the relation may serve as a null test.  The numerator is proportional to $\sqrt{q^2}\sqrt{q^2-4m_\ell^2}\,\sum_\lambda (|A^R_\lambda|^2 + |B^R_\lambda|^2 - \{R\leftrightarrow L\})$ (see Appendix~\ref{sec:pol-assym-obs-Lst} for analytical expressions), which measures the chiral asymmetry between right- and left-handed amplitudes. In the high $q^2$ bin, $m_\mu^2/q^2<10^{-3}$, so the massless limit is very accurate. Therefore, $P_L^{\ast\pm}$ is close to $\pm 1$ because in the SM $C_9\approx -C_{10}$ which makes the left-handed amplitudes much larger than the right-handed ones.

\begin{figure}[H]
    \centering
    \includegraphics[width=0.4\textwidth]{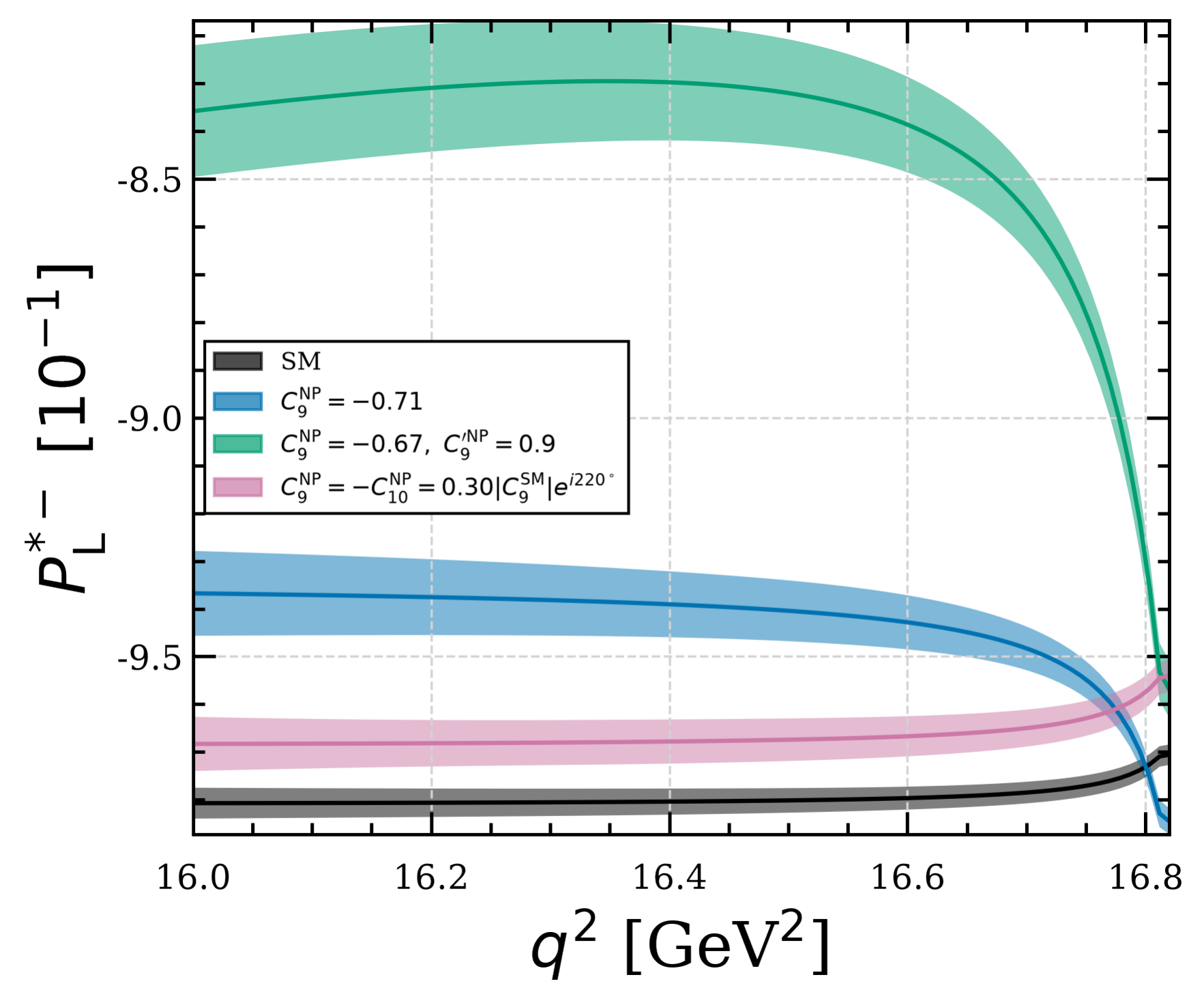}
\includegraphics[width=0.4\textwidth]{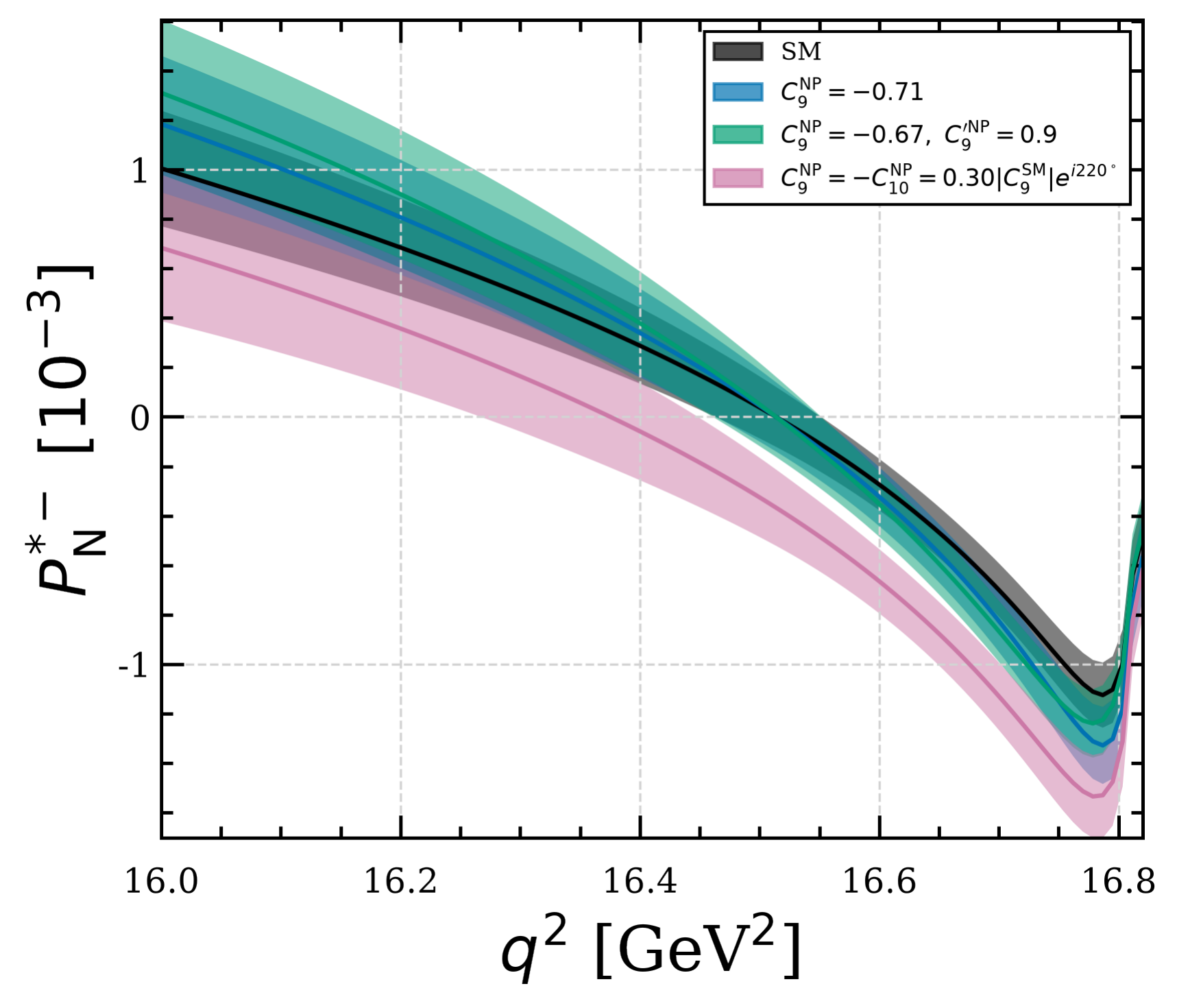}
    \caption{Single lepton polarization asymmetries $P_{L}^{\ast-}$ and $P_{N}^{\ast-}$ as a function of $q^2$ for $\Lambda_b \to \Lambda^\ast \mu^+\mu^-$ decay in the SM and beyond. }
    \label{fig:Lb2LStmumu_PLmiPNmi}
\end{figure}

The SM prediction gives $P_{L}^{\ast-} \approx -0.98$ throughout the kinematic range, as confirmed in Fig.~\ref{fig:Lb2LStmumu_PLmiPNmi}. The NP sensitivity is similar to the $\Lambda$ channel. All three benchmarks shift $P_{L}^{\ast\pm}$ toward less negative values across the whole bin, with benchmarks I and II giving the largest shifts. The uncertainty bands, however, decrease substantially near the endpoint. The $P_{L}^{\ast+}$ plot is the mirror image, with all values positive and approaching $+1$; it is therefore not shown.

The normal polarization asymmetries $P_{N}^{
\ast\pm}$ are proportional to imaginary parts of interference terms between transversity amplitudes,
\be
P_{N}^{\ast\pm} \propto m_\ell\sqrt{q^2 - 4m_{\ell}^2} \, \text{Im} \left[ \pm A_{\parallel 1}^R (A_{\perp 1}^L)^* \pm B_{\parallel 1}^R (B_{\perp 1}^L)^* + (A_{\parallel t}^L - A_{\parallel t}^R)((A_{\parallel 0}^L)^* - (A_{\parallel 0}^R)^*) + \{ \parallel \leftrightarrow \perp \} \right]\, .
\ee
In the second term $(A_{\parallel t}^L - A_{\parallel t}^R)((A_{\parallel 0}^L)^* - (A_{\parallel 0}^R)^*)$ the imaginary part cancels, since both of its factors are proportional to the same combination of $C_{10}$ and $C_{10}^\prime$. Only the first term contributes, and it changes sign between $\ell^+$ and $\ell^-$, so
\begin{equation}
P_{N}^{\ast-} = -P_{N}^{\ast+}.
\end{equation}

The SM and all three benchmark scenarios cross zero near $q^2\approx16.5\,\text{GeV}^2$. To understand the reason, note that in the SM and in benchmarks I and II, $C_{10}$ is real. Apart from a small contribution from the imaginary part of $C_7^{\rm eff}$, which comes from higher-order charm- and up-quark loops, the numerator of $P_N^{\ast\pm}$ is ${\rm Im}[(C_9^{\rm eff}\pm C_{10})(C_7^{\rm eff})^\ast]$ times a combination of form factors. So the zero crossing is fixed mainly by the form factors. Because $C_{7}^{\rm eff}$ and $C_{10}$ are the same in these three scenarios, numerators are identical, and the curves differ only through the total rate. In the benchmark III $C_{10}$ has a complex phase, the sign of ${\rm Im}(C_9^{\rm eff} C_{10}^\ast)$ is reversed. There is additional ${\rm Im}(C_{10} C_7^{\rm eff})$ which changes the zero crossing slightly. The figures for all four scenarios are shown in Fig.~\ref{fig:Lb2LStmumu_PLmiPNmi}.

\begin{figure}[H]
    \centering
    \includegraphics[width=0.4\textwidth]{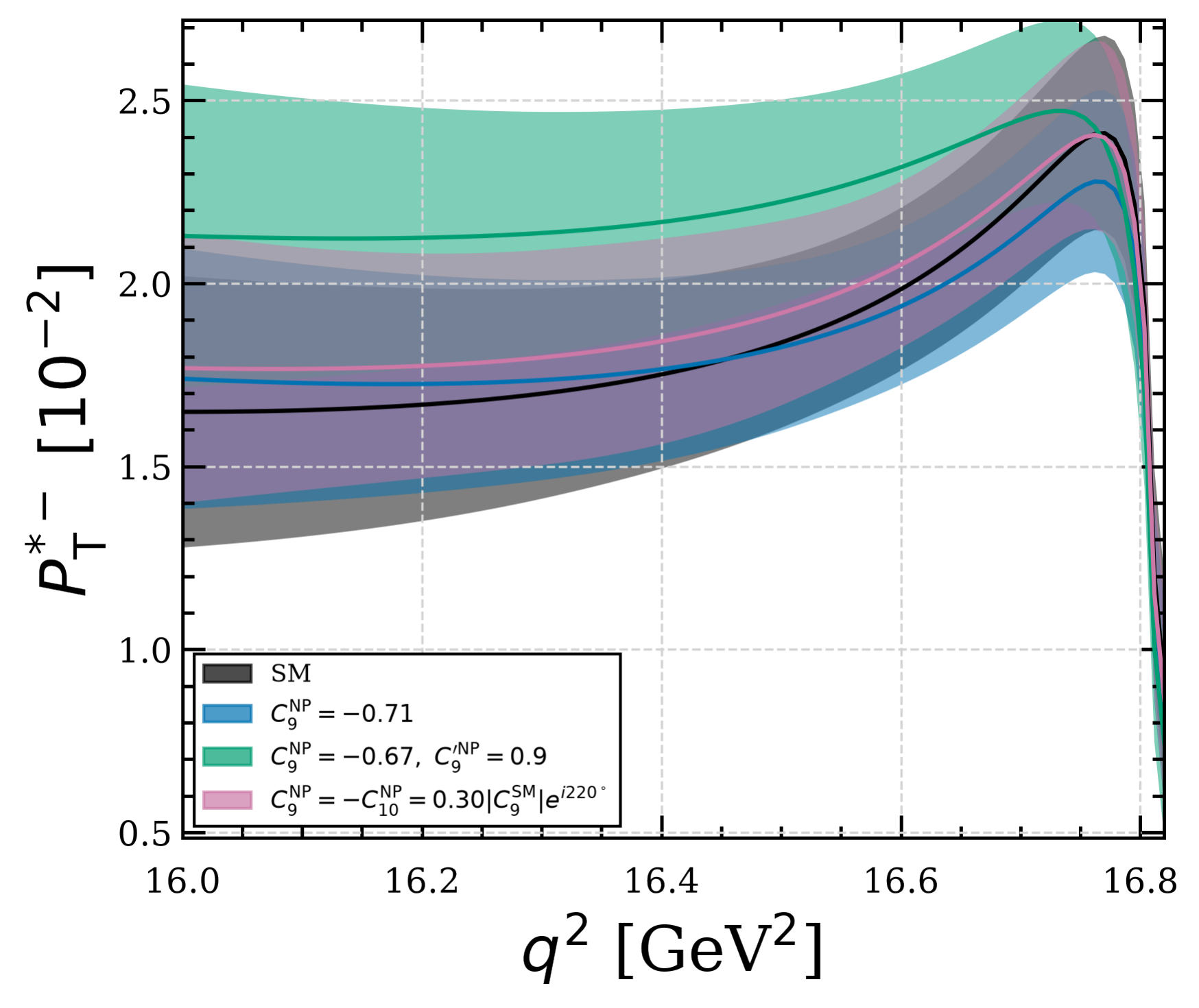}
    \includegraphics[width=0.4\textwidth]{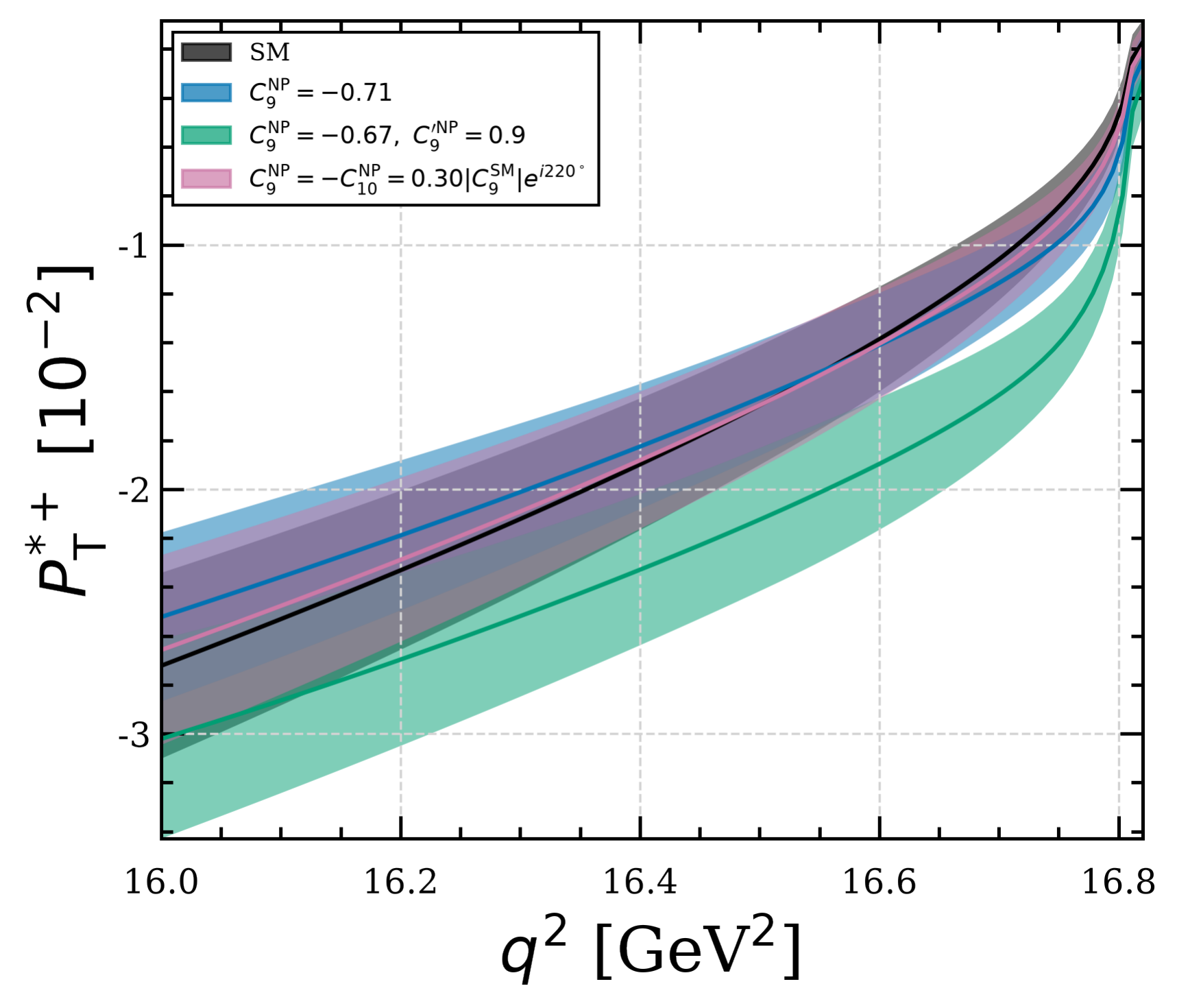}
    \caption{Single lepton polarization asymmetry $P_{T}^{\ast-}$ and $P_{T}^{\ast+}$ as a function of $q^2$ for $\Lambda_b \to \Lambda^\ast \mu^+\mu^-$ decay in the SM and beyond.}
    \label{fig:Lb2LStmumu_PTmiPTpl}
\end{figure}

The transverse asymmetries $P_{T}^{\ast\pm}$ are both lepton-mass suppressed  and are of order $\mathcal{O}(10^{-2})$, with bin values $P_{T}^{\ast-} = (+1.76\pm0.29)\times10^{-2}$ and $P_{T}^{\ast+} = (-2.08\pm0.25)\times10^{-2}$. The NP sensitivity of $P_{T}^{\ast-}$ and $P_{T}^{\ast+}$ is moderate but distinguishable between scenarios as shown in Fig.~\ref{fig:Lb2LStmumu_PTmiPTpl}. Benchmark II (green) gives the largest size in both, about $2.2\times 10^{-2}$ for $P_{T}^{\ast-}$ and $-3\times 10^{-2}$ for $P_{T}^{\ast+}$ at the low-end of the bin. In scenario III (pink), both $P_{T}^{\ast-}$ and $P_{T}^{\ast+}$ lie within the SM band. As can be seen from the expressions in Appendix~\ref{sec:pol-assym-obs-Lst}, the difference between them originates from a sign flip in the interference term between timelike and longitudinal amplitudes.

\subsection{Double lepton polarization asymmetries}
\begin{figure}[ht]
    \centering
    \includegraphics[width=0.34\textwidth]{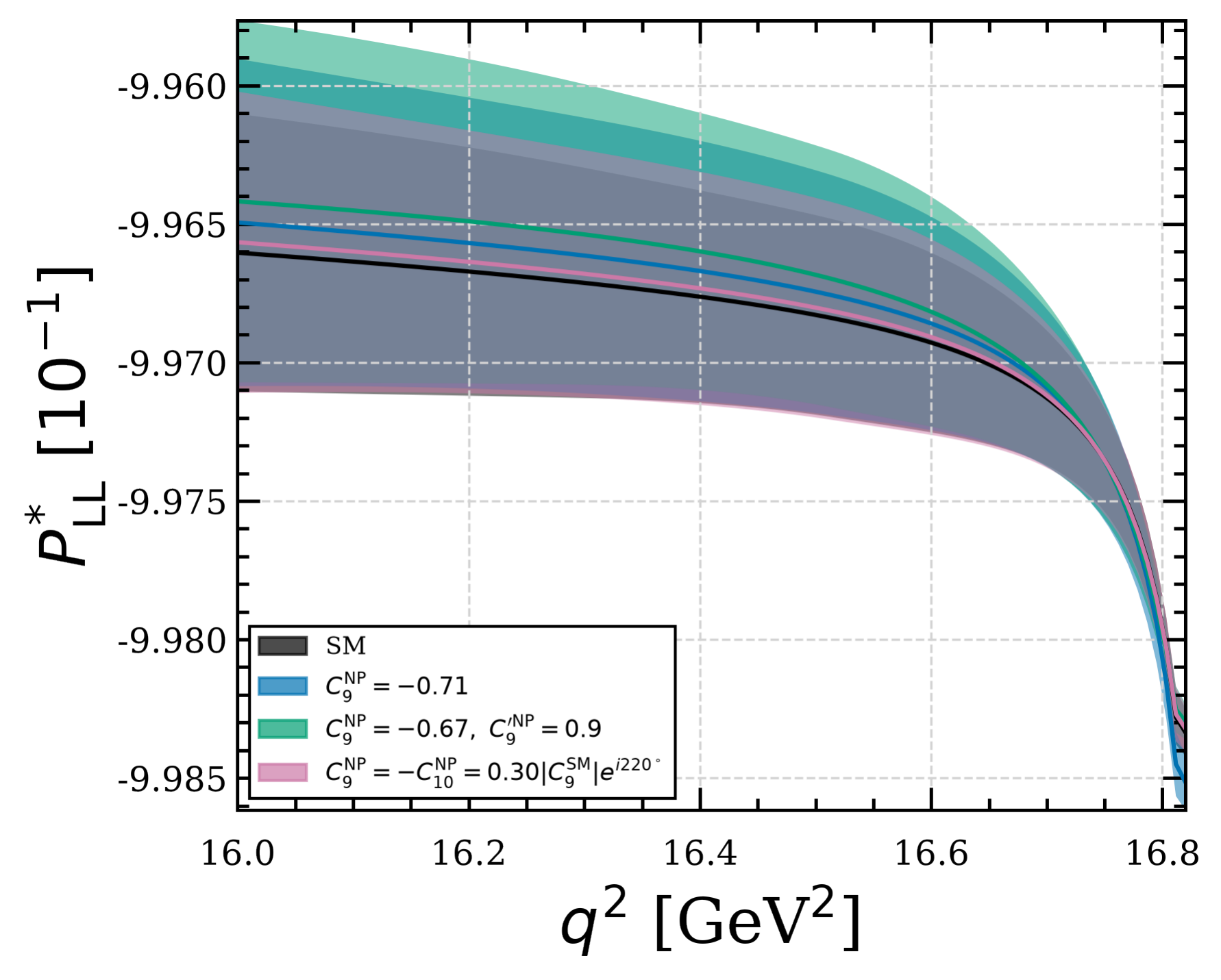}\hfill
    \includegraphics[width=0.33\textwidth]{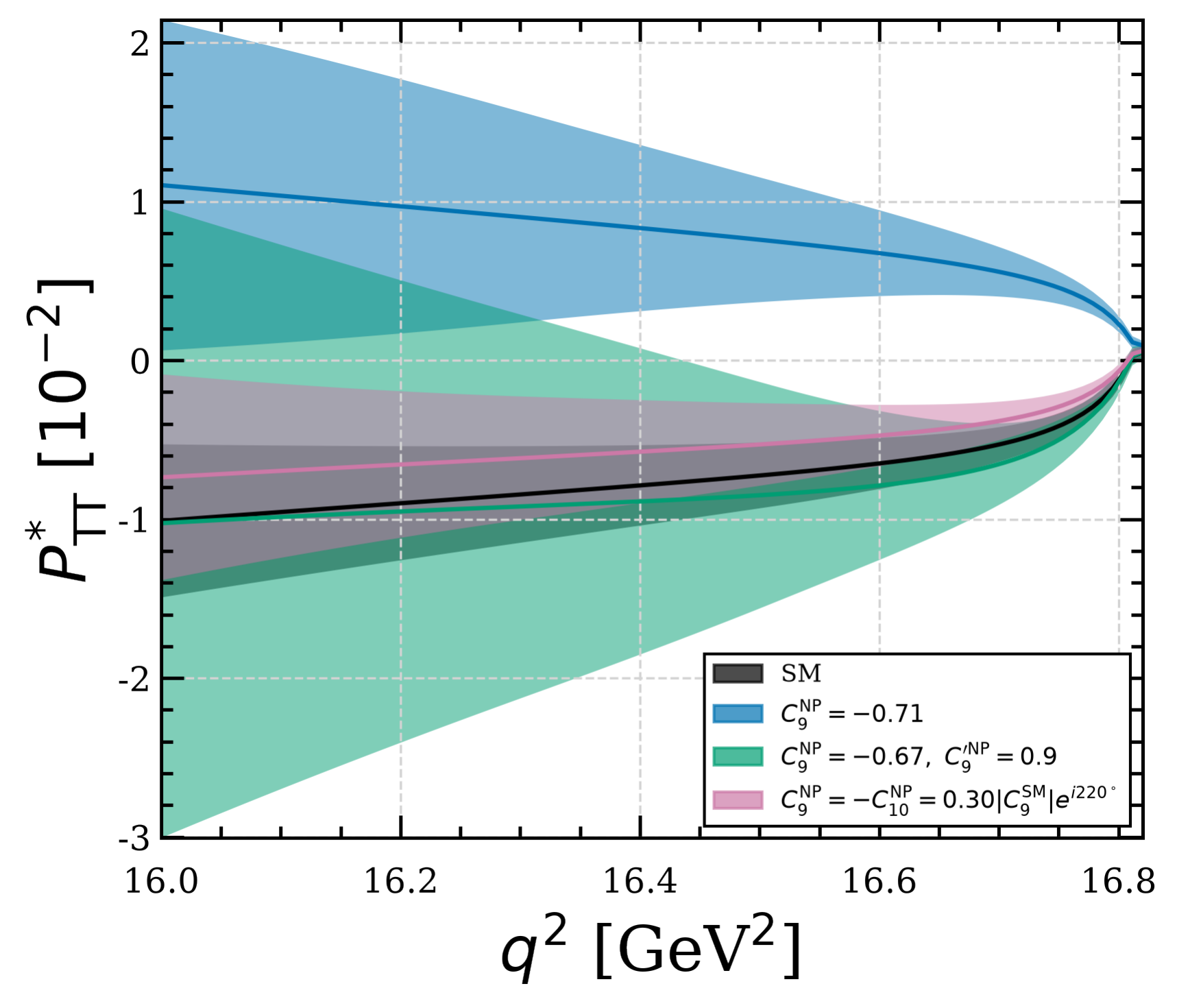}\hfill
    \includegraphics[width=0.33\textwidth]{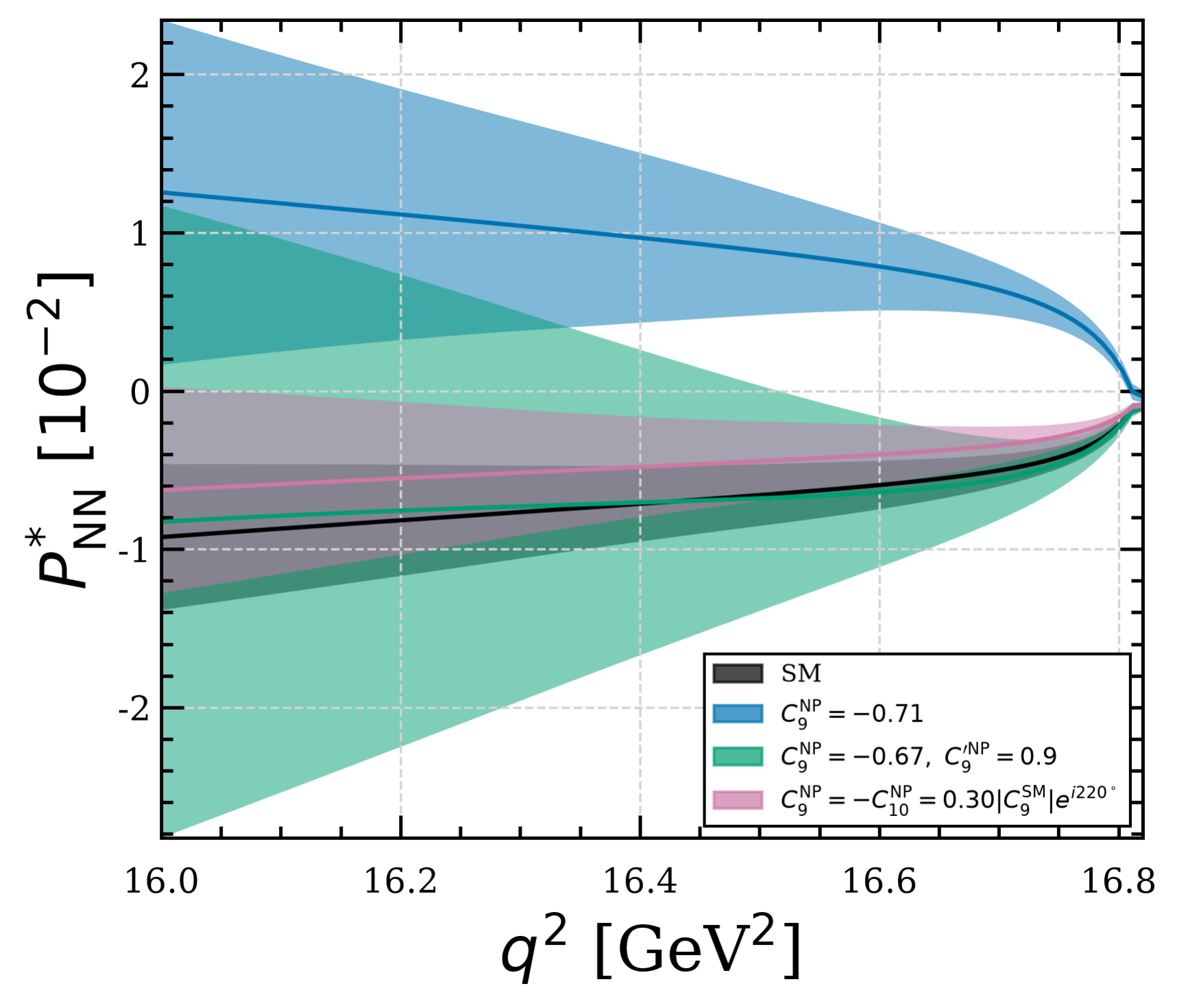}
    \caption{Double lepton polarization asymmetries $P_{LL}^{\ast}$,$P_{TT}^{\ast}$,$P_{NN}^{\ast}$ as a function of $q^2$ for $\Lambda_b \to \Lst \mu^+\mu^-$ in the SM and beyond.}
    \label{fig:Lb2LStmumu_PLLPTTPNN}
\end{figure}

The asymmetry $P_{LL}^{\ast}$ is close to $-1$ throughout the kinematic range, with bin value $(-9.97\pm0.004)\times10^{-1}$. As in the $\Lambda$ channel, $P_{LL}^\ast\to -1$ in the massless limit for any $q^2$ and independently of the Wilson coefficients, because vector and axial-vector currents always produce the two leptons with opposite helicities. The deviation from -1 comes from the $m_{\ell}^2$ terms in Eq.(\ref{eq:PLL_Lst}) and here $m_{\mu}^2/q^2\approx 7\times 10^{-4}$. The NP sensitivity is negligible: the four scenarios differ by atmost $\sim10^{-3}$ (Fig.~\ref{fig:Lb2LStmumu_PLLPTTPNN}) since the limit is independent of the Wilson coefficients and the $m_{\ell}^2$ corrections are common to all scenarios.

The numerators of both $P_{TT}^{\ast}$ and $P_{NN}^{\ast}$ contain the same left–right interference structure
\begin{equation}
P_{TT}^{\ast},\,P_{NN}^{\ast}
\;\propto\;
\operatorname{Re}\!\left(
  q^2\bigl(
    -2A^R_{\parallel 0}A^{L\ast}_{\parallel 0}
    + A^R_{\parallel 1}A^{L\ast}_{\parallel 1}+B^R_{\parallel 1}B^{L\ast}_{\parallel 1}
    + \{\parallel\leftrightarrow\perp\} \bigr) + m_\ell^2 (\cdots)
\right),
\end{equation}
(see Appendix~\ref{sec:pol-assym-obs-Lst} for analytical expressions), so that $P_{TT}^{\ast} \approx P_{NN}^{\ast}$ in the massless limit, as in the $\Lambda$ channel. The SM predictions are close to zero in the bin: $P_{TT}^{\ast} = (-8.30\pm2.82)\times10^{-3}$ and $P_{NN}^{\ast} = (-7.58\pm3.16)\times10^{-3}$. The two numbers differ by about $ 7\times 10^{-4}$, which is the order of $\mathcal{O}(m_\mu^2/q^2)$ correction. Both observables show clear NP sensitivity, comparable to the $\Lambda$ channel at similar $q^2$. The benchmark scenario I produce a well-separated band, distinct from the SM cases, for both $P_{TT}^{\ast}$ and $P_{NN}^{\ast}$ over most of the bin, with the separation closing only as all curves converge to zero at the endpoint. In this scenario, both observables are positive throughout the bin, whereas the benchmark scenarios II and III remain close to the SM and are negative.

The benchmark I scenario (blue) is strongly enhanced in the positive direction in both observables, reaching about $+1\times 10^{-2}$ at $q^2=16 $GeV$^2$ as shown in Fig.~\ref{fig:Lb2LStmumu_PLLPTTPNN}. Benchmark II (green) lies essentially on top of the SM, with a much wider uncertainty band, and is therefore not distinguishable from it in these two observables. Because the two observables agree up to $\mathcal{O}(m_{\mu}^2/q^2)$, corrections, they carry essentially the same information, and their difference (of about $7\times 10^{-4}$) is well below the quoted theoretical uncertainties. Each is individually a useful discriminator.

\begin{figure}[H]
    \centering
    \includegraphics[width=0.4\textwidth]{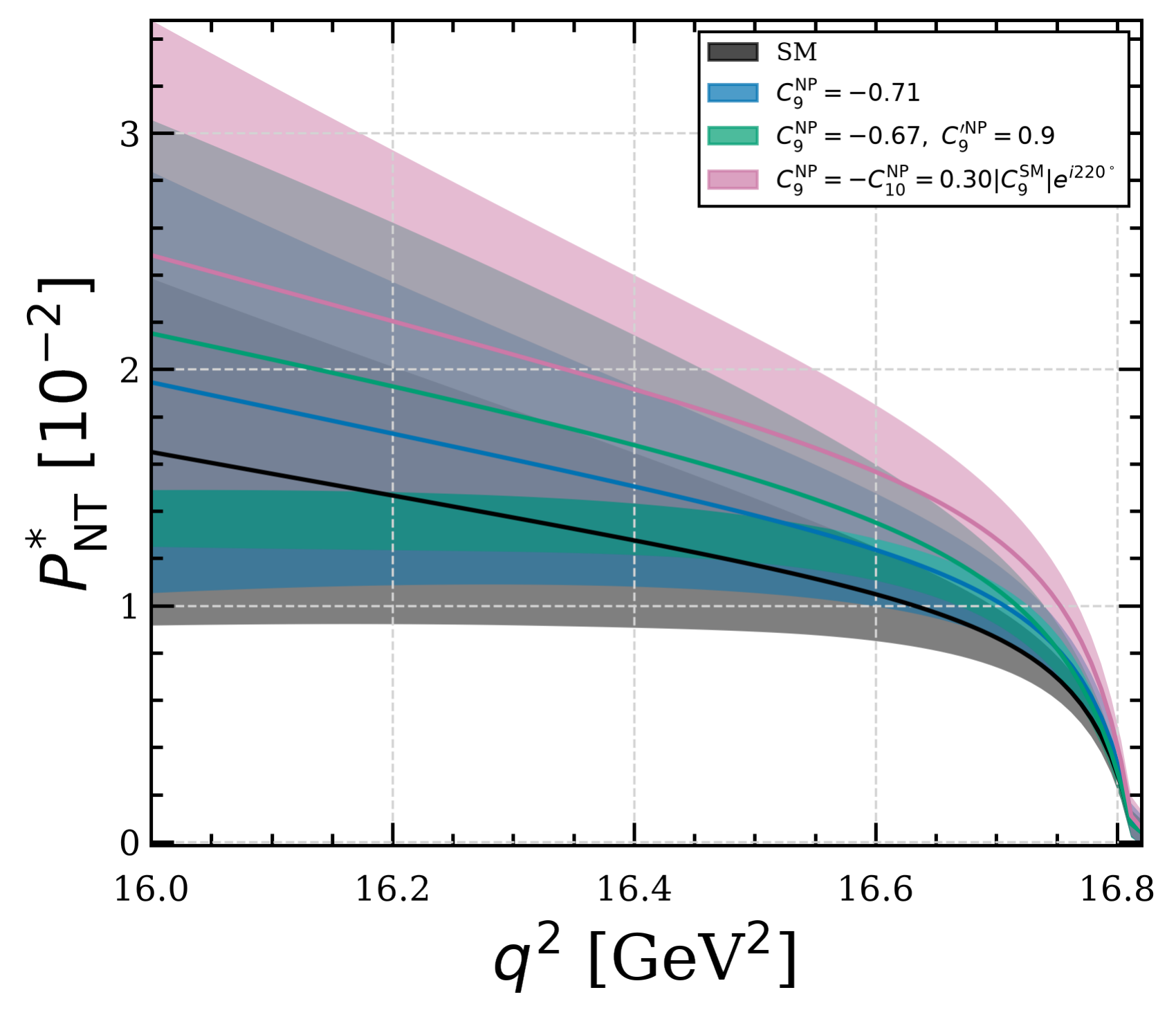}
    \includegraphics[width=0.4\textwidth]{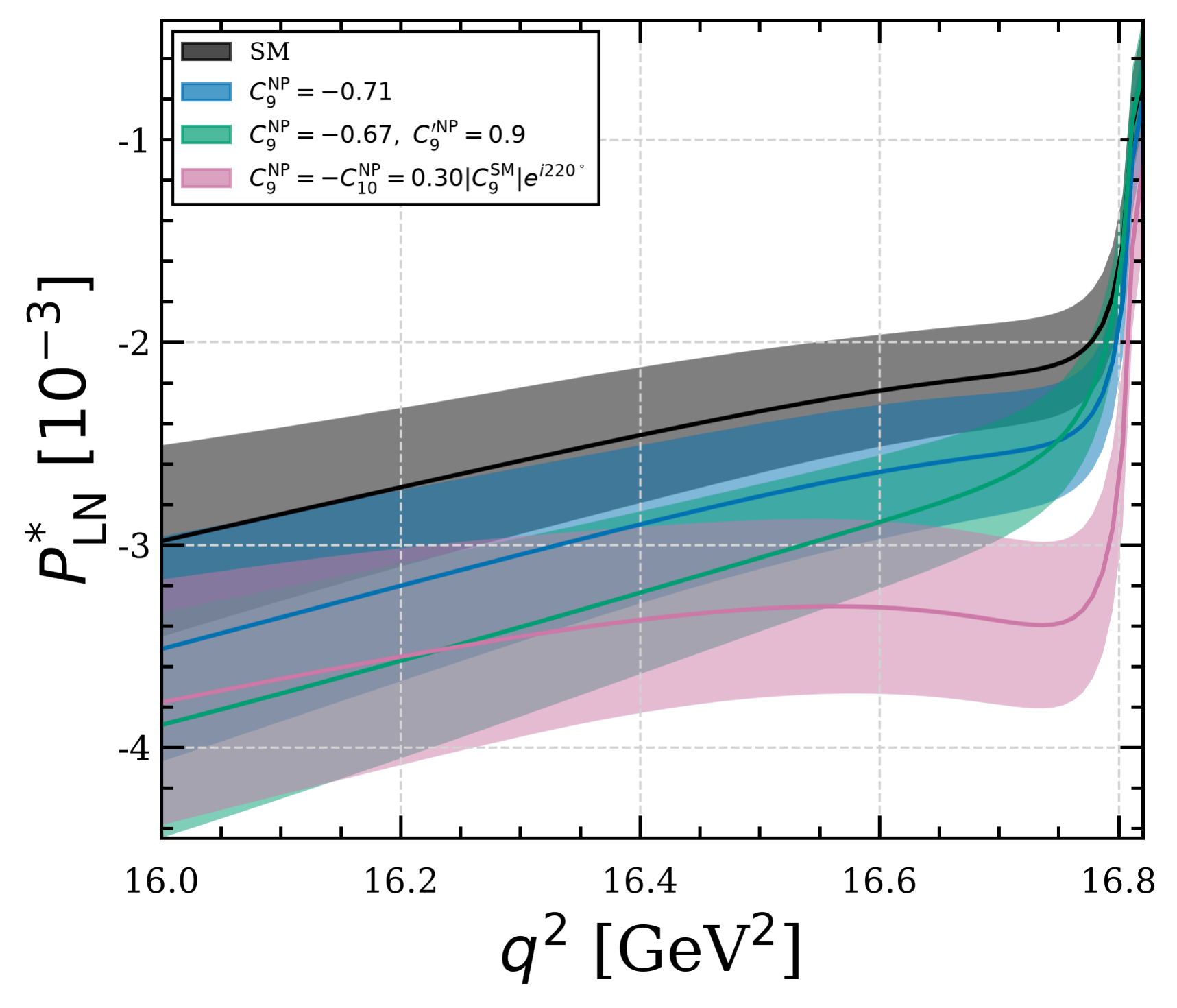}
    \caption{Double lepton polarization asymmetries $P_{NT}^{\ast}$ and $P_{LN}^{\ast}$ as a function of $q^2$ for $\Lambda_b \to \Lst \mu^+\mu^-$ in the SM and beyond.}
    \label{fig:Lb2LStmumu_PNTPLN}
\end{figure}
\begin{figure}[H]
\centering
\includegraphics[width=0.4\textwidth]{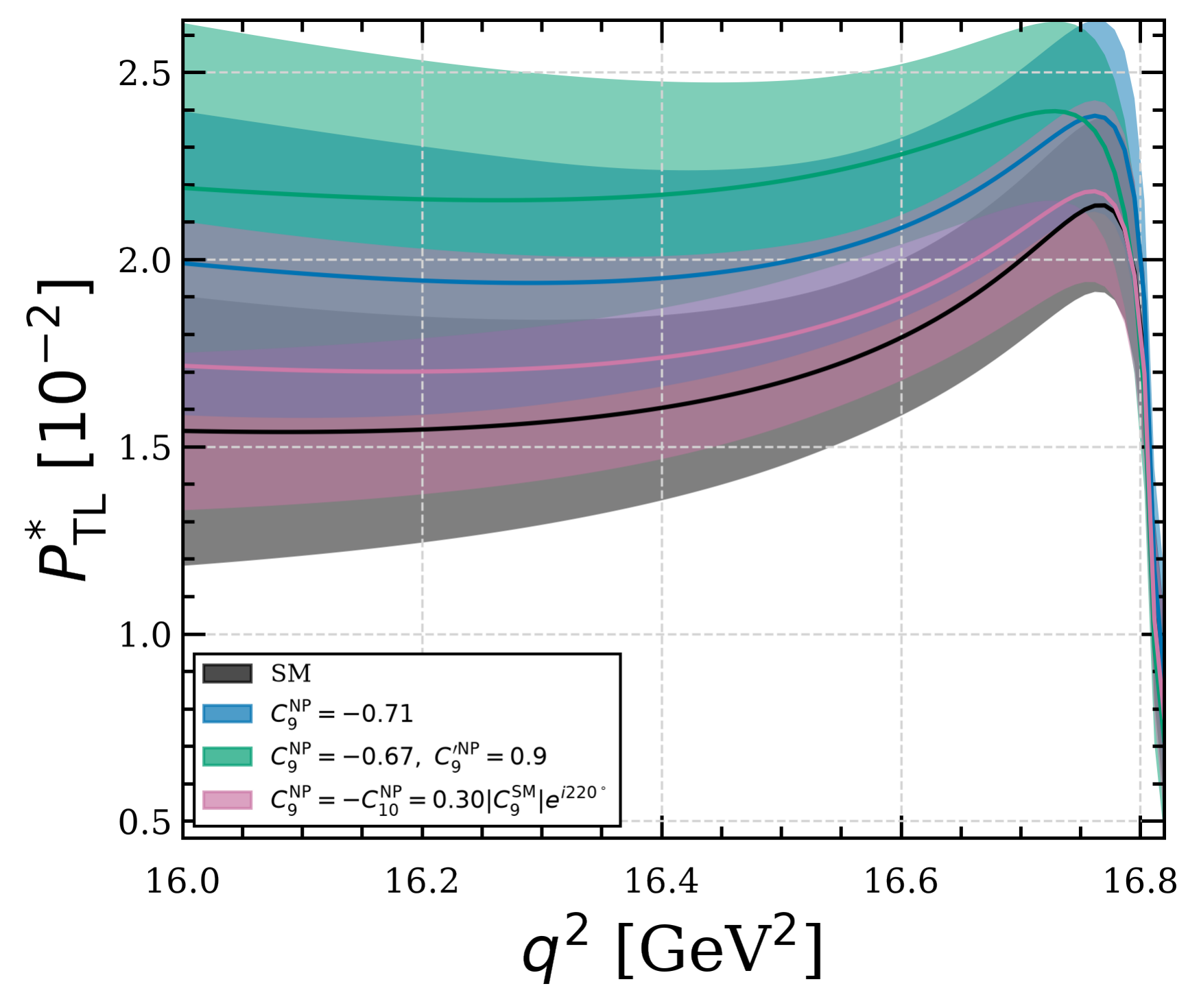}
\includegraphics[width=0.4\textwidth]{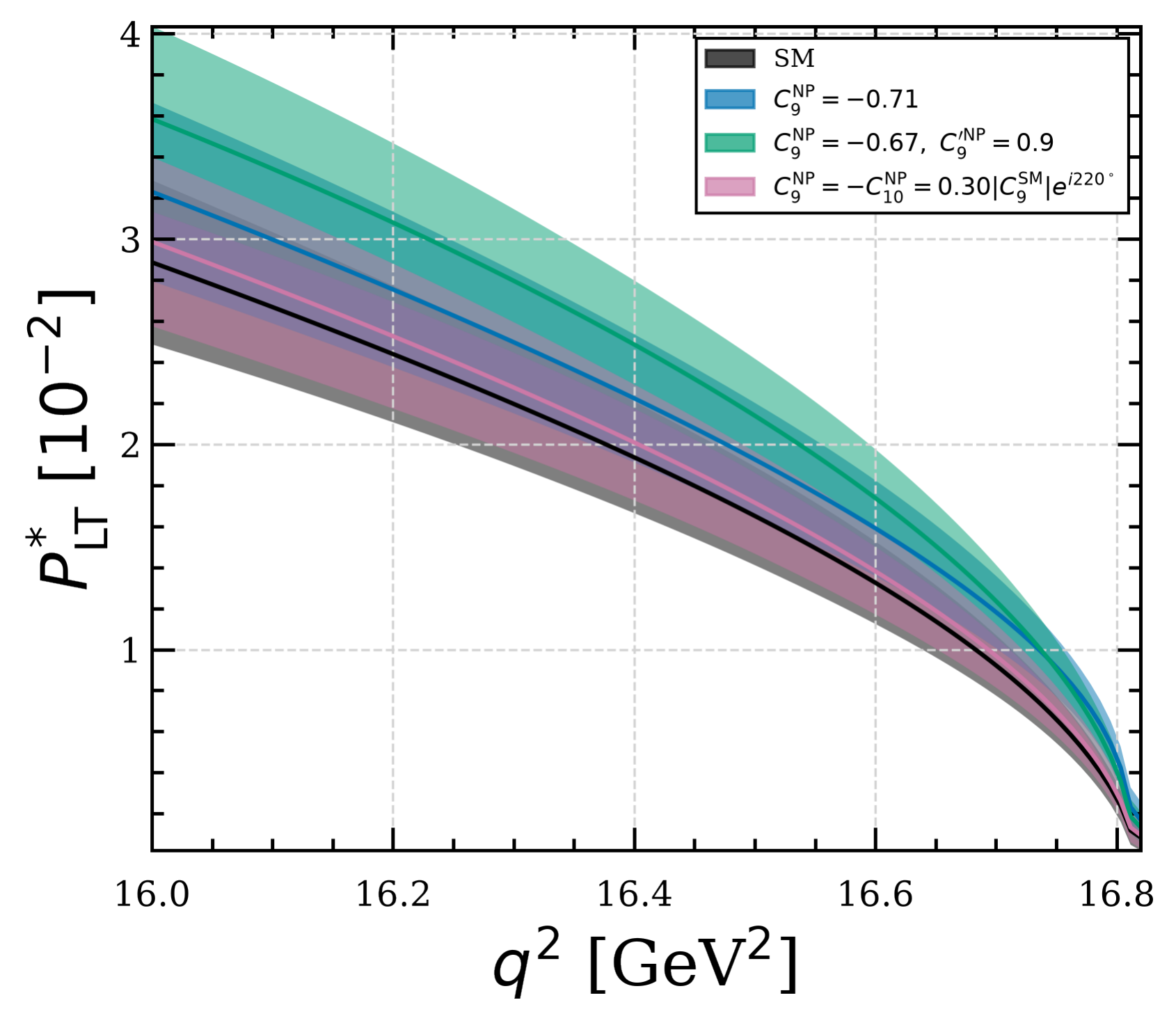}
\caption{Double lepton polarization asymmetries $P_{TL}^{\ast}$ and $P_{LT}^{\ast}$ as a function of $q^2$ for $\Lambda_b \to \Lst \mu^+\mu^-$ in the SM and beyond.}
\label{fig:Lb2LStmumu_PTLPLT}
\end{figure}

The asymmetry $P_{NT}$ is proportional to $\mathrm{Im}(A^R A^{L*})$ with no explicit $m_\ell$ suppression beyond the phase-space factor. For $\Lambda^*$ the expression includes the $B$-amplitude contribution:
\begin{equation}
P_{NT}^{\ast} \propto \sqrt{q^2}\sqrt{q^2-4m_\ell^2}\,
\operatorname{Im}\!\left(
  2A^R_{\parallel 0}A^{L\ast}_{\parallel 0}
  - A^R_{\parallel 1}A^{L\ast}_{\parallel 1}
  - B^R_{\parallel 1}B^{L\ast}_{\parallel 1}
  + \{\parallel\leftrightarrow\perp\}
\right).
\end{equation}
The SM and the three New Physics benchmark scenarios produce a small but nonzero signal at $\mathcal{O}(10^{-2})$, arising from the absorptive imaginary parts of the charm-loop contributions to the effective Wilson coefficients. The New Physics sensitivity is moderate and the benchmarks shift $P_{NT}^{\ast}$ upward, by about $6\times 10^{-3}$ for benchmark III at $q^2$=16 ${\rm GeV}^2$, but all bands overlap with the SM as shown in Fig.~\ref{fig:Lb2LStmumu_PNTPLN}. The bin value is $P_{NT}^{\ast} = (+1.35\pm0.43)\times10^{-2}$ for the SM.

The longitudinal-normal asymmetry $P_{LN}^{\ast}$ presents a qualitatively different picture from the $\Lambda$ channel. In the $\Lambda$ channel, $P_{LN}$ is small but nonzero in the SM, $(4.14 \pm 1.65)\times 10^{-3}$ and $(5.61 \pm 0.37)\times 10^{-3}$ in the low- and high-$q^2$ bins, generated by the absorptive part of $C_9^{\rm eff}$. For the $\Lambda^\ast$, the prediction is of the same order of magnitude but opposite in sign, $P_{LN}^{\ast} = (-2.59\pm0.39)\times10^{-3}$. In scenario III the weak phase enters through the dipole interference $2\kappa_i C_7^{\rm eff}\,\mathrm{Im}\,C_{10}^{\rm NP}$, as in the $\Lambda$ channel. At $q^2\gtrsim16\,\mathrm{GeV}^2$, however, $\mathrm{Im}\,C_9^{\rm eff}$ is large and the SM-like term dominates. As a result
$P_{LN}^{\ast}$ remains nearly flat and does not change sign (Fig.~\ref{fig:Lb2LStmumu_PNTPLN}), in contrast to $P_{LN}$ in the low $q^2$ bin of the $\Lambda$ channel and in line with the $\Lambda$ channel at high $q^2$. It is separated from the SM only for $q^2$ $\gtrsim 16.5\,{\rm GeV}^2$, where the SM uncertainty is smallest. The SM sign itself is opposite in the two channels, which reflects the negative parity of the $\Lambda^\ast$ and provides a consistency check across the two modes rather than a stronger constraint in either one.

The mixed asymmetries $P_{LT}^\ast$ and $P_{TL}^\ast$ involve real parts of timelike–longitudinal interference terms ${\rm Re}(A_{\|t} A_{\|0}^\ast)$ and its $\perp$ counterpart together with same-chirality transverse terms, all multiplied by $m_\ell\sqrt{q^2 - 4m_\ell^2}$. The two differ only in the sign of the same-chirality transverse terms. Two differences from the $\Lambda$ channel appear here. First, both observables change sign: they are negative for $\Lambda$ but positive for $\Lambda^\ast$, $P_{LT}^\ast = (+2.14 \pm 0.28) \times 10^{-2}$ and $P_{TL}^\ast = (+1.62\pm 0.28) \times 10^{-2}$ in the bin. Second, they are much closer to each other than in the high $q^2$ bin of $\Lambda$ where they differ by a factor of 1.8, see table \ref{tab:LbToL-bin-avg}. $P_{LT}^{\ast}$ decreases monotonically with $q^2$, while $P_{TL}^{\ast}$ is nearly flat and rises slightly up to $q^2\approx 16.7$ as shown in Fig.~\ref{fig:Lb2LStmumu_PTLPLT}, before dropping sharply at the kinematic endpoint.

\subsection{Forward-backward asymmetries}
\begin{figure}[H]
    \centering
    \includegraphics[width=0.4\textwidth]{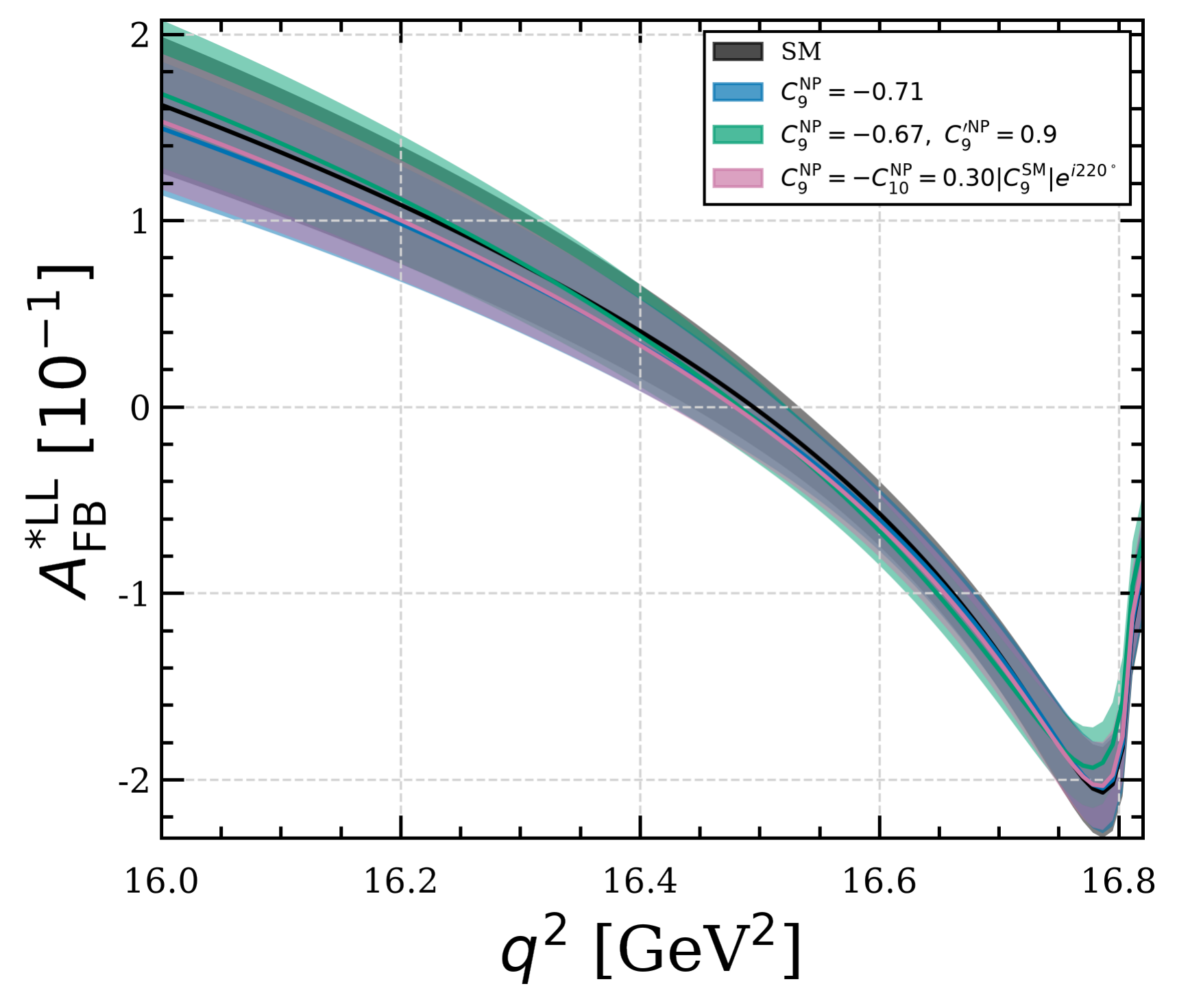}
    \includegraphics[width=0.4\textwidth]{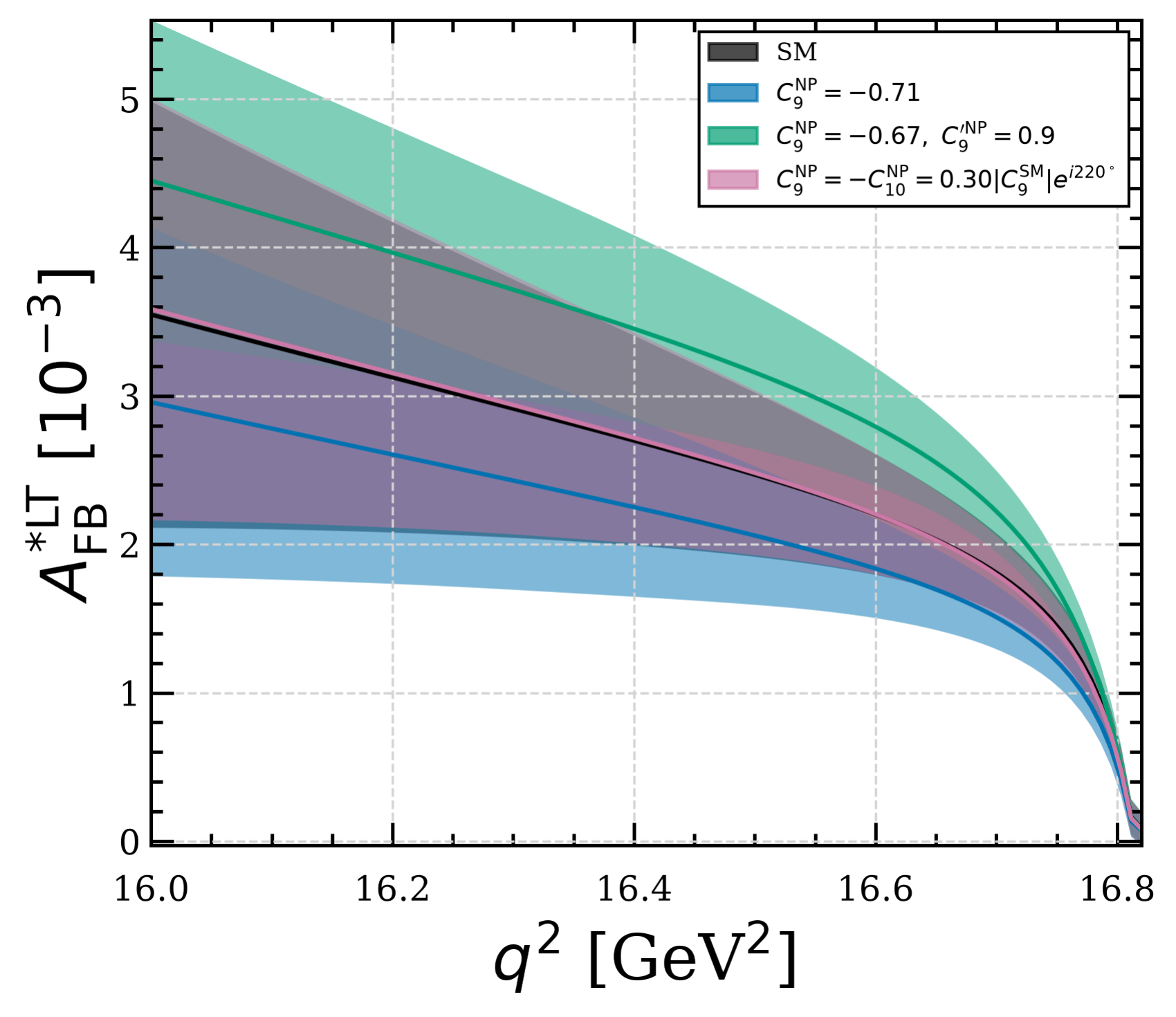}
    \caption{Doubly polarized forward-backward asymmetries $A_{FB}^{\ast LL}$ and $A_{FB}^{\ast LT}$ as a function of $q^2$ for $\Lambda_b \to \Lst \mu^+\mu^-$ in the SM and beyond.}
    \label{fig:Lb2LStmumu_AFBLLLT}
\end{figure}

The forward-backward asymmetry $A_{\small\rm FB}^{\ast LL}$ is proportional to $\mathrm{Re}(A^L_{\parallel 1}A^{L*}_{\perp 1} + B^L_{\parallel 1} B^{L*}_{\perp 1} - \{L\leftrightarrow R\})$ and coincides with the standard leptonic $A^{\ell}_{\small\rm FB}$ in the massless leptonic limit. In fact $P^{(\ast)}_{LT}-P^{(\ast)}_{TL}$=$\frac{\pi m_{\ell}}{\sqrt{q^2}}A_{\small\rm FB}^{(\ast) LL}$ , so only two of them are independent. The SM prediction starts at about $+1.72\times 10^{-1}$ at $q^2 = 16$GeV$^2$, crosses zero near $q^2\approx 16.5$GeV$^2$, reaches its most negative value of about $-2\times 10^{-1}$ near $q^2\approx 16.8$GeV$^2$ and returns to zero at the kinematic endpoint. The integrated bin value is $A_{\tiny \rm FB}^{\ast LL} = (+6.26\pm3.01)\times10^{-2}$. It stays positive despite the zero crossing because the rate weights the lower part of the bin more heavily, and the large relative uncertainty reflects the partial cancellation between the positive and negative parts of the range. The NP benchmark scenarios stay close to the SM and shift the zero crossing slightly to lower $q^2$ as shown in Fig.~\ref{fig:Lb2LStmumu_AFBLLLT}.

The asymmetry $A_{\small\rm FB}^{\ast LT}=-A_{\small\rm FB}^{\ast TL}$ is positive throughout the kinematic range and decreasing monotonically from about $\sim 4\times10^{-3}$ at $q^2 = 16$GeV$^2$ to zero at the kinematic endpoint. The bin value is $(+2.87\pm0.97)\times10^{-3}$. Benchmark I lies below the SM and benchmark II above it throughout the $q^2$ range. Only benchmark III overlaps with the SM as shown in Fig.~\ref{fig:Lb2LStmumu_AFBLLLT}.

\begin{figure}[H]
\centering
\includegraphics[width=0.4\textwidth]{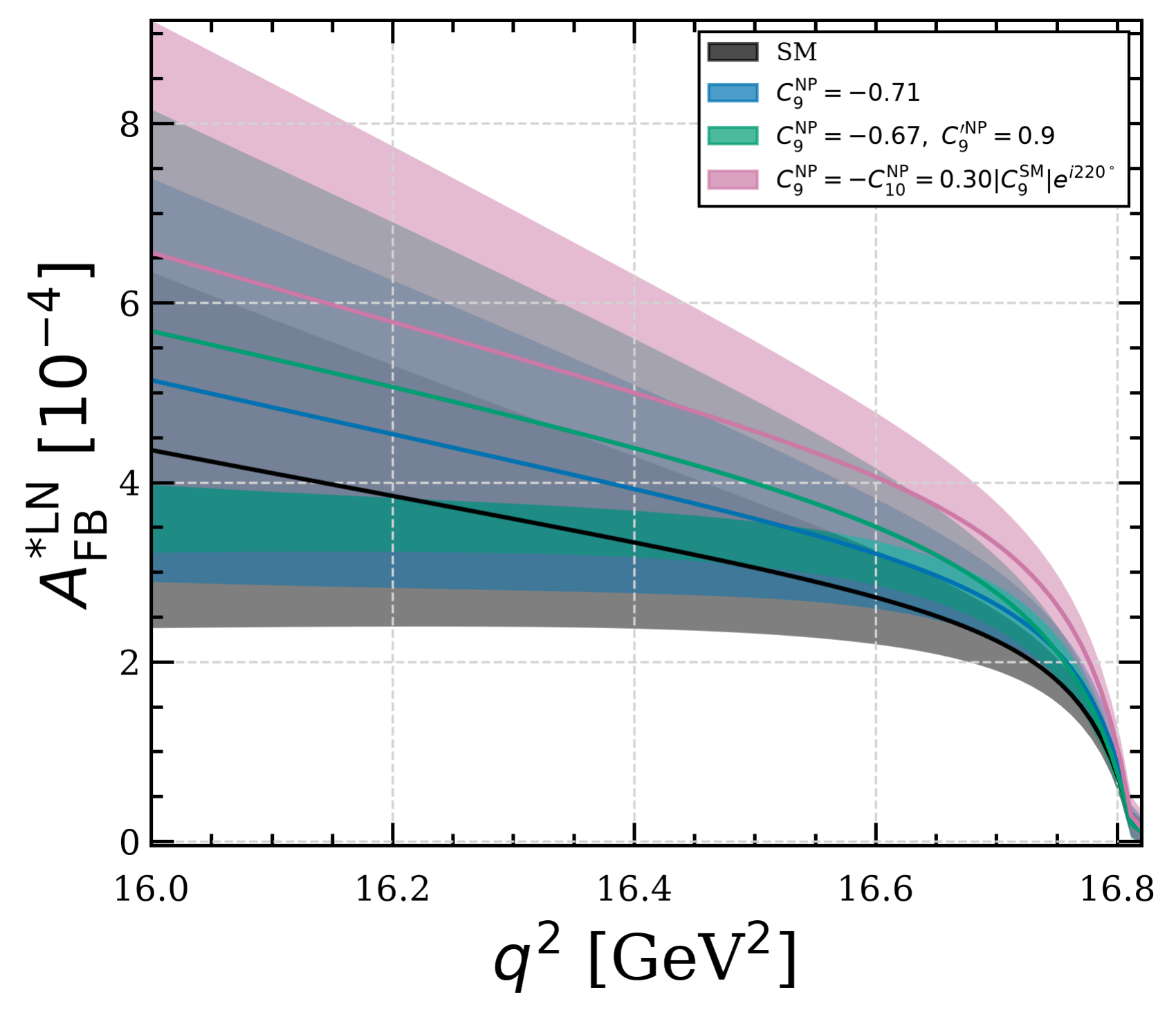}
\includegraphics[width=0.4\textwidth]{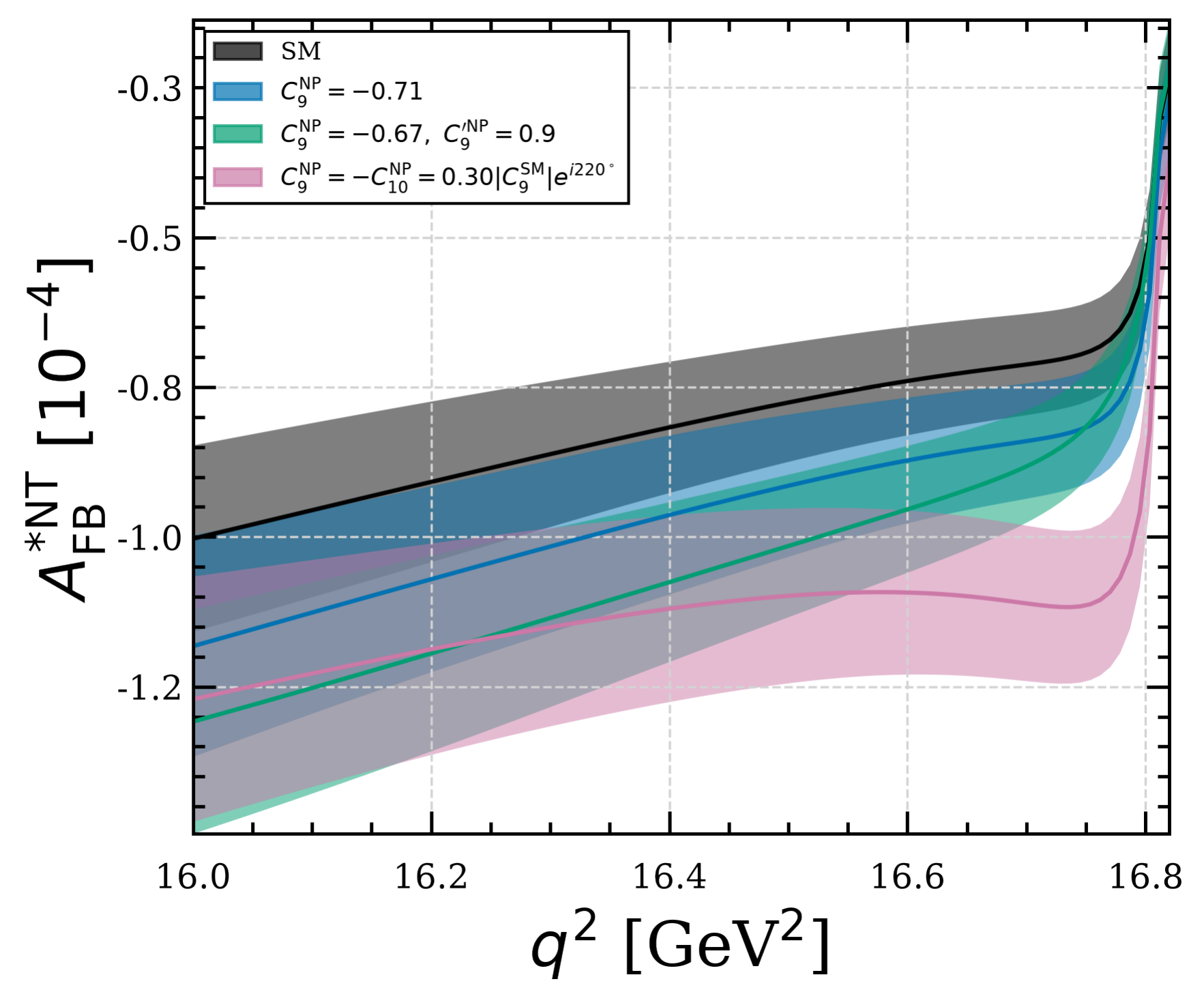}
\caption{Doubly polarized forward-backward asymmetries $A_{FB}^{\ast LN}$ and $A_{FB}^{\ast NT}$ as a function of $q^2$ for $\Lambda_b \to \Lst \mu^+\mu^-$ in the SM and beyond.}
\label{fig:Lb2LStmumu_AFBLNNT}
\end{figure}

The asymmetry $A_{\small\rm FB}^{\ast LN} =A_{\small\rm FB}^{\ast NL}$  is proportional to $\mathrm{Im}(2A^R_{\parallel 0}A^{L*}_{\parallel 0} - A^R_{\parallel 1}A^{L*}_{\parallel 1} - B^R_{\parallel 1}B^{L*}_{\parallel 1} + \ldots)$, which is the same imaginary combination as $P_{NT}^{\ast}$. In fact, $A_{\small\rm FB}^{(\ast) LN} = (m_\ell/\sqrt{q^2})P_{NT}^{(\ast)}$ exactly, so the two observables are not independent. The SM and the three New Physics benchmark scenarios produce a small positive signal of order $\mathcal{O}(10^{-4})$, and three New Physics benchmarks lie above the SM as shown in Fig.~\ref{fig:Lb2LStmumu_AFBLNNT}. The bin value is $A_{FB}^{\ast LN} = (+3.54\pm1.20)\times10^{-4}$ for the SM. The magnitude of SM value for $\Lambda^\ast$ closely matches the $\Lambda$ channel high-$q^2$ value (see Table ~\ref{tab:LbToL-bin-avg}) .

$A^{\ast NT}_{\small\rm FB}$ is nonzero in both channels, driven by the same charm-loop absorptive contribution to $C_9^{\rm eff}$.  The expression involves $\mathrm{Im}(A^R_{\parallel t}A^{L*}_{\parallel 0} + A^L_{\parallel 0}A^{L*}_{\parallel t} - A^L_{\parallel t} A^{R*}_{\parallel 0} - A^R_{\parallel 0}A^{R*}_{\parallel t} + \{\parallel\leftrightarrow\perp\})$, which is the same timelike--longitudinal imaginary structure as $P_{LN}^\ast$. In fact $A_{\rm\small FB}^{(\ast) NT} = (4m_\ell/\pi \sqrt{q^2})P_{LN}^{(\ast)}$ so the two are proportional in both channels, and the opposite signs of $A_{\small\rm  FB}^{\ast NT}$ for $\Lambda$ and $\Lambda^\ast$ follow directly from the opposite signs of $P_{LN}^{(\ast)}$.The three New Physics benchmarks lie below the SM and maintain the same shape as $P^\ast_{LN}$ as shown in  Fig.~\ref{fig:Lb2LStmumu_AFBLNNT}. However, it is proportional to $P^\ast_{LN}$ and suppressed by $\mathcal{O}(m_\ell/\sqrt{q^2})$. Therefore $P_{LN}^\ast$ is more accessible than $A_{\small\rm FB}^{\ast NT}$.


\begin{table}[htbp]
    \centering
    
    
        \centering
        \begin{tabular}{ccc} 
            \toprule
            \textbf{Observable} & \textbf{ $ [1.1,6]$ $\text{GeV}^2$} & \textbf{ $ [15, (m_{\Lb}-m_{\Lambda})^2]$ $\text{GeV}^2$} \\
            \midrule
            $P_{L}^-$ & $(-6.71\pm 1.56)\times 10^{-1}$ & $(-9.80\pm 0.02)\times 10^{-1}$ \\
            $P_{N}^-$ & $(4.67\pm 3.44)\times 10^{-4}$ & $(-1.81\pm 0.14)\times 10^{-3}$ \\
            $P_{T}^-$ & $(-2.57\pm 0.64)\times 10^{-1}$ & $(-2.88\pm 0.21)\times 10^{-2}$ \\
             $P_{T}^+$ & $(2.77\pm 0.60)\times 10^{-1}$ & $(5.08\pm 0.22)\times 10^{-2}$ \\
            $P_{LL}$ & $(-8.43\pm 0.78)\times 10^{-1}$ & $(-9.91\pm 0.006)\times 10^{-1}$ \\
            $P_{TT}$ & $(1.17\pm 6.91)\times 10^{-2}$ & $(-7.49\pm 1.44)\times 10^{-3}$ \\
            $P_{NN}$ & $(1.57\pm 0.75)\times 10^{-1}$ & $(-1.25\pm 1.46)\times 10^{-3}$ \\
            $P_{NT}$ & $(1.43\pm 0.24)\times 10^{-2}$ & $(1.16\pm 0.28)\times 10^{-2}$ \\
            $P_{LN}$ & $(4.14\pm 1.65)\times 10^{-3}$ & $(5.61\pm 0.37)\times 10^{-3}$ \\
            $P_{TL}$ & $(-3.29\pm 0.67)\times 10^{-1}$ & $(-3.08\pm 0.22)\times 10^{-2}$ \\
            $P_{LT}$ & $(-3.20\pm 0.67)\times 10^{-1}$ & $(-5.60\pm 0.25)\times 10^{-2}$ \\
            $A_{FB}^{LL}$ & $(3.17\pm 7.12)\times 10^{-2}$ & $(-3.18\pm 0.14)\times 10^{-1}$ \\
            $A_{FB}^{LT}$ & $(2.84\pm 1.18)\times 10^{-2}$ & $(2.30\pm 0.51)\times 10^{-3}$ \\
            $A_{FB}^{LN}$ & $(8.20\pm 1.48)\times 10^{-4}$ & $(2.96\pm 0.70)\times 10^{-4}$ \\
            $A_{FB}^{NT}$ & $(3.19\pm 1.33)\times 10^{-4}$ & $(1.81\pm 0.12)\times 10^{-4}$ \\
            \bottomrule
        \end{tabular}
        \caption{The binned $q^2$ predictions for the single-lepton, double-lepton, and polarized forward-backward asymmetry observables of the $\Lb \to \Lambda \mu ^+ \mu^-$ decay, evaluated in the $[1.1, 6]$ $\text{GeV}^2$ and $[15, (m_{\Lb}-m_{\Lambda})^2]$ $\text{GeV}^2$ intervals.}
        \label{tab:LbToL-bin-avg}
    
\end{table}
\newpage
\begin{table}[htbp]
    \centering
    
    
        \centering
        \begin{tabular}{cc} 
            \toprule
            \textbf{Observable}  & \textbf{ $ [16, (m_{\Lb}-m_{\Lst})^2]$ $\text{GeV}^2$} \\
            \midrule
            $P_{L}^{\ast -}$  & $(-9.80\pm 0.03)\times 10^{-1}$ \\
            $P_{N}^{\ast -}$  & $(4.20\pm 1.66)\times 10^{-4}$ \\
            $P_{T}^{\ast -}$  & $(1.76\pm 0.29)\times 10^{-2}$ \\
            $P_{T}^{\ast +}$  & $(-2.08\pm 0.25)\times 10^{-2}$ \\
            $P_{LL}^{\ast}$  & $(-9.97\pm 0.004)\times 10^{-1}$ \\
            $P_{TT}^{\ast}$  & $(-8.30\pm2.82)\times 10^{-3}$ \\
            $P_{NN}^{\ast}$  & $(-7.58\pm 3.16)\times 10^{-3}$ \\
            $P_{NT}^{\ast}$  & $(1.35\pm 0.43)\times 10^{-2}$ \\
            $P_{LN}^{\ast}$  & $(-2.59\pm 0.39)\times 10^{-3}$ \\
            $P_{TL}^{\ast}$  & $(1.62\pm 0.28)\times 10^{-2}$ \\
            $P_{LT}^{\ast}$ & $(2.14\pm 0.28)\times 10^{-2}$ \\
            $A_{FB}^{\ast LL}$  & $(6.26\pm 3.01)\times 10^{-2}$ \\
            $A_{FB}^{\ast LT}$  & $(2.87\pm 0.97)\times 10^{-3}$ \\
            $A_{FB}^{\ast LN}$  & $(3.54\pm 1.20)\times 10^{-4}$ \\
            $A_{FB}^{\ast NT}$  & $(-8.65\pm 1.09)\times 10^{-5}$ \\
            \bottomrule
        \end{tabular}
        \caption{The binned $q^2$ predictions for the single-lepton, double-lepton, and polarized forward-backward asymmetry observables of the $\Lb \to \Lst \mu ^+ \mu^-$ decay, evaluated in the $[16, (m_{\Lb}-m_{\Lst})^2]$ $\text{GeV}^2$ intervals.}
        \label{tab:LbToLSt-bin-avg}
    
\end{table}

\section{Summary}
In this paper we have studied lepton polarization observables in the rare baryonic decays $\Lambda_b \to \Lambda\ell^+\ell^-$ and $\Lambda_b \to  \Lambda^*\ell^+\ell^-$, both driven by the quark-level transition  $b \to s\ell^+\ell^-$. We have worked in the effective Hamiltonian  framework and computed single lepton polarization asymmetries, double  lepton polarization asymmetries, and polarized forward-backward asymmetries  using the helicity method. The observables are expressed in terms of transversity amplitudes and evaluated using state-of-the-art lattice QCD form factors. The new-physics sensitivity is studied using three benchmark scenarios, two of which are real Wilson coefficients one of which introduces a weak phase in the Wilson coefficients.

For the $\Lambda_b \to \Lambda\mu^+\mu^-$ channel, the single lepton polarization asymmetries exhibit a clear hierarchy in magnitude, with the longitudinal asymmetry being the largest. The normal asymmetry on the other hand is sensitive to the weak phase in the Wilson coefficients, in the presence of the SM strong phase from the absorptive part of $\mathcal{C}^{\rm eff}_9$. Among the double lepton polarization asymmetries, the longitudinal-normal asymmetry $P_{LN}$, together with $A_{\rm FB}^{NT}$, which is proportional to it is the most sensitive double-polarization probe of a weak phase. Among the polarized forward-backward asymmetries the $A^{LL}_{\rm FB}$ reproduces the standard leptonic forward-backward asymmetry in the vanishing lepton mass limit and exhibits a characteristic zero-crossing controlled by the interference of $\mathcal{C}^{\rm eff}_7$ 
and $\mathcal{C}^{\rm eff}_9$.

The $\Lambda_b \to \Lambda^*\mu^+\mu^-$ channel is analysed only in the high-$q^2$ region $[16,16.8]~\mathrm{GeV}^2$, where the lattice QCD form factors are applicable. The spin-3/2 nature of the $\Lambda^*(1520)$ introduces additional transversity amplitudes. As in the $\Lambda$ channel, the single lepton polarization asymmetries exhibit a clear hierarchy in magnitude, with the longitudinal asymmetry being the largest. In the $\Lst$ channel, the effects of CP-violating benchmarks are modest:$P^\ast_{LN}$ and $A_{\rm FB}^{*NT}$ deviate from the SM for $q^2\gtrsim 16.5$ 
$\mathrm{GeV}^2$ without changing sign. Among the polarized forward-backward asymmetries, $A^{*LL}_{\rm FB}$ reproduces the standard leptonic forward-backward asymmetry in the vanishing lepton mass limit. Some of the observables are related exactly, e.g. $A^{LN}_{\rm FB}\propto P_{NT}$ and $A^{NT}_{\rm FB}\propto P_{LN}$, so they do not carry independent information.

\section*{Acknowledgements}
The authors would like to thank Girish Kumar for fruitful discussions and for useful communications. DD acknowledges support under the SERB SRG grant (Sanction Order No. SRG/2023/001318) and the IIIT Hyderabad Seed Fund (No. IIIT/R\&D Office/Seed-Grant/2021- 22/013). 
RS is supported by the National Natural Science Foundation of China under Grant Nos.~12635005, 12475094, and 12135006, as well as the Science and Technology Innovation Leading Talent Support Program of Henan Province under Grant No.~254000510039.

\appendix
\section{Transversity amplitudes \label{eq:tas}}
The transversity amplitudes which are defined as linear combinations  of the helicity amplitudes. For VA operators we obtain twelve transversity amplitudes
\begin{eqnarray}
A^{L,(R)}_{\perp_1} &=& -\sqrt{2}N \bigg( f^V_\perp \sqrt{2s_-} \mC^{L,(R)}_{\rm VA+} + \frac{2m_b}{q^2} f^T_\perp (\mLb + \mL) \sqrt{2s_-} \mC_7^{\rm eff} \bigg)\, ,\\
A^{L,(R)}_{\|_1} &=& \sqrt{2}N \bigg( f^A_\perp \sqrt{2s_+} \mC^{L,(R)}_{\rm VA-} + \frac{2m_b}{q^2} f^{T5}_\perp (\mLb - \mL) \sqrt{2s_+} \mC_7^{\rm eff} \bigg)\, ,\\
A^{L,(R)}_{\perp_0} &=& \sqrt{2}N \bigg( f^V_0 (\mLb + \mL) \sqrt{\frac{s_-}{q^2}} \mC^{L,(R)}_{\rm VA+} + \frac{2m_b}{q^2} f^T_0\sqrt{q^2s_-} \mC_7^{\rm eff} \bigg)\, ,\\
A^{L,(R)}_{\|_0} &=& -\sqrt{2}N \bigg( f^A_0 (\mLb - \mL) \sqrt{\frac{s_+}{q^2}} \mC^{L,(R)}_{\rm VA-} + \frac{2m_b}{q^2} f^{T5}_0\sqrt{q^2s_+} \mC_7^{\rm eff} \bigg)\, ,\\
A^{L,(R)}_{\perp t} &=& \mp 2\sqrt{2}N f^V_t (\mLb - \mL) \sqrt{\frac{s_+}{q^2}} (\mC_{10} + \mC_{10}^\prime)\, ,\\
A^{L,(R)}_{\| t} &=& \pm 2\sqrt{2}N f^A_t (\mLb + \mL) \sqrt{\frac{s_-}{q^2}} (\mC_{10} - \mC_{10}^\prime) \, .
\end{eqnarray}
The Wilson coefficients $\mC^{\rm eff}_9,\mC_{10},\mC_{9,10}^{(\prime)}$ appear in the VA amplitudes in the following combinations
\begin{align}
& \mC_{\rm VA,+}^{L(R)} = (\mC^{\rm eff}_9\mp \mC_{10})+(\mC_9^\prime \mp \mC_{10}^\prime)\, ,\\
& \mC_{\rm VA,-}^{L(R)} = (\mC^{\rm eff}_9\mp \mC_{10})-(\mC_9^\prime \mp \mC_{10}^\prime)\, ,
\end{align}
and we have defined $s_\pm$ as
\begin{equation}\label{eq:spm}
s_\pm = (\mLb \pm \mL)^2 - q^2\, .
\end{equation}
The $q^2$ dependent normalization constant $N$ is given by 
\begin{equation}
N(q^2) = G_F V_{tb}V_{ts}^\ast \alpha_e \sqrt{ \frac{\sqrt{\lambda(\mmLb,\mmL,q^2)}}{2^{15} m^3_{\Lambda_b} \pi^5 }\beta_\ell}\, ,\quad \beta_\ell = \sqrt{1 - \frac{4m_\ell^2}{q^2}}\, .
\end{equation}
The transversity amplitudes for $\Lb\to\Lst$ are
\begin{align}\label{eq:TAVA1}
& B_{\perp_1}^{L(R)} = \sqrt{2}N \Bigg( f_g^V \sqrt{s_+} \mC_{\rm VA+}^{L(R)} + \frac{2m_b}{q^2} f_g^{T} \sqrt{s_+} \mC_7^{\rm eff}   \bigg)\, , \\
& B_{\|_1}^{L(R)} = \sqrt{2}N \Bigg( f_g^A \sqrt{s_-} \mC_{\rm VA-}^{L(R)} + \frac{2m_b}{q^2} f_g^{T5} \sqrt{s_-} \mC_7^{\rm eff}  \bigg)\, , \\
& A_{\perp_0}^{L(R)} = -\sqrt{2}N \bigg( f_0^V \frac{(\mLb+m_\Lst)}{\sqrt{q^2}} \frac{s_-\sqrt{s_+}}{\sqrt{6}m_\Lst} \mC_{\rm VA+}^{L(R)} + \frac{2m_b}{q^2} f^T_0 \sqrt{q^2} \frac{s_-\sqrt{s_+}}{\sqrt{6}m_\Lst} \mC_7^{\rm eff}\bigg)\, , \\
& A_{\|_0}^{L(R)} = \sqrt{2}N \bigg( f_0^A \frac{(\mLb-m_\Lst)}{\sqrt{q^2}} \frac{s_+\sqrt{s_-}}{\sqrt{6}m_\Lst} \mC_{\rm VA-}^{L(R)} + \frac{2m_b}{q^2} f^{T5}_0 \sqrt{q^2} \frac{s_+\sqrt{s_-}}{\sqrt{6}m_\Lst} \mC_7^{\rm eff}\bigg)\, , \\
& A_{\perp_1}^{L(R)} = -\sqrt{2}N \Bigg( f_\perp^V \frac{s_- \sqrt{s_+}}{\sqrt{3}m_\Lst} \mC_{\rm VA+}^{L(R)} + \frac{2m_b}{q^2} f_\perp^T (\mLb+m_\Lst) \frac{s_- \sqrt{s_+}}{\sqrt{3}m_\Lst} \mC_7^{\rm eff}  \bigg)\, , \\
& A_{\|_1}^{L(R)} = -\sqrt{2}N \Bigg( f_\perp^A \frac{s_+ \sqrt{s_-}}{\sqrt{3}m_\Lst} \mC_{\rm VA-}^{L(R)} + \frac{2m_b}{q^2} f_\perp^{T5} (\mLb-m_\Lst) \frac{s_+ \sqrt{s_-}}{\sqrt{3}m_\Lst} \mC_7^{\rm eff}  \bigg)\, , \\
& A_{\perp_t}^{L(R)} = \mp \sqrt{2}N f^V_t \frac{(\mLb-m_\Lst)}{\sqrt{q^2}} \frac{s_+\sqrt{s_-}}{\sqrt{6}m_\Lst} (\mC_{10} + \mC_{10^\prime} ) \, ,\\
\label{eq:TAVA1end}
& A_{\|_t}^{L(R)} = \pm \sqrt{2}N f^A_t \frac{(\mLb+m_\Lst)}{\sqrt{q^2}} \frac{s_-\sqrt{s_+}}{\sqrt{6}m_\Lst} (\mC_{10} - \mC_{10^\prime} )\, ,
\end{align} 
where $\lambda(a,b,c) = a^2 + b^2 + c^2 - 2(ab+bc+ca)$, and
\begin{equation}
s_\pm = (m_\Lb \pm m_\Lst)^2 - q^2\, .
\end{equation}
The Wilson coefficients $\mC_{\rm V\!A\pm}^{L(R)}$ are defined as follows
\begin{align}
\begin{split}
& \mC_{\rm V\!A+}^{L(R)} = \big( \mC^{\rm eff}_9 \mp \mC_{10} \big) + \big( \mC_{9^\prime} \mp \mC_{10^\prime} \big)\, ,\\
& \mC_{\rm V\!A-}^{L(R)} = \big( \mC^{\rm eff}_9 \mp \mC_{10} \big) - \big( \mC_{9^\prime} \mp \mC_{10^\prime} \big)\, .
\end{split}
\end{align}
\begin{equation}
N = G_F V_{tb}V_{ts}^\ast \alpha_e \sqrt{ \frac{\sqrt{\lambda(\mmLb,m^2_\Lst,q^2)}}{ 2^{15} m_\Lb^3 \pi^5} \beta_\ell}\, ,\quad \beta_\ell = \sqrt{1-\frac{4m_\ell^2}{q^2}}\, ,
\end{equation}
\subsection{Form Factors}
The $\Lambda_b\to \Lambda$ hadronic matrix element is parametrized in terms of 10 form factors which have been taken from \cite{Detmold:2016pkz}. On the other hand 14 form factors contribute to the $\Lambda_b\to \Lambda^\ast$ which are taken from the lattice QCD calculations presented in \cite{Meinel:2021mdj}. For new physics analysis we chose the best fit to NP Wilson coefficients presented \cite{LHCb:2024onj,Fleischer:2022klb,Fleischer:2025ucq} in
Table~\ref{tab:NP_benchmarks}.

\textbf{Parametrization of form factors $\Lambda_b\to \Lambda$} \\
\begin{equation}
 f(q^2) = \frac{1}{1-q^2/(m_{\rm pole}^f)^2} \big[ a_0^f + a_1^f\:z(q^2) \big]. \label{eq:nominalfitphys}
\end{equation}
\begin{equation}
z(q^2) = \frac{\sqrt{t_+-q^2}-\sqrt{t_+-t_0}}{\sqrt{t_+-q^2}+\sqrt{t_+-t_0}}.
\end{equation}
Here, $t_0$ determines the value of $q^2$ that is mapped to $z=0$; we set
\begin{equation}
t_0 = q^2_{\rm max} = (m_{\Lambda_b} - m_{\Lambda})^2.
\end{equation}
Values of $q^2$ greater than $t_+$ will be mapped onto the unit circle in the complex $z$ plane. We set $t_+$ equal
to the onset location of the branch cut associated with the $B\, \bar{K}$ threshold,
\begin{equation}
t_+ = (m_B + m_K)^2. \label{eq:tplus}
\end{equation}
\textbf{Parametrization of form factors $\Lambda_b\to \Lambda^\ast$}
\begin{equation}
 f(q^2) = \frac{1}{1-q^2/(m_{\rm pole}^f)^2} \sum_{n=0}^{1} a_n^f L_n^f \:(w-1)^n, \label{eq:LbL1520nominal}
\end{equation}
\begin{equation}
 w(q^2)=(m_{\Lambda_b}^2+m_{\Lambda^*}^2-q^2)/(2m_{\Lambda_b}m_{\Lambda^*}).
\end{equation}
The factor $L_n^f$ describe the lattice-spacing and pion-mass dependence and in the physical limit, $L_n^f=1$. \\
\textbf{$\Lambda_b \to \Lambda$ hadronic matrix elements}
\begin{align}\label{eq:VAhme1}
\langle \Lambda(k,s_k)|\bar{s}\gamma^\mu b |\Lambda(p,s_p)\rangle =& \bar{u}(k,s_k)\Bigg[f^V_t(q^2)(\mLb-\mL)\frac{q^\mu}{q^2}\nn\\ +& f^V_0(q^2) \frac{\mLb+\mL}{s_+} \{p^\mu + k^\mu  - \frac{q^\mu}{q^2}(\mmLb - \mmL) \} \nn\\ + & f^V_\perp(q^2) \{ \gamma^\mu - \frac{2\mL}{s_+}p^\mu - \frac{2\mLb}{s_+}k^\mu \} \Bigg]u(p,s_p)\, ,
\end{align}
and the axial-vector current is
\begin{align}\label{eq:VAhme2}
\langle \Lambda(k,s_k)|\bar{s}\gamma^\mu\gamma_5 b |\Lambda(p,s_p)\rangle =& - \bar{u}(k,s_k) \gamma_5 \Bigg[ f_t^A(q^2) (\mLb + \mL) \frac{q^\mu}{q^2} \nn\\ +& f_0^A(q^2) \frac{\mLb - \mL}{s_-} \{p^\mu + k^\mu - \frac{q^\mu}{q^2} (\mmLb - \mmL) \} \nn\\ + & f_\perp^A(q^2) \{\gamma^\mu + \frac{2\mL}{s_-}p^\mu - \frac{2\mLb}{s_-}k^\mu \}  \Bigg] u(p,s_p)\, .
\end{align}
For the dipole operators, we get 
\begin{eqnarray}
\langle \Lambda |\bar{s}i q_\nu\sigma^{\mu\nu}b|\Lambda_b\rangle &=& -\bar{u}(k,s_k)\Bigg[ f^T_0(q^2) \frac{q^2}{s_+}\Bigg(p^\mu + k^\mu - \frac{q^\mu}{q^2}(\mmLb - \mmL) \Bigg) \,\nn\\ &+& f^T_\perp (\mLb+\mL) \Bigg( \gamma^\mu - \frac{2\mL}{s_+}p^\mu - \frac{2\mLb}{s_+}k^\mu \Bigg) \Bigg]u(p,s_p)\, ,
\end{eqnarray}
and 
\begin{eqnarray}
\langle \Lambda |\bar{s}i q_\nu\sigma^{\mu\nu}\gamma_5 b|\Lambda_b\rangle &=& -\bar{u}(k,s_k)\gamma_5 \Bigg[f^{T5}_0 \frac{q^2}{s_-} \Bigg( p^\mu + k^\mu - \frac{q^\mu}{q^2}(\mmLb - \mmL) \Bigg) \, \nn\\ &+& f^{T5}_\perp (\mLb - \mL) \Bigg( \gamma^\mu + \frac{2\mL}{s_-}p^\mu - \frac{2\mLb}{s_-}k^\mu \Bigg) \Bigg] u(p,s_p)\, .
\end{eqnarray}\\
\textbf{$\Lambda_b \to \Lambda^\ast$ hadronic matrix elements}\\
\begin{align}\label{eq:ffVA1}
\langle\Lst | \bar s \gamma^\mu b|\Lb\rangle=
& \bar u_\alpha(k,s_\Lst)\biggl\{p^\alpha\biggl[f_t^V(q^2) (m_{\Lb}-m_\Lst)\frac{q^\mu}{q^2}\nn\\
&+ f_0^V(q^2) \frac{m_{\Lb}+m_\Lst}{s_+}(p^\mu +k^\mu-\frac{q^\mu}{q^2}(m_{\Lb}^2-m_\Lst^2))\nn\\
&+ f_\perp^V(q^2)(\gamma^\mu-2\frac{m_\Lst}{s_+}p^\mu -2\frac{m_{\Lb}}{s_+}k^\mu)\biggr]\nn\\
&+ f_g^V(q^2) \left[g^{\alpha\mu}+m_\Lst\frac{p^\alpha}{s_-} \left(\gamma^\mu - 2 \frac{k^\mu}{m_\Lst} +2 \frac{m_\Lst p^\mu +m_\Lb k^\mu}{s_+}\right)\right]\biggr\}u(p,s_{\Lb})\,,\\
\langle\Lst | \bar s \gamma^\mu \gamma ^5 b|\Lb\rangle=& -\bar u_\alpha(k,s_\Lst)\gamma^5\biggl\{p^\alpha\biggl[
f_t^A(q^2) (m_{\Lb}+m_\Lst)\frac{q^\mu}{q^2}\nn\\
&+ f_0^A(q^2) \frac{m_{\Lb}-m_\Lst}{s_-}(p^\mu +k^\mu-\frac{q^\mu}{q^2}(m_{\Lb}^2-m_\Lst^2))\nn\\
&+ f_\perp^A(q^2)(\gamma^\mu+2\frac{m_\Lst}{s_-}p^\mu -2\frac{m_{\Lb}}{s_-}k^\mu)\biggr]\nn\\
&+ f_g^A(q^2) \left[g^{\alpha\mu}-m_\Lst\frac{p^\alpha}{s_+} \left(\gamma^\mu + 2 \frac{k^\mu}{m_\Lst} -2 \frac{m_\Lst p^\mu -m_\Lb k^\mu}{s_-}\right)\right]\biggr\}u(p,s_{\Lb})\,.
\end{align}

The matrix elements corresponding to the tensor and axial-tensor currents are parametrized in terms of six form factors
%
%
\begin{align}
\langle\Lst | \bar si \sigma^{\mu\nu}q_\nu b|\Lb\rangle=& - \bar u_\alpha(k,s_\Lst)\biggl\{p^\alpha\biggl[f_0^T(q^2) \frac{q^2}{s_+}(p^\mu +k^\mu-\frac{q^\mu}{q^2}(m_{\Lb}^2-m_\Lst^2))\nn\\
&+ f_\perp^T(q^2)(m_{\Lb} + m_\Lst)(\gamma^\mu-2\frac{m_\Lst}{s_+}p^\mu -2\frac{m_{\Lb}}{s_+}k^\mu)\biggr]\nn\\
&+ f_g^T(q^2)\left[g^{\alpha\mu}+m_\Lst\frac{p^\alpha}{s_-} \left(\gamma^\mu - 2 \frac{k^\mu}{m_\Lst} +2 \frac{m_\Lst p^\mu +m_\Lb k^\mu}{s_+}\right)\right]\biggr\}u(p,s_{\Lb}),\\
\label{eq:ffT}
\langle\Lst | \bar si \sigma^{\mu\nu}\gamma^5q_\nu b|\Lb\rangle=& - \bar u_\alpha(k,s_\Lst)\gamma^5\biggl\{p^\alpha\biggl[f_0^{T5}(q^2) \frac{q^2}{s_-}(p^\mu +k^\mu-\frac{q^\mu}{q^2}(m_{\Lb}^2-m_\Lst^2))\nn\\
&+ f_\perp^{T5}(q^2)(m_{\Lb} - m_\Lst)(\gamma^\mu+2\frac{m_\Lst}{s_-}p^\mu -2\frac{m_{\Lb}}{s_-}k^\mu)\biggr]\nn\\
&+ f_g^{T5}(q^2)\left[g^{\alpha\mu}-m_\Lst\frac{p^\alpha}{s_+} \left(\gamma^\mu + 2 \frac{k^\mu}{m_\Lst} -2 \frac{m_\Lst p^\mu -m_\Lb k^\mu}{s_-}\right)\right]\biggr\}u(p,s_{\Lb}).
\end{align}

\section{Polarization asymmetry observables for $\Lambda_b \to \Lambda \mu^+ \mu^-$ decay \label{sec:pol-assym-obs-L}}

\begin{align}
\label{eq:PLmi_L}
P_{L}^{-}
&= -P_L^{+} = \frac{-16\sqrt{q^2}\sqrt{q^2 - 4 m_\ell^2}}{3\Delta}
\operatorname{Re}
\Big(
|A^L_{\parallel 0}|^2
+ |A^L_{\parallel 1}|^2
+ |A^L_{\perp 0}|^2
+ |A^L_{\perp 1}|^2-\{L\leftrightarrow R\}
\Big) \\
\label{eq:PTmi_L}
P_{T}^{-}
&= \frac{-4 m_\ell \pi \sqrt{q^2}}{\Delta}\,
\operatorname{Re}
\Big(A^L_{\parallel t}A^{L\ast}_{\parallel 0} - A^R_{\parallel t}A^{L\ast}_{\parallel 0}+A^L_{\parallel t}A^{R\ast}_{\parallel 0} - A^R_{\parallel t}A^{R\ast}_{\parallel 0}+\{\parallel\leftrightarrow\perp\} \nonumber \\
&+A^L_{\parallel 1}A^{L\ast}_{\perp 1} +A^R_{\parallel 1} A^{L\ast}_{\perp 1}+\{L\leftrightarrow R\}\Big)\\
\label{eq:PTpl_L}
P_{T}^{+}
&= \frac{-4 m_\ell \pi \sqrt{q^2}}{\Delta}\,
\operatorname{Re}
\Big(
-  A^L_{\parallel t}A^{L\ast}_{\parallel 0} + A^R_{\parallel t}A^{L\ast}_{\parallel 0}-A^L_{\parallel t}A^{R\ast}_{\parallel 0}+ A^R_{\parallel t}A^{R\ast}_{\parallel 0}+\{\parallel\leftrightarrow\perp\}\nonumber \\
&+A^L_{\parallel 1}A^{L\ast}_{\perp 1} +A^R_{\parallel 1} A^{L\ast}_{\perp 1}+\{L\leftrightarrow R\}
\Big)\\
\label{eq:PNmi_L}
P_{N}^{-}
&= \frac{4 m_\ell \pi \sqrt{q^2 - 4 m_\ell^2}}{\Delta}\,
\operatorname{Im}
\Big(A^R_{\parallel 1} A^{L\ast}_{\perp 1}+ A^L_{\parallel t}A^{L\ast}_{\parallel 0} - A^R_{\parallel t}A^{L\ast}_{\parallel 0}-A^L_{\parallel t}A^{R\ast}_{\parallel 0} +A^R_{\parallel t}A^{R\ast}_{\parallel 0}+\{\parallel\leftrightarrow\perp\}
\Big)\\
\label{eq:PNpl_L}
P_{N}^{+}
&= \frac{4 m_\ell \pi \sqrt{q^2 - 4 m_\ell^2}}{\Delta}\,
\operatorname{Im}
\Big( -A^R_{\parallel 1} A^{L\ast}_{\perp 1}
+ A^L_{\parallel t}A^{L\ast}_{\parallel 0} - A^R_{\parallel t}A^{L\ast}_{\parallel 0}-A^L_{\parallel t}A^{R\ast}_{\parallel 0} +A^R_{\parallel t}A^{R\ast}_{\parallel 0}+\{\parallel\leftrightarrow \perp\}
\Big)\\
\label{eq:PLL_L}
\text{$P_{LL}$}
&= \frac{16}{3\Delta}\operatorname{Re}\Big(-q^2\bigl(|A^L_{\parallel 0}|^2+|A^L_{\parallel 1}|^2+|A^L_{\perp 0}|^2+|A^L_{\perp 1}|^2+\{L \leftrightarrow R\}\bigr) \nn \\
&+ m_\ell^2 \bigl(3|A^L_{\parallel 0}|^2+3|A^L_{\parallel 1}|^2+3|A^L_{\parallel t}|^2+3|A^L_{\perp 0}|^2+3|A^L_{\perp 1}|^2+3|A^L_{\perp t}|^2+\{L\leftrightarrow R\}\nn \\
&- 2 A^R_{\parallel 0} A^{L\ast}_{\parallel 0}-2 A^R_{\parallel 1} A^{L\ast}_{\parallel 1}-6 A^L_{\parallel t} A^{R\ast}_{\parallel t}+\{\parallel\leftrightarrow\perp\}\bigr)\Big)
\\
\label{eq:PTT_L}
\text{$P_{TT}$} &= \frac{16}{3\Delta}\operatorname{Re}\Bigl( q^{2}\bigl(- 2 A^{R}_{\parallel 0} A^{L^\ast}_{\parallel 0} 
 + A^{R}_{\parallel 1} A^{L^\ast}_{\parallel 1} 
 +\{\parallel\leftrightarrow \perp \}\bigr) \nn \\
& - m_{\ell}^{2} \bigl( 
   |A^{L}_{\parallel 0}|^{2} - 2 |A^{L}_{\parallel 1}|^{2} + 3 |A^{L}_{\parallel t}|^{2} 
 + |A^{L}_{\perp 0}|^{2} - 2 |A^{L}_{\perp 1}|^{2} + 3 |A^{L}_{\perp t}|^{2} +\{L \leftrightarrow R\}\nn \\
& - 6 A^{R}_{\parallel 0} A^{L^\ast}_{\parallel 0}-6A^{L}_{\parallel t} A^{R^\ast}_{\parallel t}+\{\parallel \leftrightarrow \perp \}\bigr)\Big) \\
\label{eq:PNN_L}
P_{NN}
&= \frac{16}{3\Delta}\operatorname{Re}\Big(
q^2 \,
\big(
-2 A^R_{\parallel 0} A^{L\ast}_{\parallel 0}
+ A^R_{\parallel 1} A^{L\ast}_{\parallel 1}
+\{\parallel \leftrightarrow \perp\}\big) \nn \\
&- m_\ell^2 \,
\big(
3 |A^L_{\parallel 0}|^2+3 |A^R_{\parallel 0}|^2-3|A^R_{\parallel t}|^2
-3 |A^L_{\parallel t}|^2+\{\parallel \leftrightarrow\perp\} \nn \\ 
&+ 4 A^R_{\parallel 1} A^{L\ast}_{\parallel 1}
-2 A^R_{\parallel 0} A^{L\ast}_{\parallel 0}
+6 A^L_{\parallel t}
 A^{R\ast}_{\parallel t}+\{\parallel \leftrightarrow\perp\}
\big)
\Big) \\
\label{eq:PNT_L}
P_{NT}
&= -P_{TN} =\frac{16}{3\Delta}\sqrt{q^2}\,
\sqrt{q^2 - 4 m_\ell^2}
\operatorname{Im}
\Big(
2 A^R_{\parallel 0} A^{L\ast}_{\parallel 0}
- A^R_{\parallel 1} A^{L\ast}_{\parallel 1}
+ \{\parallel \leftrightarrow \perp\}
\Big) \\
\label{eq:PLT_L}
P_{LT}
&= \frac{4\pi m_\ell\sqrt{q^2 - 4 m_\ell^2}}{\Delta}
\operatorname{Re}\Big(- A^L_{\parallel t} A^{L\ast}_{\parallel 0}+ A^R_{\parallel t} A^{L\ast}_{\parallel 0}+ A^L_{\parallel t} A^{R\ast}_{\parallel 0}- A^R_{\parallel t} A^{R\ast}_{\parallel 0}+\{\parallel\leftrightarrow\perp\}\nn \\
&+A^L_{\parallel 1} A^{L\ast}_{\perp 1}-\{L \leftrightarrow R\}
\Big) \\
\label{eq:PTL_L}
P_{TL}
&= \frac{4\pi m_{\ell}\sqrt{q^2 - 4 m_\ell^2}}{\Delta}\,
\operatorname{Re}\Big(- A^L_{\parallel t} A^{L\ast}_{\parallel 0}+ A^R_{\parallel t} A^{L\ast}_{\parallel 0}+ A^L_{\parallel t} A^{R\ast}_{\parallel 0}- A^R_{\parallel t} A^{R\ast}_{\parallel 0}+ \{\parallel\leftrightarrow\perp\} \nn \\
&- A^L_{\parallel 1} A^{L\ast}_{\perp 1}-\{L \leftrightarrow R\}
\Big) \\
\label{eq:PLN_L}
P_{LN}
&= -P_{NL} =-\frac{4\pi m_{\ell} \sqrt{q^2}}{\Delta}
\operatorname{Im}\Big(A^L_{\parallel t} A^{L\ast}_{\parallel 0}-A^R_{\parallel t} A^{L\ast}_{\parallel 0}+A^L_{\parallel t} A^{R\ast}_{\parallel 0}-A^R_{\parallel t} A^{R\ast}_{\parallel 0}+\{\parallel\leftrightarrow\perp\}
\Big) \\
\label{eq:AFBLL_L}
\text{$A_{FB}^{LL}$} &= \frac{8 \sqrt{q^2}}{\Delta}\sqrt{q^2-4 m_{\ell}^2} \operatorname{Re} \Big( A^L_{\perp 1} A^{L^\ast}_{\parallel 1} - A^R_{\perp 1} A^{R^\ast}_{\parallel 1}\Big) \\
\label{eq:AFBLT_L}
A_{FB}^{LT}
&=-A_{FB}^{TL}= \frac{8m_\ell\sqrt{q^2}}{3\Delta}
\operatorname{Re}\Big(2|A^L_{\parallel 0}|^2+2|A^R_{\parallel 0}|^2-|A^L_{\parallel 1}|^2-|A^R_{\parallel 1}|^2+4A^R_{\parallel 0}A^{L^\ast}_{\parallel 0} \nn \\
&-2A^R_{\parallel 1}A^{L^\ast}_{\parallel 1}+\{\parallel\leftrightarrow\perp\}
\Big) \\
\label{eq:AFBLN_L}
A_{FB}^{LN} &=A_{FB}^{NL}= \frac{16m_\ell \sqrt{q^2-4m_{\ell}^2}}{3\Delta}
\, \text{Im} \bigl(
2 A^R_{\parallel 0} A^{L^\ast}_{\parallel 0}
- A^R_{\parallel 1} A^{L^\ast}_{\parallel 1}
+\{\parallel \leftrightarrow \perp\}\bigr)\\
\label{eq:AFBTN_L}
A_{FB}^{TN} &=A_{FB}^{NT}=-\frac{16 \, m_{\ell}^2}{\Delta} \,
\text{Im} \bigl(
A^L_{\parallel t} A^{L^\ast}_{\parallel 0}
- A^R_{\parallel t} A^{L^\ast}_{\parallel 0}+ A^L_{\parallel t} A^{R^\ast}_{\parallel 0}
- A^R_{\parallel t} A^{R^\ast}_{\parallel 0}+\{\parallel \leftrightarrow \perp\}\bigr)\\
\Delta &= 
\frac{16}{3} \operatorname{Re}\Bigl( q^2\bigl( |A^L_{\parallel 0}|^2 + |A^L_{\parallel 1}|^2 + |A^L_{\perp 0}|^2 + |A^L_{\perp 1}|^2  +\{L \leftrightarrow R\}\bigr)  \nn\\
& - m^2_{\ell}\bigl( |A^L_{\parallel 0}|^2 + |A^L_{\parallel 1}|^2 - 3 |A^L_{\parallel t}|^2 + |A^L_{\perp 0}|^2 + |A^L_{\perp 1}|^2 - 3 |A^L_{\perp t}|^2+\{L \leftrightarrow R\} \nn \\
&-  6 A^R_{\parallel 0} A^{L\ast}_{\parallel 0} -6 A^R_{\parallel 1} A^{L\ast}_{\parallel 1} +6 A^L_{\parallel t} A^{R\ast}_{\parallel t} +\{\parallel\leftrightarrow \perp\} \bigr) \Bigr)
\end{align}
Where
\begin{equation}
\Delta= \sum_{\lambda^-,\lambda^+}\sum_{s_\Lb,s_\Lambda} |\mathcal{M}_{\Lambda_b}^{\lambda^-,\lambda^+}(s_{\Lambda_b},s_{\Lambda})|^2 \nn
\end{equation}
\section{Polarization asymmetry observables for $\Lambda_b \to \Lst \mu^{+} \mu^{-}$ decay} \label{sec:pol-assym-obs-Lst}
\begin{align}
\label{eq:PLmi_Lst}
P_{L}^{\ast -}
&= -P_{L}^{\ast +}=-\frac{\sqrt{q^2}\sqrt{q^2 - 4 m_\ell^2}}{3\Delta^\ast}
\,
\operatorname{Re}
\Big(
|A^L_{\parallel 0}|^2
+ |A^L_{\parallel 1}|^2
+ |A^L_{\perp 0}|^2
+ |A^L_{\perp 1}|^2 \nn \\
&+ |B^L_{\parallel 1}|^2
+ |B^L_{\perp 1}|^2-\{L \leftrightarrow R\}
\Big) \\
%
%
\label{eq:PTmi_Lst}
P_{T}^{\ast -}
&= -\frac{ m_\ell \pi \sqrt{q^2}}{4\Delta^\ast}
\operatorname{Re}
\Big(
 -A^R_{\parallel t} A^{L\ast}_{\parallel 0}+ A^L_{\parallel 0} A^{L\ast}_{\parallel t}+A^L_{\parallel t} A^{R\ast}_{\parallel 0}- A^R_{\parallel 0} A^{R\ast}_{\parallel t}+ A^R_{\perp 1} A^{L\ast}_{\parallel 1}+B^R_{\parallel 1} B^{L\ast}_{\perp 1}+\{\parallel\leftrightarrow\perp\}\nn \\
&+ A^L_{\parallel 1} A^{L\ast}_{\perp 1}+ B^L_{\parallel 1} B^{L\ast}_{\perp 1}+\{L\leftrightarrow R\}\Big)
\\
\label{eq:PTpl_Lst}
P_{T}^{\ast +}
&= -\frac{ m_\ell \pi \sqrt{q^2}}{4\Delta^\ast}
\operatorname{Re}
\Big(
 A^R_{\parallel t} A^{L\ast}_{\parallel 0}- A^L_{\parallel 0} A^{L\ast}_{\parallel t}-A^L_{\parallel t} A^{R\ast}_{\parallel 0}+ A^R_{\parallel 0} A^{R\ast}_{\parallel t}+ A^R_{\perp 1} A^{L\ast}_{\parallel 1}+B^R_{\parallel 1} B^{L\ast}_{\perp 1}+\{\parallel\leftrightarrow\perp\}\nn \\
&+ A^L_{\parallel 1} A^{L\ast}_{\perp 1}+ B^L_{\parallel 1} B^{L\ast}_{\perp 1}+\{L\leftrightarrow R\}\Big)\\
\label{eq:PNmi_Lst}
P_{N}^{\ast -}
&= -\frac{m_\ell \pi \sqrt{q^2 - 4 m_\ell^2}}{4\Delta^\ast}
\operatorname{Im}
\Big(
A^R_{\parallel t} A^{L\ast}_{\parallel 0}
+ A^L_{\parallel 0} A^{L\ast}_{\parallel t}
- A^R_{\parallel 1} A^{L\ast}_{\perp 1}
+ A^L_{\parallel t} A^{R\ast}_{\parallel 0}
+ A^R_{\parallel 0} A^{R\ast}_{\parallel t}\nn \\
&- B^R_{\parallel 1} B^{L\ast}_{\perp 1}+\{\parallel \leftrightarrow \perp\}
\Big) \\
\label{eq:PNpl_Lst}
P_{N}^{\ast +}
&= -\frac{m_\ell \pi \sqrt{q^2 - 4 m_\ell^2}}{4\Delta^\ast}
\operatorname{Im}
\Big(
A^R_{\parallel t} A^{L\ast}_{\parallel 0}
+ A^L_{\parallel 0} A^{L\ast}_{\parallel t}
+ A^R_{\parallel 1} A^{L\ast}_{\perp 1}
+ A^L_{\parallel t} A^{R\ast}_{\parallel 0}
+ A^R_{\parallel 0} A^{R\ast}_{\parallel t}\nn \\
&+ B^R_{\parallel 1} B^{L\ast}_{\perp 1}+\{\parallel \leftrightarrow \perp\}
\Big) \\
\label{eq:PLL_Lst}
P_{LL}^{\ast}
&= \frac{1}{3\Delta^\ast}\operatorname{Re}\Big(-q^2\bigl(|A^L_{\parallel 0}|^2+|A^L_{\parallel 1}|^2+|A^L_{\perp 0}|^2+|A^L_{\perp 1}|^2+|B^L_{\parallel 1}|^2+|B^L_{\perp 1}|^2+\{L\leftrightarrow R\}\bigr)\nn \\
&+m_\ell^2\bigl(3|A^L_{\parallel 0}|^2+3|A^L_{\parallel 1}|^2+3|A^L_{\parallel t}|^2+3|A^L_{\perp 0}|^2+3|A^L_{\perp 1}|^2+3|A^L_{\perp t}|^2+3|B^L_{\parallel 1}|^2\nn \\
&+3|B^L_{\perp 1}|^2+\{L\leftrightarrow R\}- 2A^R_{\parallel 0} A^{L\ast}_{\parallel 0}-2A^R_{\parallel 1} A^{L\ast}_{\parallel 1}-6A^R_{\parallel t} A^{L\ast}_{\parallel t}-2B^R_{\parallel 1} B^{L\ast}_{\parallel 1}+\{\parallel\leftrightarrow \perp\}\bigr)\Big)
\\
\label{eq:PTT_Lst}
P_{TT}^{\ast}
&= \frac{1}{3\Delta^\ast}\operatorname{Re}\Big(
q^2\,
\big(
-2 A^R_{\parallel 0} A^{L\ast}_{\parallel 0}
+ A^R_{\parallel 1} A^{L\ast}_{\parallel 1}
+ B^R_{\parallel 1} B^{L\ast}_{\parallel 1}
+\{\parallel \leftrightarrow \perp\}
\big)
\nonumber\\
&- m_\ell^2\,
\big(
|A^L_{\parallel 0}|^2- 2 |A^L_{\parallel 1}|^2
+ 3 |A^L_{\parallel t}|^2+ |A^L_{\perp 0}|^2- 2 |A^L_{\perp 1}|^2
+ 3 |A^L_{\perp t}|^2- 2 |B^L_{\parallel 1}|^2\nn \\
&- 2 |B^L_{\perp 1}|^2+\{L \leftrightarrow R\}
- 6 A^R_{\parallel 0} A^{L\ast}_{\parallel 0}
-6 A^R_{\parallel t} A^{L\ast}_{\parallel t}+\{\parallel\leftrightarrow
\perp\}
\big)
\Big) \\
%
\label{eq:PNN_Lst}
P_{NN}^{\ast}
&= \frac{1}{3\Delta^\ast}\operatorname{Re}\Big(
q^2\,
\big(
-2 A^R_{\parallel 0} A^{L\ast}_{\parallel 0}
+ A^R_{\parallel 1} A^{L\ast}_{\parallel 1}
+ B^R_{\parallel 1} B^{L\ast}_{\parallel 1}
+ \{\parallel \leftrightarrow \perp\}
\big)
\nonumber\\
&- m_\ell^2\,
\big(
3 |A^L_{\parallel 0}|^2+ 3 |A^R_{\parallel 0}|^2
- 3 |A^L_{\parallel t}|^2- 3 |A^R_{\parallel t}|^2+\{\parallel\leftrightarrow\perp\}
\nonumber\\
&- 2 A^R_{\parallel 0} A^{L\ast}_{\parallel 0}
+ 4 A^R_{\parallel 1} A^{L\ast}_{\parallel 1}
+ 6 A^R_{\parallel t} A^{L\ast}_{\parallel t}
+ 4 B^R_{\parallel 1} B^{L\ast}_{\parallel 1}
+\{\parallel \leftrightarrow \perp\}
\big)
\Big) \\
\label{eq:PNT_Lst}
P_{NT}^{\ast}
&= -P_{TN}^{\ast}= \frac{\sqrt{q^2}\sqrt{q^2 - 4 m_\ell^2}}{3\Delta^\ast}
\operatorname{Im}
\Big(
2 A^R_{\parallel 0} A^{L\ast}_{\parallel 0}
- A^R_{\parallel 1} A^{L\ast}_{\parallel 1}
- B^R_{\parallel 1} B^{L\ast}_{\parallel 1}
+\{\parallel \leftrightarrow \perp \}
\Big) \\
\label{eq:PLT_Lst}
P_{LT}^{\ast}
&= \frac{m_{\ell}\pi}{4\Delta^\ast}\,
\sqrt{q^2 - 4 m_\ell^2}
\operatorname{Re}\Big( A^R_{\parallel t} A^{L\ast}_{\parallel 0}-A^L_{\parallel 0} A^{L\ast}_{\parallel t}-A^R_{\parallel 0} A^{R\ast}_{\parallel t}+A^L_{\parallel t} A^{R\ast}_{\parallel 0}+\{\parallel\leftrightarrow\perp\}\nn \\
&+ A^L_{\parallel 1} A^{L\ast}_{\perp 1}+B^L_{\parallel 1} B^{L\ast}_{\perp 1}-\{L\leftrightarrow R\}\Big) \\
\label{eq:PTL_Lst}
P_{TL}^{\ast} &= \frac{m_{\ell}\pi}{4\Delta^\ast}\,
\sqrt{q^2 - 4 m_\ell^2}
\operatorname{Re}\Big( A^R_{\parallel t} A^{L\ast}_{\parallel 0}-A^L_{\parallel 0} A^{L\ast}_{\parallel t}-A^R_{\parallel 0} A^{R\ast}_{\parallel t}+A^L_{\parallel t} A^{R\ast}_{\parallel 0}+\{\parallel\leftrightarrow\perp\}\nn \\
&- A^L_{\parallel 1} A^{L\ast}_{\perp 1}-B^L_{\parallel 1} B^{L\ast}_{\perp 1}-\{L\leftrightarrow R\}\Big)\\
\label{eq:PLN_Lst}
P_{LN}^{\ast}
&= -P_{NL}^{\ast}= \frac{\pi m_{\ell}\sqrt{q^2}}{4\Delta^\ast}
\operatorname{Im}\Big(A^R_{\parallel t} A^{L\ast}_{\parallel 0}+A^L_{\parallel 0} A^{L\ast}_{\parallel t}-A^L_{\parallel t} A^{R\ast}_{\parallel 0}-A^R_{\parallel 0} A^{R\ast}_{\parallel t}+\{\parallel\leftrightarrow \perp\}\Big)
\\
%
\label{eq:AFBLL_Lst}
\text{$A_{FB}^{\ast LL}$} &= \frac{1}{2\Delta^{\ast}} \sqrt{q^2}\sqrt{q^2-4m^2_{\ell}} \operatorname{Re} \Big( A^L_{\parallel 1} A^{L^\ast}_{\perp 1}  +B^L_{\parallel 1} B^{L^\ast}_{\perp 1} -\{L\leftrightarrow R\}  \Big)  \\
\label{eq:AFBTL_Lst}
\text{$A_{FB}^{\ast TL}$} &= -\text{$A_{FB}^{\ast LT}$}=\frac{m_\ell\sqrt{q^2}}{6\Delta^{\ast}} \operatorname{Re}\big( -2 |A^{L}_{\parallel 0}|^2 +|A^{L}_{\parallel 1}|^2-2 |A^{L}_{\perp 0}|^2+|A^{L}_{\perp 1}|^2+|B^{L}_{\parallel 1}|^2  \nn \\
&+ |B^{L}_{\perp 1}|^2+\{L\leftrightarrow R\}-4 A^R_{\parallel 0}A^{L\ast}_{\parallel 0}+2 A^R_{\parallel 1}A^{L\ast}_{\parallel 1}+2 B^R_{\parallel 1}B^{L\ast}_{\parallel 1}+\{\parallel\leftrightarrow \perp\}\big)\\
\label{eq:AFBLN_Lst}
\text{$A_{FB}^{\ast LN}$} &= \text{$A_{FB}^{\ast NL}$}=\frac{ m_\ell \sqrt{q^2-4m^2_\ell}}{3\Delta^{\ast}} \operatorname{Im} \bigl( 2 A^R_{\parallel 0} A^{L^\ast}_{\parallel 0} - A^R_{\parallel 1} A^{L^\ast}_{\parallel 1} 
- B^R_{\parallel 1} B^{L^\ast}_{\parallel 1} +\{\parallel\leftrightarrow \perp\} \bigr)  \\
\label{eq:AFBTN_Lst}
\text{$A_{FB}^{\ast TN}$} &= \text{$A_{FB}^{\ast NT}$}=\frac{m_{\ell}^2}{\Delta^{\ast}}\operatorname{Im} \bigl( A^R_{\parallel t} A^{L^\ast}_{\parallel 0} + A^L_{\parallel 0} A^{L^\ast}_{\parallel t}  - A^L_{\parallel t} A^{R^\ast}_{\parallel 0} - A^R_{\parallel 0} A^{R^\ast}_{\parallel t}+\{\parallel\leftrightarrow\perp\}\bigr) \\
\Delta^\ast &= 
\frac{1}{3 }\operatorname{Re}\Bigl(q^2 \bigl(
|A^L_{\parallel 0}|^2 + |A^L_{\parallel 1}|^2 + |A^L_{\perp 0}|^2 + |A^L_{\perp 1}|^2
+ |B^L_{\parallel 1}|^2 + |B^L_{\perp 1}|^2 + \{L \leftrightarrow R\}\bigr)\nonumber \\
&- m_\ell^2 \bigl(
|A^L_{\parallel 0}|^2 + |A^L_{\parallel 1}|^2 - 3|A^L_{\parallel t}|^2
+ |A^L_{\perp 0}|^2 + |A^L_{\perp 1}|^2- 3|A^L_{\perp t}|^2
+ |B^L_{\parallel 1}|^2 + |B^L_{\perp 1}|^2 +\{L \leftrightarrow R\}\nonumber \\
&- 6
A^R_{\parallel 0}A^{L\ast}_{\parallel 0}
-6 A^R_{\parallel 1}A^{L\ast}_{\parallel 1}
+6 A^R_{\parallel t}A^{L\ast}_{\parallel t}-6 B^{R}_{\parallel 1}B^{L\ast}_{\parallel 1}+\{\parallel \leftrightarrow \perp\}\bigr)\Bigr)\nn \\
\end{align}
Where
\begin{equation}
\Delta^\ast = \sum_{\lambda^-,\lambda^+}\sum_{s_\Lb,s_\Lst} |\mathcal{M}_{\Lambda_b}^{\lambda^-,\lambda^+}(s_{\Lambda_b},s_{\Lst})|^2 \nn
\end{equation}

%

\section{Lorentz Boost \label{sec:LB}}
We now boost this orthogonal system to the $\ell^+\ell^-$ rest-frame by the boost matrix
 \begin{center}
 $\begin{pmatrix}
 \gamma & -\gamma v_x &  -\gamma v_y &  -\gamma v_z \\
  -\gamma v_x & 1+(\gamma -1)\frac{v_x^2}{v^2} & (\gamma -1)\frac{v_x v_y}{v^2} &  (\gamma -1)\frac{v_x v_z}{v^2} \\
   -\gamma v_y &  (\gamma -1)\frac{v_x v_y}{v^2} &  1+(\gamma -1)\frac{v_y^2}{v^2} & (\gamma -1)\frac{v_y v_z}{v^2} \\
    -\gamma v_z & (\gamma -1)\frac{v_x v_z}{v^2} & (\gamma -1)\frac{v_z v_y}{v^2} & 1+(\gamma -1)\frac{v_z^2}{v^2}
    \end{pmatrix}$ \\
    \end{center} 
where $\gamma = 1/\sqrt{1-v^2}$ is the Lorentz boost factor, $v$ is the velocity of the lepton in the dilepton rest frame in the $c=1$ unit, and $v_{x,y,z}$ are the components. As the leptons $\ell^-$,$\ell^+$ are in the $x-z$ plane by choice (see kinematics above) then $v_y=0$. Also, according to our choice, $\ell^-$ makes an angle of $\theta_\ell$ with the positive $z$-axis in the first quadrant. 
%
From the 4-momentum $q^\mu_-$ of the $\ell^-$ we get $\gamma = {\sqrt{q^{2}}}/{2m_{\ell}}$, $v_x = ({2|q_{2\ell}|}/{\sqrt{q^2}})\sin\theta_\ell $ and $v_z = ({2|q_{2\ell}|}/{\sqrt{q^2}})\cos\theta_\ell$. Here $v$ is the velocity of the lepton as seen from the dilepton rest frame, so the boost applied is $\Lambda(-v)$ relative to the general matrix above; this accounts for the sign of the off-diagonal terms in what follows. Hence the boost matrix from $\ell^{-}$ rest frame to the dilepton rest frame is 
 \begin{center}
   $\begin{pmatrix}
   \frac{\sqrt{q^{2}}}{2m_{\ell}} & \frac{|q_{2\ell}|}{m_{\ell}}\sin\theta_{\ell} & 0 &  \frac{|q_{2\ell}|}{m_{\ell}}\cos\theta_{\ell} \\
    \frac{|q_{2\ell}|}{m_{\ell}}\sin\theta_{\ell} &  \frac{\sqrt{q^{2}}}{2m_{\ell}}\sin^{2}\theta_{\ell}+\cos^{2}\theta_{\ell} & 0 & ( \frac{\sqrt{q^{2}}}{2m_{\ell}}-1 )\sin\theta_{\ell}\cos\theta_{\ell} \\
    0 & 0 & 1 & 0 \\
   \frac{|q_{2\ell}|}{m_{\ell}}\cos\theta_{\ell} &  ( \frac{\sqrt{q^{2}}}{2m_{\ell}}-1 )\sin\theta_{\ell}\cos\theta_{\ell} & 0 &  \frac{\sqrt{q^{2}}}{2m_{\ell}}\cos^{2}\theta_{\ell}+\sin^{2}\theta_{\ell}
   \end{pmatrix}$ \\
   \end{center}

For the positively charged lepton:
In the $\ell^{+}$ rest frame the spin vectors are
     \begin{eqnarray}
       s_{L}^{+\mu} &=& [0,-\sin\theta _{\ell},0,-\cos \theta_{\ell}] \\
     s_{T}^{+\mu} &=& [0,-\cos\theta_{\ell},0,\sin\theta_{\ell}]  \\
   s_{N}^{+\mu} &=& [0,0,1,0] 
   \end{eqnarray}
The matrix required to boost the $s^{+\mu}_{L,N,T}$ from the lepton $\ell^{+}$ rest frame to the dilepton rest frame is 
\begin{center}
$\begin{pmatrix}
\frac{\sqrt{q^{2}}}{2m_{\ell}} & -\frac{|q_{2\ell}|}{m_{\ell}}\sin\theta_{\ell} & 0 &  -\frac{|q_{2\ell}|}{m_{\ell}}\cos\theta_{\ell} \\
    -\frac{|q_{2\ell}|}{m_{\ell}}\sin\theta_{\ell} &  \frac{\sqrt{q^{2}}}{2m_{\ell}}\sin^{2}\theta_{\ell}+\cos^{2}\theta_{\ell} & 0 & ( \frac{\sqrt{q^{2}}}{2m_{\ell}}-1 )\sin\theta_{\ell}\cos\theta_{\ell} \\
    0 & 0 & 1 & 0 \\
   -\frac{|q_{2\ell}|}{m_{\ell}}\cos\theta_{\ell} &  ( \frac{\sqrt{q^{2}}}{2m_{\ell}}-1 )\sin\theta_{\ell}\cos\theta_{\ell} & 0 &  \frac{\sqrt{q^{2}}}{2m_{\ell}}\cos^{2}\theta_{\ell}+\sin^{2}\theta_{\ell}
   \end{pmatrix}$ \\
   \end{center}
%

%

\end{document}